\documentclass[twoside,12pt]{article}
\usepackage{epsfig}
\usepackage{epsfig,graphicx,float,pst-all}
\usepackage[utf8]{inputenc}
\usepackage[caption = false]{subfig}
\usepackage{cite}
\usepackage{authblk}
\usepackage{bm}
\usepackage{multirow}
\usepackage{booktabs}
\usepackage{orcidlink}
\usepackage{threeparttable}
\usepackage{lineno,hyperref}
\usepackage[normalem]{ulem}
\usepackage{devanagari}
\usepackage{amsmath}
\usepackage{braket}
\usepackage{footnote}
\usepackage{float}
\newcommand{\be}{\begin{equation}}
\newcommand{\ee}{\end{equation}}
\newcommand{\bea}{\begin{eqnarray}}
\newcommand{\eea}{\end{eqnarray}}

\newcommand{\nucl}[2]{^{#1}\mathrm{#2}}

\begin{document} 
\newcommand{\gs}[1]{\textcolor{red}{#1}}
\newcommand{\gsout}[1]{\textcolor{red}{\sout{#1}}}
\newcommand{\rcc}[1]{\textcolor{violet}{\it #1}}
\newcommand{\rcout}[1]{\textcolor{violet}{\sout{#1}}}
\newcommand{\js}[1]{\textcolor{blue}{#1}}
\newcommand{\jsout}[1]{\textcolor{blue}{\sout{#1}}}
\newcommand{\sh}[1]{\textcolor{purple}{#1}}
\newcommand{\shout}[1]{\textcolor{green}{\sout{#1}}}
\newcommand{\RB}[1]{\textcolor{magenta}{#1}}
\newcommand{\RBout}[1]
{\textcolor{magenta}{\sout{#1}}}
\newcommand{\WH}[1]{\textcolor{teal}{#1}}
\title{ \vspace{1cm} Deformation, halo, and bubble structure:
        A paradigm shift of exotic phenomena in light to medium mass nuclei}
        
%\author{R. Chatterjee$^{1}$,  W. Horiuchi$^{2,3,4,5}$, and M. Kimura$^{4,5}$} 
\author[1,2]{R. Barman}
\author[1]{R. Chatterjee}
\author[3,4,2]{W. Horiuchi}
\author[2]{ M. Kimura}
\author[6]{G. Singh}
\author[7,8,9]{Jagjit Singh}
\author[10]{Shubhchintak}
\affil[1]{ \it Department of Physics, Indian Institute of Technology Roorkee, Roorkee 247 667, India }
\affil[2]{\it RIKEN Nishina Center, Wako 351-0198, Japan}
\affil[3]{\it Department of Physics, Osaka Metropolitan University, Osaka 558-8585, Japan}
\affil[4]{\it Nambu Yoichiro Institute of Theoretical and Experimental Physics (NITEP) , Osaka Metropolitan University, Osaka 558-8585, Japan}
%\affil[5]{\it  Department of Physics, Hokkaido University, Sapporo 060-0810, Japan}
\affil[6]{\it National Centre for Nuclear Research, ul. Andrzeja So\l tana 7, 05-400 Otwock, Poland.}
\affil[7]{\it Department of Physics, University of Manchester, Manchester M13 9PL, UK}
\affil[8]{\it Department of Physics, Akal University, Talwandi Sabo, Bathinda, Punjab 151302, India}
\affil[9]{\it Research Centre for Nuclear Physics (RCNP), Osaka University, Ibaraki 567-0047, Japan}
\affil[10]{\it Department of Physics, Dr. B. R. Ambedkar National Institute of Technology, Jalandhar, Punjab 144008, India}

\maketitle

\begin{abstract} 
The emergence of exotic nuclear structures, such as deformation, one- and two-neutron halos, and bubble configurations, marks a paradigm shift in our understanding of light- to medium-mass nuclei far from stability, particularly near and within the island of inversion extending across
$N=20-28$. In this review, we integrate microscopic structure calculations using the antisymmetrized molecular dynamics method with reaction theories such as the Glauber model for high-energy collisions, and highlight the use of the fully quantum mechanical finite-range distorted wave Born approximation for calculating both inclusive and exclusive Coulomb breakup observables for these medium mass systems. These theoretical frameworks enable precise probing of nuclear density profiles through observables such as total reaction cross sections, neutron removal cross sections, relative energy spectra, parallel momentum distributions, and angular distributions. Applications to several nuclei in the island of inversion reveal enhanced halo extensions, neutron-neutron correlations in Borromean nuclei, and central density depletions in bubbles, challenging traditional shell-model paradigms. Furthermore, the sensitivity of astrophysical reaction rates to these exotic inputs is explored, demonstrating their role in the refinement of r-process nucleosynthesis models and elemental abundance predictions. This unified approach not only bridges nuclear structure and reactions, but also highlights the driplines as frontiers for unraveling nuclear matter under extreme conditions, with implications for rare-isotope beam experiments and beyond.
\end{abstract}

%PACS numbers: 24.10.-i, 25.60.-t, 24.10.Eq, 25.60.Gc \\
%KEYWORDS: 
%\newpage
\tableofcontents

\section{Introduction}
At the heart of every atom lies the nucleus, a minuscule, ultra-dense region where protons and neutrons (nucleons) are held by the strong nuclear force. These nuclei form the building blocks of all matter, from the stars that light our skies to the elements that compose our bodies, the air we breathe, and even the materials embedded in everyday life, such as cell phones. Most of this matter we encounter consists of stable nuclei, which are combinations of nucleons that can stay together (bound) for billions of years. Despite their ubiquity and central role in nature, stable nuclei occupy only a narrow strip on the present day nuclear Segrè chart, comprising a handful ($\sim250$-$300$) of known isotopes. This slender region of stability is flanked by vast areas of instability on both the neutron and proton rich sides \cite{HBNP2023,Nazarewicz2025,Filomena2021}. As nucleons are added to a given nucleus (say, for a fixed proton and neutron number), the resulting nuclei eventually reach a limit beyond which they cannot remain bound. These boundaries, known as the neutron and proton \textit{driplines}\footnote{Driplines are the loci of points where the nucleon separation energy goes to zero.}, mark the edges of nuclear existence \cite{Erler2012} beyond which the extra nucleons begin to {drip} away, much like excess water spilling from a brimming glass. The nuclear landscape is notably asymmetric: the region between the line of stability and the driplines is much larger on the neutron rich side than on the proton rich side due to the absence of Coulomb repulsion among the neutrons \cite{Adamian2020}. %\jsout{Additionally, shell effects, akin to electron shells in atoms, create regions of enhanced stability at specific magic numbers of protons or neutrons. These effects shape the structure of exotic nuclei and influence the contours of nuclear stability.} \jsout{as they define the true borders of nuclear existence and offer \gs{insights} into matter under extreme conditions.}}

Mapping the driplines, however, remains a major challenge in nuclear physics. Despite decades of experimental advances, the neutron dripline has only been firmly established up to Neon ($Z = 10$) \cite{Ahn2019, Ahn2022}, leaving much of the neutron-rich nuclear chart unexplored. Yet, in this exotic frontier, around $6700$ isotopes are theoretically predicted, but fewer than half have been observed. Many of these are highly unstable, decaying in fractions of a second, and are produced either in extreme astrophysical environments or at advanced rare-isotope beam (RIB) facilities on Earth \cite{Balantekin2014,Crawford2024, Brown2025_FRIB, Ye2025}. Investigating such ephemeral nuclei aligns with the fundamental questions outlined in the 2024 NuPECC long-range plan \cite{NuPECC2025} and national roadmaps from India \cite{MSV2035-NP}, Japan \cite{RIKENupgrade}, UK \cite{STFC2024UKNP}, {Canada \cite{CanadaLRP}}, and elsewhere, such as what are the ultimate limits of nuclear existence? How does nuclear structure evolve far from stability? Which mechanisms govern the emergence of new structural phenomena at the edges?

In an effort to answer some of these questions, a significant number of experimental as well as theoretical studies have been undertaken near the driplines. They have revealed striking deviations from patterns seen in stable nuclei, including the erosion of magic numbers \cite{Dobaczewski1994, Sorlin2008}, shape coexistence \cite{Heyde2011,Nishibata2019}, changes in spin–orbit splittings \cite{Kay2017,Chen2025}, and abrupt limits of binding \cite{Kahlbow2024}. These observations underscore the complex interplay between core-nucleon interactions, nucleon-nucleon interactions, weak binding effects, and the role of the continuum. The resultant exotic systems, such as nuclear halos (where one or two nucleons orbit far from a compact core) \cite{HJ87,Jensen04RMP} and bubbles (nuclei with depleted central densities) \cite{Mutschler2017NatPhys, Duguet17PRC, Perera2022PRC}, offer true testing grounds in our quest to understand how matter organizes itself at the edge of stability, as well as its behavior under extreme conditions.
\begin{figure}[htbp]
\centering
%\resizebox{\textwidth}{!}{\includegraphics{NuclearChart.pdf}}
\makebox[\textwidth]{\includegraphics[width=1.0\columnwidth]{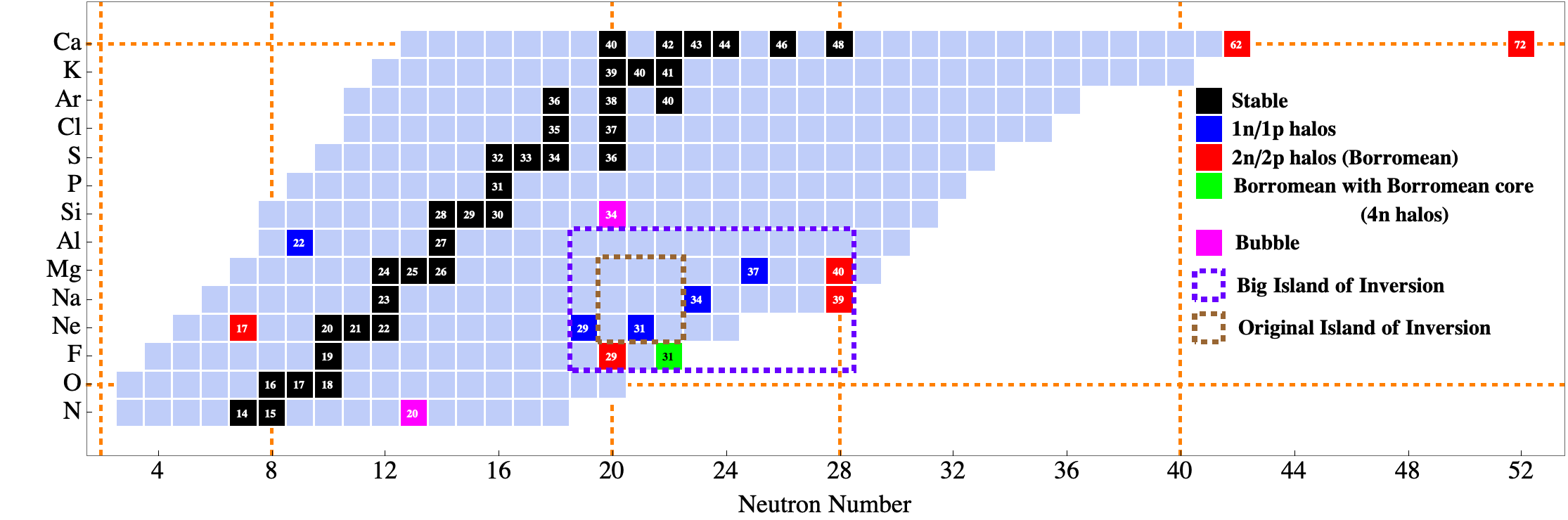}}
\caption{A selected area of the nuclear Segr\`e chart for the $7 \le Z \le 20$ isotopes. The orange lines correspond to traditional proton and neutron magic numbers at $2$, $8$, $20$, $28$, and $40$. Black, blue, red, green, and magenta squares mark the stable, one-neutron/proton halos, two-neutron/proton {Borromean} halos, Borromean with Borromean core (four-neutron halos), and bubble systems, respectively. The original and modern versions of the `island of inversion' in the $N$ = 20 - 28 region are also marked. For further details, please see the text.} 
\label{Fig1.0a}
\end{figure}
These halos and bubbles have been observed or predicted in various light to medium-mass nuclei. A bird's eye view of the picture depicting these phenomena from Nitrogen to Calcium isotopes is shown in Fig. \ref{Fig1.0a}. In this selected portion of the Segrè chart, the color of each square indicates the type of nucleus: stable nuclei are displayed in black, while their unstable siblings are in light blue. Further, one-neutron/proton halos (dark blue), two-neutron/proton {Borromean} halos (dark red), Borromean nuclei with a Borromean core (green), and bubble systems (magenta) are also highlighted. Horizontal and vertical (orange) dashed lines mark the proton and neutron magic numbers $2$, $8$, $20$, $28$, and $40$. Special attention is given to the island of inversion (IoI), spotlighted with brown and purple outlines. 

The IoI, a term coined by Warburton, Becker, and Brown in 1990 \cite{Warburton90}, designates regions of the nuclear chart where the expected ordering of shell-model orbitals is inverted and the traditional magic numbers lose their stabilising effect. The shell structure in nuclei was thought to be universal, until anomalies in mass measurements of neutron-rich Na isotopes near \( N=20 \) challenged this view~\cite{thibault_1975,Huber1978,Guillemaud1984}. Hartree-Fock calculations attributed these deviations to deformation driven by the occupation of the intruding $f$-orbital~\cite{CAMPI1975193}. Further evidence came from the low-lying first \( 2^+ \) state in \( ^{32}\mathrm{Mg} \), indicating strong collectivity~\cite{detraz_1979}. All these led to the idea of the island of inversion, and was first identified in nine neutron-rich Ne, Na, and Mg isotopes around $N\simeq20$ \cite{Warburton90}. Essentially, as one moves away from the line of stability, changes in the relative energies of single-particle orbits, driven by the evolving balance of neutron–proton interactions, reduce the shell gaps between $sd$- and $pf$-shells. When the gap becomes small enough, orbitals from the $pf$-shell drop below those of the $sd$-shell, favoring cross-shell particle-hole excitations that dominate the ground state configurations. This inversion produces strongly deformed nuclei in place of the spherical structures expected near the closed shells \cite{Caurier1998, CaurierRMP2005,OtsukaRMP2020}.

\begin{figure}[htbp]
\centering
\includegraphics[width=0.9\linewidth]{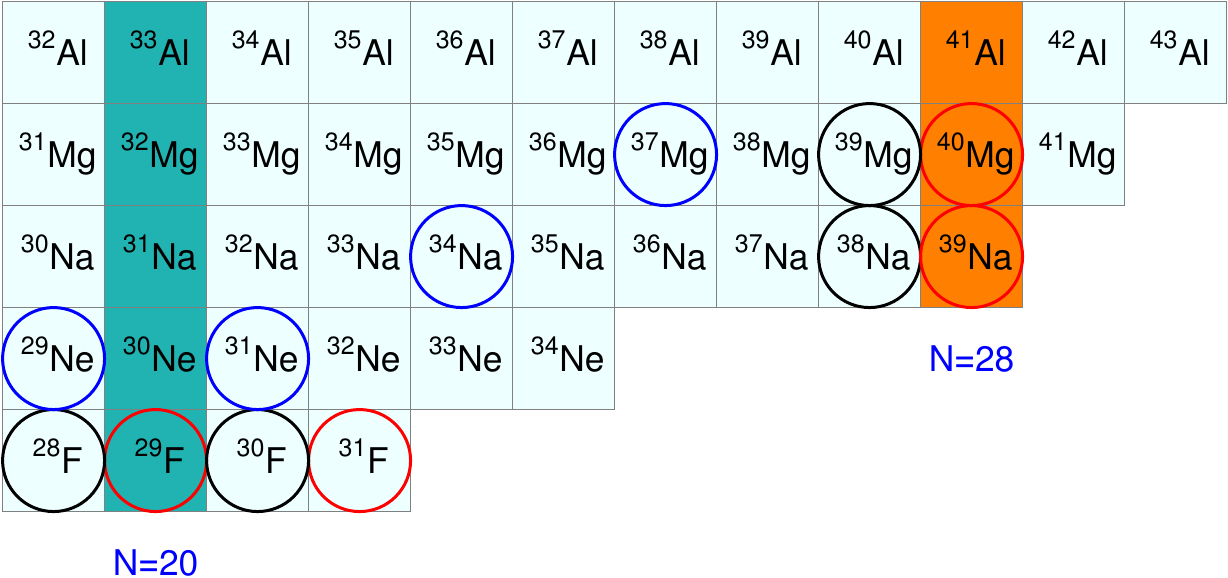}
\caption{A selected area of the nuclear Segr\`e chart for the $9 \le Z \le 13$ isotopes. The green and orange blocks correspond to neutron magic numbers $N=20$ and $28$, respectively. Black, blue, and red circles mark the unbound, one-neutron halo and two-neutron {Borromean} halo systems, respectively.} 
\label{Fig1.0}
\end{figure}

Since then, the IoI has become a key pilot zone for modern shell-model interactions and the broader concept of shell evolution away from the valley of stability. Over the past three decades, experimental and theoretical work has demonstrated that the original $N=20$ IoI extends and merges with the $N=28$ region, leading to the concept of a Big island of inversion (B-IoI) \cite{Caurier2014,Doornenbal2013}. Concentrating on the N=20 and N=28 regions, as well as the transitional nuclei between them, is therefore crucial for understanding how shell gaps collapse, why magic numbers disappear in exotic nuclei, and how advanced shell-model interactions capture the mechanisms of deformation and inversion far from stability. Thus, Fig.~\ref{Fig1.0} presents an enlarged segment of the nuclear Segr\`e chart, emphasizing the B-IoI, which constitutes the primary subject of this review.

%The few-body results based on various reaction theories guided by few-body and many-body structure models covering a wide range of observables for several one-neutron halo nuclei (indicated by blue circles) and two-neutron halo/Borromean nuclei (indicated by red circles) will be extensively reviewed. 
All nuclei typically need to be studied from three main perspectives: i) their structure and composition; ii) their reaction dynamics; and iii) their astrophysical significance. Consequently, this article aims to examine the phenomenon of halo formation as well as phenomena such as bubble structure formation in medium-mass nuclei. Furthermore, we intend to demonstrate how nuclear reaction theories, when integrated with microscopic structure models, can effectively investigate nuclear density profiles. The behavior of various reaction observables for nuclei lying within the island of inversion is also explored using both few-body and many-body structure models. Finally, the consequences of the inclusion of exotic structure in nuclear reaction rates, relevant for nuclear astrophysics, are considered.
%This review concentrates on few-body results derived from various reaction theories, bolstered by both few-body and many-body structure models, and encompasses a broad spectrum of observables for numerous one-neutron halo and Borromean nuclei.
The write-up is organized as follows. Section 2 contains a brief review of some structure and reaction models that have been used to study nuclei within or near the IoI. We primarily focus on the microscopic antisymmetrized molecular dynamics (AMD) method, followed by a description of the Glauber model and the fully quantal theory of the finite-range distorted wave Born approximation as applied to Coulomb breakup.
The applications of these theories to exotic nuclei in the medium-mass region are described in Section 3. Further applications, especially in sourcing accurate nuclear physics inputs for astrophysics, form the basis of Section 4, followed by conclusions and our perspectives of the field in Section 5. 

\section{Structure and reaction theories}
\subsection{Formalism of antisymmetrized molecular dynamics (AMD) method}\label{subsection:AMD}
The antisymmetrized molecular dynamics (AMD) method was originally developed for heavy-ion collisions, and was later extended to nuclear structure studies~\cite{Ono_1992, ono1_1992, Enyo_2001, Enyo_2003, Enyo_2012}. In AMD, nucleons are described as independent Gaussian wave packets forming a Slater determinant, without any \textit{a priori} assumptions. This allows the description of cluster correlations, shell-model-like configurations, and exotic deformations within a unified approach. Combining the AMD with the generator coordinate method (GCM) enables the incorporation of configuration mixing, thus allowing a comprehensive treatment of ground and excited states.
To discuss the formalism, we start with the microscopic Hamiltonian for an A-nucleon system,
 \begin{align}
   \hat{H} = \sum_{i=1}^{A} \hat{T}_i - \hat{T}_{\rm cm} + \sum_{i<j}^{A} \hat{V}^{NN}_{ij} + \sum_{i<j}^{A} \hat{V}^{\rm Coul}_{ij}.
 \end{align}
 Here, $\hat{T}_i$ is the single-particle kinetic energy, $\hat{T}_{\rm cm}$ is the energy of the centre of mass motion which can be exactly removed, $\hat{V}^{NN}_{ij}$ is the effective two-body interaction with Gogny D1S parameterization~\cite{BERGER_1991}, and $\hat{V}^{\rm Coul}_{ij}$ is the Coulomb interaction. The intrinsic AMD wave function is given by the antisymmetrized product of the single-particle wave packets,
  \begin{equation}
   \Phi_{\rm int} = \mathcal{A} \{\varphi_1, \varphi_2,..., \varphi_A\},
  \end{equation}
with,
\begin{equation}
    \varphi_i(\bm r) = \phi_i(\bm r)\otimes \chi_i \otimes \tau_i.
\end{equation}
Here, $\phi_i(\bm r)$ are the spatial wave functions of the Gaussian form, while $\chi_i$, and $\tau_i$ are the spin, and isospin wave functions, respectively, and are given by the following expressions:
 \begin{align}
 \phi_i(\bm r)&=\exp\left\{-\sum_{\sigma=x,y,z}\nu_{\sigma}\left(r_{\sigma}-\frac{Z_{i\sigma}}{\sqrt{\nu_{\sigma}}}\right)^2\right\},\\
 \chi_i &= a_i \chi_{\uparrow} + b_i \chi_{\downarrow},\\
 \tau_i &= \mathrm{proton\ or\ neutron}, 
 \end{align}
 where the Gaussian centroid $Z_{i\sigma}$, the spin direction parameters $a_i$ and
 $b_i$, and the width parameters $\nu_\sigma$ are the variational parameters. The intrinsic wave function $\Phi_{int}$ is projected to the eigenstates of parity $\pi$ to obtain $\Phi^{\pi}=\hat{P}^{\pi}\Phi_{\rm int}$, where $\hat{P}^{\pi}= (1\pm \hat{P}_r)/2$ is the parity projection operator, with $\hat{P}_r$ being the space-inversion operator. The energy variation is then done by the frictional cooling method~\cite{Enyo_2001}, where the variational parameters are optimized by minimizing the energy of the system. This is usually performed with a constraint condition. In most cases, the constraint is applied over the quadrupole deformation parameters, $\beta$ and $\gamma$. For the constraint over the parameter $\beta$, the minimum energy is achieved by the relation,
\begin{equation}
\Tilde{E} = \frac{\braket{\Phi^{\pi}|\hat{H}|\Phi^{\pi}}}{\left<\Phi^{\pi}|\Phi^{\pi}\right>} + v_{\beta}\left(\left<\hat{\beta}\right>-\beta\right)^2.  
\end{equation}
The parameter $\gamma$ takes an optimal value after energy variation. Thus, the optimized wave function, $\Phi^{\pi}(\beta)$ is obtained, which has a minimum energy for a given $\beta$.
 The wave function $\Phi^{\pi}(\beta)$ is projected to the eigenstate of the angular momentum,
 \begin{align}
 \Phi_{MK}^{J\pi}(\beta) = \int d\Omega D^{J*}_{MK}(\Omega) \hat{R}(\Omega)\Phi^{\pi}(\beta),
 \end{align}
where, $\Omega=\{u, v, w\}$ are the Euler angles, while $D^{J*}_{MK}(\Omega)$ and $\hat{R}(\Omega)$, are Wigner D-function and the rotation operator, respectively.
 
Then we employ GCM, where the angular momentum projected wave functions  $\Phi_{MK}^{J\pi}(\beta)$ with same $J^{\pi}$ but different $K$ and $\beta$ are superposed, which gives,
 \begin{align}\label{eq:9}
 \Psi_{Mn}^{J\pi} = \sum_{i}\sum_{K}g_{iKn} \Phi_{MK}^{J\pi}(\beta_i).
 \end{align}
In the above equation, the quantum numbers other than the total angular momentum and parity are denoted by $n$. The coefficients $g_{iKn}$ and the eigen-energies are determined by solving Hill-Wheeler equation \cite{hw1953}
 \begin{align}
  \sum_{i'}\sum_{K'} H_{iKi'K'}g_{i'K'n} = E_n \sum_{i'}\sum_{K'} N_{iKi'K'} g_{i'K'n},
 \end{align}
where the matrix elements of the Hamiltonian and norm are given by
\begin{align}
   H_{iKi'K'}&= \langle\Phi^{J\pi}_{MK}(\beta_{i})|\hat{H}|\Phi^{J\pi}_{MK'}(\beta_{i'})\rangle,\\
   N_{iKi'K'}&= \langle\Phi^{J\pi}_{MK}(\beta_{i})|\Phi^{J\pi}_{MK'}(\beta_{i'})\rangle.
\end{align}
The wave function obtained from Eq. \ref{eq:9} is used for the calculation of various physical properties, such as excitation spectra, density distributions and $B(E2)$ transition strengths, which are discussed in Sec. \ref{sec:3.1}.

\subsection{Cross section calculations within the Glauber reaction model}
\label{Glauber.sec}

The total reaction cross section ($\sigma_R$) is a fundamental observable in nuclear physics, providing insights into the size and structure of exotic nuclei, which quantifies the total probability of all non-elastic processes occurring when a projectile collides with a target. In case of an unstable or exotic projectile impinging on a stable target (such as $^{12}$C or $^{208}$Pb), an enhanced $\sigma_R$ provides a compelling signature of a halo structure in the projectile.
%\js{It is expressed as the sum of the interaction cross section ($\sigma_I$), which includes all processes that alter the identity of the projectile (\textit{i.e.}, changes in its neutron number (N) or proton number (Z) or both), and the inelastic scattering cross section ($\sigma_{inel}$), which involves excitations of the projectile nucleus to particle-bound excited states, given by 
%\begin{equation}
%    \sigma_R=\sigma_I+\sigma_{inel}.
%\end{equation}} 
%\js{ At sufficiently high beam energies (such as greater than 200 MeV/A in this review), experiments with light unstable nuclei have shown that $\sigma_{inel}$ is typically very small, often less than 1 millibarn (mb), making it negligible compared to $\sigma_I$, which can be of the order of 1 barn (b). Therefore, at high incident energies, the reaction cross section is often approximated by the interaction cross section ($\sigma_R \approx \sigma_I$) \cite{Kohama2008, Kanungo2023}.}
Typically, $\sigma_R$ is expressed as the sum of the interaction cross section ($\sigma_I$) and the inelastic scattering cross section ($\sigma_{inel}$). The $\sigma_I$ includes all processes that alter the identity of the projectile (\textit{i.e.}, changes in its neutron number ($N$) or proton number ($Z$) or both), while $\sigma_{inel}$ corresponds to processes which involve excitations of the projectile nucleus to particle-bound excited states. At sufficiently high beam energies (such as greater than 200 MeV/nucleon), experiments with light unstable nuclei have shown that $\sigma_{inel}$ is typically small, making it negligible compared to $\sigma_I$. Therefore, at high incident energies, the reaction cross section is often approximated by the interaction cross section ($\sigma_R \approx \sigma_I$) \cite{Kohama2008, Kanungo2023}.                                    

On the experimental side, the interaction cross section ($\sigma_I$) is typically measured using the \textit{transmission technique}, a straightforward and effective method that quantifies the attenuation of an unstable projectile as it passes through a target \cite{Tanihata1985, Tanihata1988, Tanihata2013, Kanungo2023}. The technique boasts a long and successful history, beginning with the groundbreaking observation of the first {two-neutron} Borromean halo nucleus, $^{11}$Li, in 1985 \cite{Tanihata1985}, and extending to the recent identification of the heaviest Borromean halo, $^{29}$F, in 2020 \cite{Bagchi2020}. For a historical overview of how things started experimentally and the current status of the field, see Ref. \cite{Tanihata2016HIS}.

From a theoretical frame of reference, the formalism used to extract matter radii from these measured cross sections and the theoretical estimates of the $\sigma_R$ from the density distributions \cite{Masui2020, JSingh2024} is primarily based on the Glauber reaction model \cite{Glauber}. The Glauber model~\cite{Glauber} is a robust microscopic multiple scattering theory that is widely used to study high energy nucleus-nucleus collisions. There are three main approximations that exist in the literature for evaluating phase shifts within the Glauber framework. These are the optical limit approximation \cite{Ozawa2001}, the nucleon-target Glauber approximation \cite{Abu2000, Horiuchi2007, Horiuchi2007E}, and the few-body Glauber approximation \cite{Ogawa1992, Jim1996, Tostevin1999}. In this review, we will briefly detail the recipe of the nucleon-target formalism with nucleon-nucleon profile function \cite{Ibrahim08, Ibrahim2009, Ibrahim09, Ibrahim10}. This approach has proven effective in various high energy nucleus-nucleus collision reactions, particularly those with unstable nuclei, and it successfully replicates the isotope dependence of the total reaction cross sections with appropriate density distributions \cite{Horiuchi10, Horiuchi2006,Horiuchi2007,Horiuchi2007E,Horiuchi12,Naga2018}. {This framework is further supported by early systematic Glauber model studies by Andrea Vitturi and collaborators \cite{LenziVitturi1989, LenziVitturi1990}, which investigated intermediate-energy nucleus-nucleus reaction cross sections and their sensitivity to nuclear density distributions.}
%In addition, the work by Andrea Vitturi and collaborators \cite{LenziVitturi1989, LenziVitturi1990} demonstrated that the Glauber framework can successfully describe elastic, inelastic, and coupled-channel heavy-ion reactions over a broad range of projectile energies.}% Their studies also clarified important issues related to multiple-scattering effects, convergence of the expansion, and center-of-mass correlations, thereby providing valuable guidance on how the standard approximations may be extended beyond their simplest formulation.}

We consider a high energy collision between a projectile nucleus ($P$) and a target nucleus ($T$), with mass numbers $A_P$ and $A_T$, respectively. 
By applying the adiabatic and eikonal approximations, the final state wave function of the combined projectile-target system, $\Phi_f$, can be significantly simplified. It is expressed as a product of the ground state (g.s.) wave functions of the projectile ($\Phi_0^P$) and the target ($\Phi_0^T$), along with a product of phase-shift functions of a nucleon-nucleon collision, $e^{i\chi_{NN}}$, as
\begin{align}
  \left|\Phi_f\right>=\exp
\left[i\sum_{j=1}^{A_P}\sum_{k=1}^{A_T}\chi_{NN}(\bm{b}+\hat{\bm{s}}_j^P-\hat{\bm{s}}_k^T)\right]
\left|\Phi_0^P\Phi_0^T\right>.
\end{align}
Here, $\bm{b}$ is the impact parameter vector perpendicular to the beam direction $\hat{z}$ and $\hat{\bm{s}}_j^P$ and $\hat{\bm{s}}_k^T$ represent the two-dimensional
single-particle coordinate operator projected onto the $xy$-plane of the $j$th projectile and $k$th target nucleons relative to their respective centers of mass.

%With the adiabatic and eikonal approximations, we only need to evaluate 
The optical phase-shift function,
%or the Glauber amplitude,
$e^{i\chi(\bm{b})}$, which includes all information of the elastic processes in the high energy nuclear collision, can be evaluated as
\begin{equation}
e^{i\chi(\bm{b})}=\left<\Phi_0^P\Phi_0^T\right|
\prod_{j=1}^{A_P}\prod_{k=1}^{A_T}
\left[1-\Gamma_{NN}(\bm{b}+\hat{\bm{s}}_j^P-\hat{\bm{s}}_k^T)\right]
\left|\Phi_0^P\Phi_0^T\right>,
\label{optphase.eq}
\end{equation}
where the profile function
$\Gamma_{NN}(\bm{b})=1-e^{i\chi_{NN}(\bm{b})}$ is introduced for the sake of convenience.

With this optical phase-shift function,
the total reaction cross section is evaluated 
by integrating the reaction probability, 
$1-|e^{i\chi(\bm{b})}|^2$, over $\bm{b}$ as
\begin{equation}
\sigma_R=\int d\bm{b}\, (1-|e^{i\chi(\bm{b})}|^2).
\label{sigr.eq}
\end{equation}
The elastic scattering differential cross section is given by
\begin{align}
\frac{d\sigma}{d\Omega}(\theta)=|F(\theta)|^2
\end{align}
with the scattering amplitude
\begin{align}
F(\theta)=F_C(\theta)+\frac{iK}{2\pi}\int d\bm{b} 
e^{-i\bm{q}\cdot\bm{b}+2i\eta\ln(Kb)}(1-e^{i\chi(\bm{b})}),
\end{align}
where $F_C$ is the Rutherford scattering amplitude, 
$K$ is the wave number in the relativistic kinematics,
and $\eta$ is the Sommerfeld parameter.

The profile function is usually parametrized as~\cite{Ray79},
\begin{equation}
  \Gamma_{NN}(\bm{b})=\frac{1-i\alpha_{NN}}{4\pi\beta_{NN}}
  \sigma_{NN}^{\rm tot}\exp\left[-\frac{\bm{b}^2}{2\beta_{NN}}\right],
\label{profile.eq}
\end{equation}
where $\sigma_{NN}^{\rm tot}$, 
$\alpha_{NN}$, and $\beta_{NN}$
are the total nucleon-nucleon ($NN$) cross section,
the ratio between the real and imaginary parts
of the scattering amplitude at the forward angles,
and the so-called slope parameter, respectively.
Parameter sets for various incident energies listed in Refs.~\cite{Ibrahim08,Ibrahim09,Ibrahim10}
for proton-proton ($pp$) and proton-neutron ($pn$) 
are employed for the calculations shown in this review. Charge independence is assumed, with $nn$ ($np$) taken to be the same as $pp$ ($pn$). 
We do not consider the Coulomb breakup contributions
since the effects are negligible in systems
involving small $Z$ nuclei~\cite{Horiuchi10, Horiuchi16}.
The other inputs to the theory are
the wave functions of the projectile and target nuclei.
The evaluation of Eq.~(\ref{optphase.eq}) includes multi-dimensional
integration, which is, in general, quite involved \cite{Varga2002, Naga2018,HatakeyamaNPA2019}.
{Very recently, extensive Glauber calculations have been reported in the recent preprints: Glauber-theory calculations of high-energy nuclear scattering observables using variational Monte Carlo wave functions \cite{Horiuchi_2025_b_arx} and Glauber-theory analysis of nuclear reactions on $^{12}$C target with variational Monte Carlo wave functions \cite{Horiuchi2025_a_arx}. These works examine in detail the range of validity of Glauber theory and provide a systematic discussion of the formal expressions employed in contemporary applications. For details of the various expressions, the reader is referred to the references cited above.}
%{\gs{Very recently, extensive Glauber calculations have been carried out, 
%and the validity of the theory has been examined in detail [*,**].
%*[2512.20095] Ab initio Glauber-theory calculations of high-energy nuclear scattering observables using variational Monte Carlo wave functions
%**[2512.20100] Glauber-theory analysis of nuclear reactions on 12C target with variational Monte Carlo wave functions
%For details of the various expressions appearing in this work, the reader is referred to these references.}}
%\WH{(Refs: K. Varga, S. C. Pieper, Y. Suzuki, and R. B.Wiringa, Phys. Rev. C 66, 034611 (2002)\cite{Naga2018}, S. Hatakeyama and W. Horiuchi, Nucl. Phys. A 985, 20 (2019).}

The optical limit approximation (OLA) has often been applied
as it only needs the density distributions of the projectile and target~\cite{Glauber,SuzukiCRC2003}. 
%\WH{Please cite Y. Suzuki, R. G. Lovas, K. Yabana, and K. Varga, Structure and reactions of light exotic nuclei (Taylor \& Francis, London, 2003) as well.)} 
The OLA works well for nucleus-proton scattering.
The approximated phase-shift function is given by
\begin{align}
e^{i\chi_{\rm{OLA}}(\bm{b})}=\exp\left[-\int d\bm{b}\,\rho_P(\bm{r})\Gamma_{NN}(\bm{b}+\bm{s})\right],
\end{align}
where $\rho_P$ is the one-body density of the projectile.
For nucleus-nucleus scattering, it is efficient to use the truncation based on 
the nucleon target profile function (called NTG)~\cite{Abu2000}:

\begin{align}
e^{i\chi_{\rm{NTG}}(\bm{b})}=\exp\left\{-\int d\bm{b}\,\rho_P(\bm{r})
\left[1-\exp\left(-\int d\bm{r}^\prime \rho_T(\bm{r}^\prime)\Gamma_{NN}(\bm{b}+\bm{s}-\bm{s}^\prime)\right)\right]\right\}.
\end{align}
The NTG approximation includes the multiple scattering effects between the projectile and target nucleons explicitly. Note that the symmetric expression of the projectile and target is used in actual calculations and the expression only requires the one-body density distributions of
the projectile and target ($\rho_T$) nuclei.
Once these inputs are set, the theory has no adjustable parameter.
The validity of this approach has already been confirmed
in a number of examples~\cite{Ibrahim2009, Horiuchi10, Horiuchi12, Horiuchi14, Horiuchi16, Horiuchi17, Naga2018,Horiuchi_JPS_2015} 
%\WH{(Please also cite W. Horiuchi, T. Inakura, T. Nakatsukasa, and Y. Suzuki, JPS Conf. Proc. 6, 030079 (2015).)} 
and it has become a standard method to extract nuclear size properties \cite{Kanungo_35Mg_2011,Kanungo_23O_2011,Estrade2014,Kanungo_C_2016,Bagchi_PLB_2019,Bagchi2020,Kaur2022}. 

\subsection{Finite range distorted wave Born approximation theory}\label{frdwba}
{Having discussed a structure model and a reaction approach, it is now time to delve into the workings of a reaction theory that can, in principle, utilise a microscopic approach. The finite range distorted wave Born approximation (FRDWBA) theory used in the post form is an ideal candidate that uses the microscopic structure information of the projectile in its reaction formalism, thereby highlighting the nuances in nuclear properties for different nuclear systems.}

We contemplate that a beam of a weakly bound projectile `$a$' impinges on a stable, heavy target `$t$', and under its strong Coulomb influence breaks up into a core `$b$' and a valence nucleon `$c$', i.e., the following reaction occurs:
\textit{a} + \textit{t} $\longrightarrow$ \textit{b} + \textit{c} + \textit{t}. 
%$^{34}$Na + $^{208}$Pb $\longrightarrow$ $^{33}$Na + n + $^{208}$Pb.\\

\begin{figure}[htbp] 
\centering
\includegraphics[trim={5cm 5cm 5cm 5cm},clip,width=0.75\linewidth]{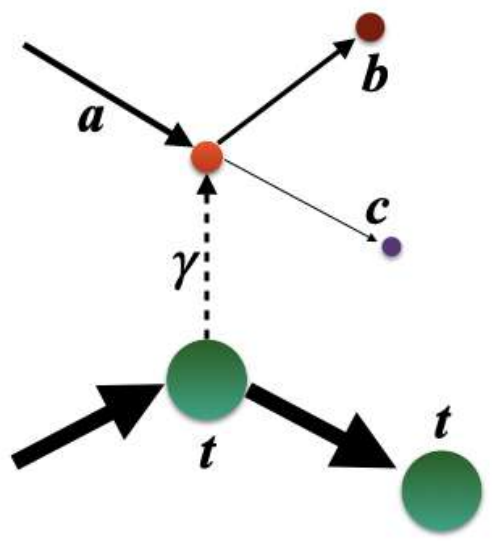}
\caption{\label{fig: CD} Schematic representation of an elastic Coulomb dissociation where the dynamic {Coulomb field} of a heavy target `$t$' breaks up a weakly bound projectile `$a$' (=$b$+$c$) into a core `$b$' and a valence nucleon `$c$', while the target itself remains in its ground state.}
\end{figure}

\begin{figure}[htbp]
\centering
\includegraphics[width=11.0cm]{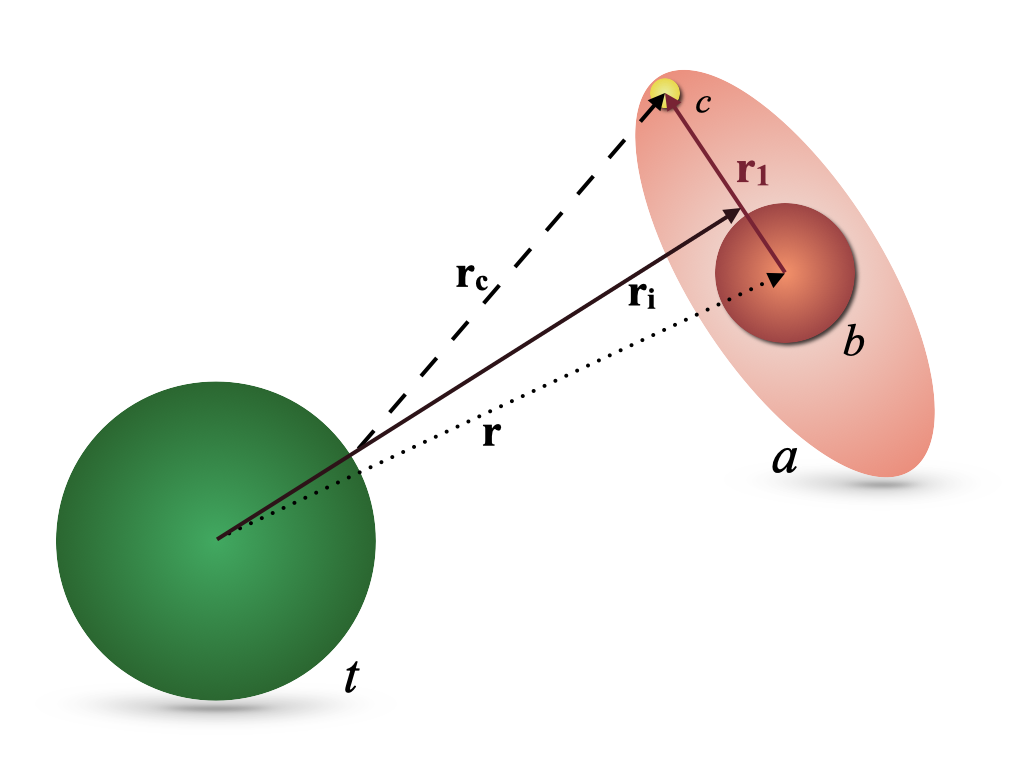}
\caption{\label{fig: Jacobi} The three-body Jacobi coordinate system for a `possibly' deformed projectile `$a$' impinging on a target `$t$'.}
\end{figure}

{Fig. \ref{fig: CD} presents a schematic representation of such an \textit{elastic breakup} process}, which is one where the target remains in its g.s. Using the \textit{post form} FRDWBA theory we formulate a plan to compute the various reaction observables that can highlight the structure of the projectile as well as calculate its neutron capture reaction rates to study its astrophysical applications and relevance.
The first step in the process is to evaluate the triple differential cross section, which can then be integrated over the different relevant variables to obtain the desired observables \cite{Chatterjee2018}. For this purpose, we employ a Jacobi coordinate system depicted in Fig. \ref{fig: Jacobi}. The Jacobi coordinate system is useful for simplifying mathematical description of many body problems and is based on binary geometric correlations, where $P$ bodies are connected to each other via $P-1$ translationally invariant coordinate vectors and a center of mass coordinate. %In the three-body system under consideration, for example, vector $\vec{r_i}$ (shown by the solid black line in the figure) connects the center of the target to the center of mass of the projectile. Similarly, the dashed line connects the valence nucleon to the center of mass of the two-body projectile target system. The core and the valence nucleon are connected within the projectile by $\vec{r_1}$ as depicted by the brown solid line. \gs{Add a description of the Jacob coordinate system too.}

%first calculate the triple differential cross section and then integrate it to find different reaction observables for the breakup mentioned above. 

Observing Fig. \ref{fig: Jacobi}, it is rather trivial to verify that the position vectors \textbf{r$_{1}$}, \textbf{r$_{i}$}, \textbf{r$_{c}$} and \textbf{r} satisfy
\begin{eqnarray}
{\bf{r}}={\bf{r}}_{i}-{\bar{\alpha}}{\bf{r}}_{1} ;\hspace{0.2in} {\bf{r}}_{c}={\bar{\beta}}{\bf{r}}_{1}+{\bar{\gamma}}{\bf{r}}_{i}\label{a1},
\end{eqnarray}
with $\bar{\alpha}, \bar{\beta}$ and $\bar{\gamma}$ being the mass factors defined as,
\begin{eqnarray}
{\bar{\alpha}} =\frac{{m_{c}}}{m_{c}+m_{b}};\hspace{0.2in}  \bar{\gamma} =\frac{m_{t}}{m_{b}+m_{t}};\hspace{0.2in} \bar{\beta}=(1-\bar{\alpha}\bar{\gamma}),\label{massfac}
\end{eqnarray}
where, \textit{m$_{t}$}, \textit{m$_{b}$} and \textit{m$_{c}$} are the masses of the target \textit{t}, core \textit{b} and the valence nucleon \textit{c}, respectively.

Generically, the triple differential cross-section is defined as,
\begin{eqnarray}
\frac{d^3\sigma}{dE_{b}d\Omega_{b}d\Omega_{c}} = \frac{2\pi}{\hbar v_{at}}\rho{(E_{b},\Omega_{b},\Omega_{c})}\frac{1}{\hat{j_{a}^{2}}}\sum_{\mu_{a}\mu_{b}\mu_{c}}|T^{(+)}_{fi}|^{2},\label{tdcrs}
%\frac{d^3\sigma}{dE_{bc}d\Omega_{bc}d\Omega_{at}} &=& \frac{2\pi}{\hbar v_{at}}\frac{\mu_{bc}\mu_{at}p_{bc}p_{at}}{h^6}\nonumber\\
%&\times& \sum_{\ell m}\frac{1}{(2\ell+1)}|\beta_{\ell m}|^2
\end{eqnarray}
where $E_{b}$ is the energy of fragment \textit{b}, $\Omega$'s are the solid angles corresponding to fragments \textit{b} and \textit{c}, $v_{at}$ is the \textit{a-t} relative velocity in the initial channel, $j_{a}$ is the total spin of the projectile with $\hat{j_{a}^{2}} = (2j_a + 1)$. $\mu$'s are the projections of the angular momenta of the corresponding particle and $\rho{(E_{b},\Omega_{b},\Omega_{c})}$ is the three-body final state phase space factor \cite{Fuchs82NIM}. $T_{fi}^{(+)}$ is the transition or the $T$-matrix. {It is worth mentioning here briefly that before FRDWBA there were other, simpler approaches that were developed in the market to describe this $T$-matrix. Two of these, the zero-range approximation \cite{SatchlerBook83} and the alternative approximation by Baur and Trautmann \cite{Baur72NPA} deserve special mention. Details about these approaches can be found in a previous review \cite{Chatterjee2018}.} For the \textit{post form} in the FRDWBA, the $T$-matrix is defined in the integral form as \cite{Chatterjee2018}:
\begin{eqnarray}
T_{fi}^{(+)} = \int\int\int d\zeta d\textbf{r}_{1}d\textbf{r}_{i}\chi_{b}^{(-)*}(\textbf{q}_{b},\textbf{r})\phi_{b}^{*}(\zeta_{b})
\chi_{c}^{(-)*}(\textbf{q}_{c},\textbf{r}_{c})\phi_{c}^{*}(\zeta_{c}) V_{bc}(\textbf{r}_{1})
\phi_{a}^{lm}(\zeta_{a},\textbf{r}_{1})\chi_{a}^{(+)}(\textbf{q}_{a},\textbf{r}_{i}) \label{Tmat}
\end{eqnarray}
In Eq. (\ref{Tmat}), the $\chi$'s represent the pure Coulomb distorted waves with incoming [( - )] and outgoing [(+)] wave boundary conditions and $\textbf{q}_{a}, \textbf{q}_{b}$ and $\textbf{q}_{c}$ are the Jacobi momenta corresponding to vectors $\textbf{r}_{1}, \textbf{r}$ and $\textbf{r}_{c}$, respectively. It is assumed in DWBA that the inelastic excitations of the projectile are quite weak, and hence, need only be treated up to the first order. This enables the three-body scattering wave function in the final channel (in the post form) to be expressed as a product of $\phi_{a}^{lm}(\zeta_{a},\textbf{r}_{1})\chi_{a}^{(+)}(\textbf{q}_{a},\textbf{r}_{i})$, where $\phi_{a}^{\ell m}(\zeta_{a}, \textbf{r}_{1})$ is the ground state wave function of the projectile \textit{a}, that has an angular momentum $\ell$ (with projection \textit{m}) and internal coordinate $\zeta_{a}$. This definition ensures that three-body DWBA wave function has vanishing overlap with all the inelastic excitation channels of the projectile. Naturally, $\phi_{a}^{\ell m}(\zeta_{a}, \textbf{r}_{1})$ includes the radial and angular parts of the projectile wave function ($\phi_{a}^{\ell m} = u_{\ell}(r_{1})Y^{m}_{\ell}(\hat{\textbf{r}}_{1})$). $\zeta_{b}$ and $\zeta_{c}$ represent the internal degrees of freedom of systems $b$ and $c$.

Integration over internal coordinates ($\zeta$'s) \textit{reduces} the $T$-matrix, and yields,

%\begin{eqnarray}
%\frac{d\sigma}{dE_{bc}}&=&\int_{\Omega_{bc}\Omega_{at}}d\Omega_{bc}d\Omega_{at}
%\nonumber\\
%&\times& \frac{2\pi}{\hbar v_{at}}
%\frac{\mu_{bc}\mu_{at}p_{bc}p_{at}}{h^6}\sum_{\ell m}\frac{1}{(2\ell+1)}|\beta_{\ell m}|^2. \label{a4.3}
%\end{eqnarray}
%Then, the photodisintegration cross section ($\sigma_{\gamma, n}^{\pi\lambda}$) for the reaction $a+\gamma\rightarrow b+c$ can be related to the relative energy spectra as,
\begin{eqnarray}
T_{fi}^{(+)} = \sum_{\ell mj\mu}\left\langle \ell mj_{c}\mu_{c}|j\mu\right\rangle \left\langle j_{b}\mu_{b}j\mu|j_{a}\mu_{a}\right\rangle i^{\ell}\hat{\ell}\beta_{\ell m}.
\end{eqnarray}
The integration ensures that the norm is still preserved, such that 
\begin{equation}
|T^{(+)}_{fi}|^{2} = |\hat{\ell}\beta_{\ell m}|^{2},
\label{a4.3}
\end{equation}

where, $\hat{\ell} = \sqrt{(2\ell + 1)}$, and we define the \textit{reduced transition amplitude}, $\beta_{\ell m}$, as \cite{Chatterjee2018}:
\begin{eqnarray}
\hat{\ell}\beta_{\ell m} = \int\int d\textbf{r}_{1}d\textbf{r}_{i}\chi_{b}^{(-)*}(\textbf{q}_{b},\textbf{r})
\chi_{c}^{(-)*}(\textbf{q}_{c},\textbf{r}_{c})
 V_{bc}(\textbf{r}_{1})
\phi_{a}^{lm}(\textbf{r}_{1})\chi_{a}^{(+)}(\textbf{q}_{a},\textbf{r}_{i}).\label{eq: beta}
\end{eqnarray}

Let us take a moment to analyze this form of the $T$-matrix. The post-form $T$-matrix, by its very construction, has the explicit presence of the bound short-range fragment-fragment (bound state) potential, which goes a long way in accelerating the convergence of the $T$-matrix even though it may contain three distorted waves. On the other hand, the prior- or the alternate-prior form of the $T$-matrix explicitly contains the fragment-target potentials, which invariably involve more complicated coordinate transformations, although they may contain only two distorted waves. It is also worthwhile to note that with exclusive observables such as triple and double differential cross sections being calculated in the theory, in principle, all inclusive reaction observables such as total cross sections, relative energy spectra, momentum and angular distributions are easily calculated by multiplying the triple differential cross sections with an appropriate Jacobian \cite{Fuchs82NIM}. For more details, one is referred to Refs. \cite{Chatterjee2018,BAUR1984}.

Going back to Eq. (\ref{eq: beta}), a cursory look and it is easy to perceive that the integral is a cumbersome six-dimensional entity which does not easily converge. As alluded to above, part of the issue is caused by the presence of Coulomb distorted waves which are inherently long range and highly oscillatory in nature. One can circumvent the problem by Taylor expanding $\chi_{b}^{(-)*}(\textbf{q}_{b},\textbf{r})$ about \textbf{r$_i$} and replacing the `Del' operator by an \textit{effective local momentum}\footnote{The two things are analogous, but not exactly the same from a mathematics point of view.} under the `local momentum approximation' (LMA) \cite{Shyam85AP, Ban98PRC, Chatterjee2018}. This enables us to write

%From the neutron capture cross section $\sigma_{n, \gamma}(E_{bc})$, one can find the reaction rate per mole, $N_A\langle\sigma v\rangle$ ($N_A$ being Avogadro constant) as:
%\begin{eqnarray}
%N_A\langle\sigma v\rangle_{nr} &=& N_A \sqrt{\frac{8}{(k_BT)^{3} \pi\mu_{bc}}} \nonumber\\
%&\times& \int^\infty_0{\sigma_{n, \gamma}(E_{bc})\hspace{0.1cm} E_{bc} \hspace{0.1cm} exp(-\frac{E_{bc}}{k_BT})\hspace{0.1cm}dE_{bc}}, \label{a4.7}
%\end{eqnarray}
\begin{eqnarray}
\chi_{b}^{(-)*}(\textbf{q}_{b},\textbf{r}) = e^{\bar{\alpha}\nabla_{r_{i}}.\textbf{r}_{1}}\chi_{b}^{(-)*}(\textbf{q}_{b},\textbf{r}_{i})\nonumber\\ \xrightarrow{LMA}e^{i\bar{\alpha}\textbf{K}.\textbf{r}_{1}}\chi_{b}^{(-)*}(\textbf{q}_{b},\textbf{r}_{i}). \label{LMA}
\end{eqnarray}

Here, \textbf{K} is the local momentum of the core \textit{b}, the magnitude of which is given by
\begin{equation}
K = \sqrt{\frac{2m_{bt}}{\hbar^{2}}(E_{bt} - V(R))} \label{a4.7}
\end{equation}
where $m_{bt}$ is the reduced \textit{b-t} mass and $E_{bt}$ is the relative energy of the \textit{b-t} system. The Coulomb potential between \textit{b} and \textit{t} at a distance \textit{R} is represented by \textit{V(R)}. Note that this Coulomb potential warrants the evaluation of \textbf{K} at some finite distance $R$ and that its magnitude be fixed for all values of \textbf{r}. It has been shown that the magnitude of \textbf{K} remains effectively constant for all \textbf{r} $>$ 10\,fm \cite{Chatterjee00NPA,Chatterjee2018,Neelam15PRC}. For peripheral reactions, which are a dominant mode of the reaction mechanism especially for weakly bound nuclei, this is the region that contributes the most to the cross section. Thus, a constant magnitude for \textbf{K}, evaluated at $R$ = 10\,fm, is taken for all the values of \textbf{r}. In fact, the results are seen to be independent of the direction in which \textbf{K} is taken \cite{Chatterjee00NPA}, and therefore, it is prudent to consider it in the same direction as that of $\textbf{q}_{b}$. Note that here we are applying the LMA to the distorted wave of the core. It is also possible to apply this approximation to the Coulomb distorted wave of the projectile itself \cite{Ban98PRC}, however, in that case the results seem to not be fully independent on the direction of \textbf{K}. 

In case the particle \textit{c} is a neutron, $\chi_{c}^{(-)*}(\textbf{q}_{c},\textbf{r}_{c})$ is further replaced by a plane wave ($e^{-i\textbf{q}_{c}.\textbf{r}_{c}}$) since then there is no Coulomb interaction between the outgoing neutron \textit{c} and the target \textit{t}. This actually makes life simpler, as then the above integral converges faster than if $c$ were a proton\footnote{The case of a proton valence nucleon or a proton halo breakup offers significant differences from a neutron breakup primarily due to the Coulomb field. Some of these have been discussed in Refs. \cite{Bonac2004, Shubh2013}.}.

The application of LMA and replacement of the distorted wave for particle $c$ by a plane wave allows us to then factorize the intimidating reduced transition amplitude of Eq. (\ref{eq: beta}) into two parts: one, the dynamics part and two, the structure part,

\begin{eqnarray}
\hat{\ell}\beta_{\ell m} = \int d\textbf{r}_{i}e^{-i\delta\textbf{q}_{c}.\textbf{r}_{i}} \chi_{b}^{(-)*}(\textbf{q}_{b},\textbf{r}_{i})\chi_{a}^{(+)}(\textbf{q}_{a},\textbf{r}_{i}) 
\int d\textbf{r}_{1}e^{-i\textbf{W}.\textbf{r}_{1}}V_{bc}(\textbf{r}_{1})\phi_{a}^{lm}(\textbf{r}_{1}). \label{eq:betafac}
\end{eqnarray} 
Here, $\textbf{W} = \gamma\textbf{q}_{c} - \alpha\textbf{K}$ as the plane wave for the neutron is surrogated by $e^{-i\textbf{q}_{c}.(\gamma\textbf{r}_{1} + \delta\textbf{r}_{i})}$ using the defined relations for position vectors. As advocated by the nomenclature used, the first integral in Eq. (\ref{eq:betafac}) controls the dynamics part in the breakup reaction and can be, with convenience, expressed analytically in terms of the famous Bremsstrahlung integral \cite{Bremsstrahlung}. The second integral in Eq. (\ref{eq:betafac}) holds information about the structure of the projectile and therefore, about any effects that deformation (if present) might produce. This separation of the reduced transition matrix integral into two parts is very advantageous, as with any changes in the shape of the nucleus only the structure part gets modified, while the dynamics part of the reduced transition amplitude remains unaffected.

Eq.~\eqref{eq: beta} now contains structure information of the projectile via the potential. It is through this potential that deformation enters our theory. We consider the axially symmetric quadrupole-deformed potential $V_{bc}$ of the form \cite{Hama04PRC},

\begin{eqnarray}
V_{bc}(\textbf{r}_{1}) = V_{s}(r_{1}) - \beta_{2}V_{ws}
R\left[  \frac{d{g(r_{1})}}{dr_{1}} \right] Y_{2}^{0}(\hat{\textbf{r}}_{1}), \label{eq:WSPot}
\end{eqnarray}
where $\beta_{2}$ is the quadrupole deformation parameter and \textit{g}(r$_{1}$) = $\left[{1 + exp(\frac{r_{1} - R}{d})}\right]^{-1}$. In defining $g(r_1)$, we have used potential radius $R = r_{0}A^{1/3}$; $r_{0}$ and \textit{d} being the radius and diffuseness parameters, respectively, while \textit{A} is the mass number of the projectile. \textit{V$_{s}$}(r$_{1}$) $=\textit{V$_{ws}$}\textit{g}(r_{1})$ in the above equation, where $V_{ws}$ is the Woods-Saxon (WS) potential depth and represents the spherical part of the potential. Notice that the potential does not contain any spin-orbit term. From a mathematical perspective, Eq. (\ref{eq:WSPot}) is essentially the pruned Taylor expansion of the potential $V_{bc}(\textbf{r}_{1})$, in which we have neglected the spin-orbit term \cite{BOHR}. In its present form, the WS potential described via Eq. \ref{eq:WSPot} has just two dependent parameters apart from $\beta_2$, i.e., $r_0$ and $d$, which can be adjusted according to the nucleus chosen for study. 

%\gs{*********Modify below this}

A deformed potential also affects the mixing of the wave functions. Thus, the radial component of the g.s. wave function, i.e., $u_{\ell}(r_{1})$ should ideally be obtained from coupled equations which result due to the presence of deformation in the axially symmetric potential \cite{BOHR}:

\begin{eqnarray}
\left\{ \frac{d^2}{dr_1^2} -\frac{\ell(\ell+1)}{r_1^2} +\frac{2\mu}{\hbar^2}
[E - V_{s}(r_1)]\right\} u_{\ell m}(r_1) = 
\frac{2\mu}{\hbar^2}\sum_{\ell^\prime} \langle Y_{\ell}^m(\hat {\bf r}_1)|- \beta_2 k(r_1) 
Y^{0}_{2}(\hat {\bf r}_1)|Y_{\ell^\prime}^m(\hat {\bf r}_1)\rangle u_{\ell^\prime m}(r_1) \label{radial}.
~
\end{eqnarray}
{Eq.~(\ref{radial}) represents the general coupled-channel description for a deformed single-particle potential.} For a given $\ell$ value, these will have an admixture of other $\ell$ components of matching parity. {In a fully self-consistent treatment, one should solve the coupled equations and use the resulting deformed intrinsic wave functions containing admixtures of different $\ell$ components. Such intrinsic states are analogous to Nilsson-type states and, in the absence of Coriolis coupling, are characterized by the projection quantum number on the symmetry axis rather than by good total angular momentum.}
In other words, this would mean that positive parity levels like \textit{s}-states can have admixtures from other \textit{s}- and \textit{d}-, \textit{g}-, etc., orbitals while the negative parity levels like the \textit{p}-states can have mixing contributions from other \textit{p}- as well as \textit{f}-, \textit{h}-, levels. {This admixture of other similar parity orbitals renders the orbital angular momentum $\ell$ to no longer be a good quantum number.} %{\color{red} This admixture of orbitals with the same parity implies that the orbital angular momentum quantum number $\ell$ is no longer a good quantum number.}\js{BUT IT MEANS THE SAME, WHY CHANGE?}

However, it is known that for weak components or contributions of higher $\ell$ states, the admixtured or mixed states will be ruled by pure states of the lowest $\ell$ values \cite{Hama04PRC, Misu97NPA}. In fact, it is clearly verified in Ref. \cite{Hama04PRC} that for very low binding energy of the valence neutron ($S_{n}$ tending to zero), the lowest $\ell$ component dominates in the neutron orbits of a realistic deformed potential, and this is independent of the magnitude of deformation that may be present. Thus, if the nucleus is very weakly bound, as is the case for most of the nuclear systems near the drip line, even in the presence of deformation, the majority contribution to the mixed states would be by $s$-states in case of positive parity and $p$-state in case of negative parity components. {For example, with $\ell$ = 1 (\textit{p}-wave), all the Nilsson orbits with $\Omega^{\pi} = 1/2^{-}$ and $3/2^{-}$ become {progressively less sensitive to deformation} (and also pair correlation) as the levels approach the continuum, which is reached very soon in drip line nuclei.} %\sout {This, in principle, means that the summation in Eq. \ref{radial} vanishes and one is left with a single second-order differential equation that can be easily solved.}
{This implies that, for very weakly bound systems, the lowest-$\ell$ component strongly dominates the asymptotic behavior of the wave function, while the admixtures from higher-$\ell$ components remain comparatively small. In such cases, one may approximately retain only the dominant weakly bound component in Eq.~(\ref{radial}) for practical calculations.}

%\sout{Thus, low separation energy for medium mass nuclei lying in the island of inversion gives us a leeway to make this not too unreasonable approximation of deriving the radial wave functions from a spherical WS potential for a single $\ell$ and then use it in conjunction with an axially symmetric quadrupole deformed WS potential for further calculations} \cite{Shubh14NPA, GSingh16PRC, Shubh15NPA} \sout{(see also, Sec. \ref{wfn})}. 
{Thus, the low separation energies of medium-mass nuclei lying in the island of inversion provide a reasonable basis for approximating the projectile ground-state wave function by its dominant weakly bound component obtained from a spherical WS potential for a single $\ell$ value, which is subsequently used together with an axially symmetric quadrupole deformed WS potential in the reaction calculations \cite{Shubh14NPA, GSingh16PRC, Shubh15NPA}. Consequently, the present approach does not incorporate static deformation effects in the projectile structure in a fully self-consistent manner, since the admixtures generated by Eq.~(\ref{radial}) are not explicitly included in the final bound-state wave function. The deformation effects enter primarily through the interaction potential and therefore correspond mainly to deformation-induced dynamical couplings during the reaction process.}

{The present framework therefore differs from particle-rotor or XCDCC-type approaches \cite{Diego2014PRC,Pesudo2017PRL,MORO2020} in which both the projectile structure and reaction dynamics consistently include explicit core excitations and construct projectile states with good total angular momentum. Nevertheless, for halo systems with extremely small separation energies, the breakup observables are expected to be governed predominantly by the asymptotic tail of the lowest-$\ell$ component, which motivates the approximation adopted here.} %\js{REPOSITION, MOVE TO THE END OF THE SECTION?!}

We should also emphasize that the FRDWBA theory can seamlessly incorporate microscopic structure inputs like the quark-meson coupling (QMC) model \cite{Chatterjee2013NPA} or the AMD approach \cite{Dan2021, Choudhary2024EPJ}. It thus, provides a convenient way to study the impact of these structure inputs to various reaction observables.

A note is in order here. The form of FRDWBA discussed above is only for a pure Coulomb breakup reaction. The inclusion of the nuclear contribution as well the Coulomb-nuclear interference terms requires the use of optical potentials. More details about the FRDWBA theory can be found in Refs. \cite{Chatterjee2018, GSingh16PRC,Shubh14NPA,Chatterjee00NPA,Shubh15NPA} among others, and about the description of Coulomb-nuclear interference effects in Refs. {\cite{Nunes1998,Garrido2000,Tarutina2004,Hussein2006,JSingh2021,Mukeru2022,MARGUERON2002,DCH2017}}. {One must also note that reaction applications with microscopic structure inputs to describe experiments have also been done in the past \cite{Feshbach1958AP,Satou2008PLB,Satchler1979PR,Gray1966PRC,Johnson1966PRC}, with Satchler and Love laying the foundations for double folding models using realistic potentials for heavy ion scattering at low energies \cite{Satchler1979PR}. Love and Franey \cite{Love1981PRC,Franey1985PRC} extended this approach to higher energies with implementations via the Glauber model using nuclear densities with realistic nucleon-nucleon interactions. While originally developed for nucleon-nucleus scattering applications, the use to light ion nucleus-nucleus scatterings through its generalised $T$-matrix is very common. Applications to heavier nuclei through the lens of microscopic DWBA, especially halos, have been more recent \cite{Satou2008PLB}. Lately, other microscopic and semi-microscopic methods merged with reaction frameworks, like using \textit{ab initio} overlaps for transfer reactions \cite{GSinghMoro,Xie2026PRC}, the Nilsson+AMD model \cite{Punta23PRC,Punta25PRC} or the halo-effective field theory \cite{Capel23EPJA} have also been used to describe halo nuclei.}

{It also demands that it be pointed out that there are several other methods in the literature for analyzing direct nuclear reactions, starting with the semi-classical approach of the Alder-Winther theory \cite{AlderWinther56RMP}. Works by Gerhard Baur and Carlos Bertulani in the eighties also deserve a special mention \cite{Ber88,Ber88b,Bert86NPA}, specifically as they led to the development of the virtual photon method for indirect applications to nuclear astrophysics \cite{Bert86NPA} (see also Section \ref{sec: astro}). The treatment of direct reactions at comparatively higher beam energies led to the evolution of eikonal treatments like the Coulomb corrected eikonal (CCE) method, the dynamical eikonal approximation (DEA) and the simplified DEA \cite{Baye2005PRL,Baye2009,Pinilla2012PRC,Goldstein2006PRC,Hebborn20PRC,Capel2005PRC}. The time dependent Schr\"{o}dinger equation (TDSE) is another post-form-like dynamical framework that allows the treatment of reactions in a non-perturbative regime \cite{Melezhik1999PRC,Melezhik2001PRC}. Since weakly bound nuclei require that the continuum be treated exactly, techniques have also been developed to discretise it and include the couplings within the continuum. The most famous of these is perhaps the continuum discretised coupled channels (CDCC) approach \cite{Rawitscher74PRC,AUSTERN1987,Kami86PTPS,Kawai86PTPSa,Yahiro86PTPS,Iseri86PTPS,Kawai86PTPSb,Sakuragi86PTPS,Yahiro2012PTEP,Moro12PRC,MORO2020,Hagino22PPNP,Nunes1999,Rasoanaivo89,moro06,DRUET2010,Descouvemont_2010,JSinghCDCC2021}, which works well also at lower beam energies, but is computationally expensive and involved. The continuum is usually discretized by binning, i.e., creating partitions of energy/momentum giving an average energy/momentum to the respective partition \cite{Rawitscher74PRC,Kawai86PTPSa,Nunes1999}, or by pseudostates where the Hamiltonian is represented approximately by a finite, square-integrable basis set \cite{Rasoanaivo89,Kawai86PTPSa,moro06,DRUET2010,Descouvemont_2010,Hagino22PPNP}. The supplement series prepared in the eighties by a Japanese group provides a nice overview of the CDCC method, including some applications to Li isotopes \cite{Kami86PTPS,Kawai86PTPSa,Yahiro86PTPS,Iseri86PTPS,Kawai86PTPSb,Sakuragi86PTPS}. Recently, core-excitations have also been incorporated within the CDCC \cite{Diego2014PRC} using the transformed harmonic oscillator (THO) basis as well as via an extension of the DWBA \cite{Moro2012PRL}. However, the dynamic core-excitations effects are known to have larger impact only for light targets. They are almost negligible for a heavier target where a standard single-particle approach is usually sufficient \cite{Pesudo2017PRL}.} %\gsout{While a treatment of core-excitation effects is desirable, for most weakly bound neutron rich systems, it is often sufficient to treat the core as inert since the Fermi surface for the protons is far below that of the neutrons}. 
{Other methods to tackle nuclear reactions from a theoretical viewpoint are also available and have been reasonably successful \cite{Sagawa2025EPJA,SHYAM1992,Shyam01PRC,Bertulani1992b,Bertulani1993PR,Esbensen1995,bertulani05}. For more comprehensive comparisons within the different reaction theories, one can see Refs. \cite{Chatterjee2018,Moro2025EPJA,Sagawa2025EPJA,Hagino22PPNP}, among others.}

{A critical difference among many of these theories is the form of the $T$-matrix. The form of the $T$-matrix discussed here, as mentioned above, is the \textit{post form} $T$-matrix, which is equivalent to the \textit{prior form}, but not equal to the \textit{alternate prior form} that is used in many of the other approaches. 
%While the \textit{post-prior} equivalence is well established, recent works suggest that the Ichimura-Austern-Vincent (IAV) \cite{Ichimura85PRC} prescription of the \textit{post form} presents a more suitable description of experimental data for inclusive breakup as well as transfer reactions \sh{I have doubt here}\cite{Lei2015PRC,Lei2018PRC,Potel2015PRC} than the \textit{prior form} of Udagawa-Tamura (UT) \cite{Udagawa80PRL,Li1984PRC} without the orthogonality term in the latter.
While the \textit{post-prior} equivalence is well established, to the best of our knowledge \cite{Chatterjee2018}, most calculations involving the CDCC and the eikonal trace their origin to the \textit{alternate prior form} $T$-matrix. However, the present case of the FRDWBA is with the \textit{post-form}. Therefore, any comparison for the approximation of the three-body wave function with the DWBA (FRDWBA), CDCC, is best done keeping the \textit{form} of the $T$-matrix the same. Indeed, this needs to be addressed in the future.}

After drawing a picture of the structure and reaction frameworks, the next section shows some applications of these approaches.

\section{Applications of structure and reaction frameworks}
\subsection{Deformation in medium mass nuclei}\label{sec:3.1}
%[Increase of radii, variation of diffuseness, and their impact on the total reaction cross section and diffractions.]

%\textbf{References addition - DONE}\\

The density distribution of nuclei has traditionally been investigated by systematic analysis of electron scattering {for stable nuclei}, and is known to have saturated internal densities and diffused nuclear surfaces. {However, electron scattering measurements for unstable nuclei remain experimentally challenging and are not yet routinely available.} Systematic measurements of total reaction (interaction) cross sections using high energy radioactive isotope beams have revealed various exotic nuclear structure characteristics of proton/neutron rich nuclei (see recent review \cite{Tanaka2024}). Measurements of elastic scattering cross sections of unstable nuclei have also become possible. These observables directly reflect the density distribution of nuclei and contain information on various nuclear structural parameters. Systematic measurements of total reaction cross sections and elastic scattering cross sections can be a probe of nuclear structure, and a new spectroscopic method based on related theoretical research results.
The nuclear deformation phenomenon is directly related to the change of density distribution on the nuclear surface. The surface density distribution of a deformed nucleus is diffused compared to the spherical system. %(and might result in exotic structures, such as bubble nuclei). 
As a result, the radius increases. Such deformation effects to the nuclear radius were established by comparison of the measured interaction cross sections \cite{Takechi2012, Takechi2014} and theoretical calculations \cite{Minomo2011, Minomo2012, Sumi2012, Horiuchi2012, Watanabe2014} for neutron rich Ne and Mg isotopes. The enhancement of the nuclear radius associated with large nuclear deformation is typically found in the island of inversion region about $N=20$, while also appearing in the $N=40$ region. The characteristic change of density distribution and cross section increase have been theoretically predicted \cite{Horiuchi2022b}. Deformation of a nucleus greatly affects the structure of the Fermi surface, and it is manifested by the change of the nuclear surface density distribution \cite{Choudhary2021, Horiuchi2021}. This is quantified as “diffuseness” of the nuclear surface density distribution. The nuclear diffuseness can be an observable physical quantity by measuring the forward diffraction peak of the elastic differential cross section using a proton target \cite{Hatakeyama2018}. Recently, a novel mechanism of the enlargement of the deformation region has been pointed out in Zr isotopes. The measurement of the total reaction cross section and the elastic differential cross section is desired to clarify the mechanism \cite{Horiuchi2023}.
Nuclear deformation phenomena are also related to cluster formation on the nuclear surface. In Ref. \cite{Horiuchi2022c, Okada2024}, competition between shell-like and cluster-like configurations has been discussed for short-lived titanium isotopes, $^{44,48,52}$Ti, which is relevant to astrophysical reactions. The difference in density distribution due to these different configurations appears significantly in the elastic scattering differential cross sections. The appearance of such cluster states is also seen in the density distribution of $^{12}$C, $^{16}$O \cite{Horiuchi2023b}, and $^{20}$Ne \cite{Yamaguchi2023}. Measurement of elastic scattering cross sections is desired for universal understanding of clustering phenomena on nuclear surfaces.

\begin{figure}[ht]
    \centering
    \includegraphics[width=0.5\linewidth]{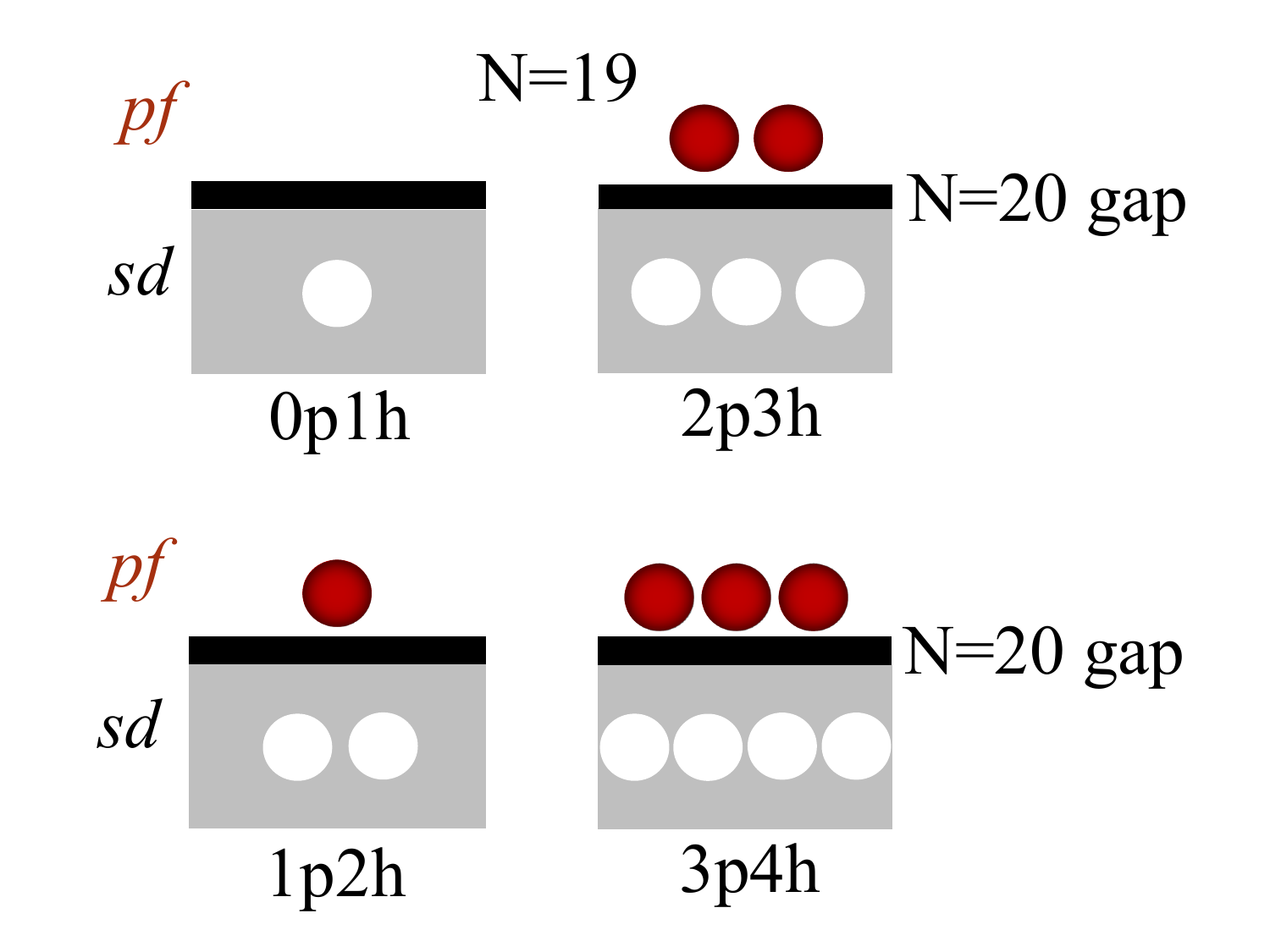}
    \caption{A schematic illustration of particle-hole configurations for N=19 nuclei.}
    \label{fig:p-h_config}
\end{figure}  

Nuclear deformation is a prominent feature in $N$=20 island of inversion. The intruder configurations result in strong deformation in these systems, generating different particle-hole ($m$p$n$h) configurations. As illustrated in Fig. \ref{fig:p-h_config}, a nucleus is said to have $m$p$n$h configuration if $m$ number of particles occupy the intruder $pf$ orbits, creating $n$ holes in the $sd$ shell. Various such configurations often coexist within small excitation energy, and the ground state often becomes difficult to predict.

\begin{figure}[htbp]
    \centering
    \includegraphics[width=\linewidth]{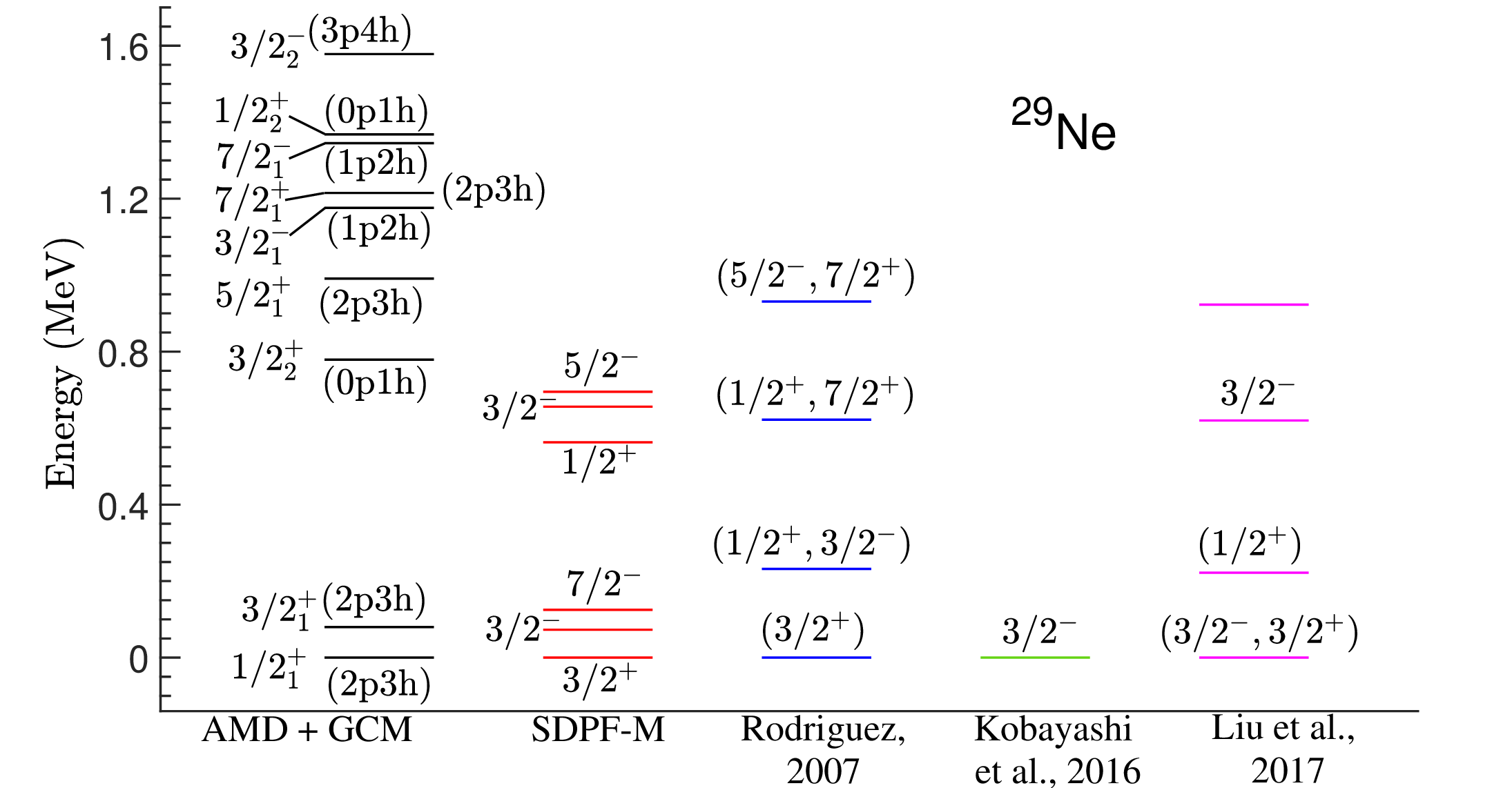}
    \caption{The excitation spectra of $^{29}$Ne. The experimental data are taken from Refs.~\cite{rodriguez_2007,Kobayashi16PRC, LIU_2017}.}
    \label{fig:spectra29Ne}
\end{figure}

To get a better view of the deformation and shape coexistence in these nuclei, we discuss the energy spectra of the few low-lying states of  $^{29}$Ne in Fig. \ref{fig:spectra29Ne} calculated using the AMD+GCM formalism. After the GCM calculation, each state is described by the mixing of different particle-hole configurations. However, we find there is small mixing between different configurations and a single configuration dominates, which are labeled for each state in Fig. \ref{fig:spectra29Ne}. The AMD calculations give a strongly deformed $1/2^+$ (2p3h) ground state, which arises due to the occupation of $[2, 0, 0, 1/2]$ Nilsson orbit emerging from the spherical $1d_{3/2}$ orbit, with a $3/2^+$ (2p3h) state lying just a few keV above the ground state~\cite{rbarman_2025}. The experiments propose the ground state to be either $3/2^+$ or $3/2^-$, with the two states lying very close to each other~\cite{rodriguez_2007, Kobayashi16PRC, LIU_2017}. This makes it one of the most complex and widely studied nuclei in this region.

The coexistence of these configurations can be confirmed by the $B$(E2) values for inter- and intra-band transitions, as shown in Table \ref{Ne29-BE2}. The largely deformed 2p3h configuration forms a $K^{\pi}=1/2^+$ rotational band with the band-head $1/2_1^+$ as the ground state, and $3/2_1^+$, $5/2_1^+$, and $7/2_1^+$ states at 0.1\,MeV, 1\,MeV, and 1.2\,MeV, respectively. This is evident from the relatively large $B$(E2) values between these states. In contrast, the smaller $B$(E2) values for transitions between states with different particle-hole configurations reinforce the presence of configuration coexistence.
\begin{table}[ht]
    \centering
\begin{tabular}{cc}
    \hline \hline
    $B(E2, J_i\rightarrow J_f)$ & Strength ($\mathrm{e^2} \mathrm{fm^4}$) \\
    \hline
    $B(E2, 3/2_{1}^{+}\rightarrow 1/2_{1}^{+})$ & 68.65  \\
    $B(E2, 3/2_{2}^{+}\rightarrow 1/2_{1}^{+})$ & 0.37  \\

    $B(E2, 5/2_{1}^{+}\rightarrow 1/2_{1}^{+})$ & 69.74 \\
    $B(E2, 5/2_{1}^{+}\rightarrow 3/2_{1}^{+})$ & 19.22 \\

    $B(E2, 7/2_{1}^{+}\rightarrow 3/2_{1}^{+})$ & 81.29 \\
    $B(E2, 7/2_{1}^{+}\rightarrow 5/2_{1}^{+})$ & 9.07 \\
    $B(E2, 7/2_{1}^{-}\rightarrow 3/2_{1}^{-})$ & 51.05 \\
    $B(E2, 7/2_{1}^{-}\rightarrow 3/2_{2}^{-})$ & 0.02 \\
    \hline
    
    \end{tabular}
    \caption{Electric quadrupole transition strengths for $^{29}$Ne.}
    \label{Ne29-BE2}
\end{table}

The nuclear surface density provides valuable insights into such complicated structures of nuclei. Surface properties such as radii and diffuseness reflect different particle-hole configurations and can be probed through total reaction and elastic scattering cross sections. These observables offer a useful means to explore ground-state configurations.
%\RBout{ To elucidate this, we study the low-lying states with different configurations in the Ne and Mg nuclei within the AMD framework. We particularly focus on the odd-mass systems since their valence particles or holes significantly influence the overall configuration. We explore how the particle-hole configurations affect their density properties and cross sections.} 
{To elucidate this, we employ AMD+GCM framework to investigate low-lying states with different particle-hole configurations and establish the relationship of their density profile with particle-hole configurations. Using the densities from AMD as input, we calculate the total reaction and proton-elastic scattering cross sections using the Glauber model to examine how these configurations manifest these cross sections. We particularly focus on the odd-mass systems since their valence particles or holes significantly influence the overall configuration.}

\renewcommand{\arraystretch}{1.3}
\begin{table}[ht!]
\centering
\begin{threeparttable}
\begin{tabular}{ccccccccccc}
\hline\hline
Nuclei & $J^{\pi}$ ($m$p$n$h) & $E_x$ & $ \beta$& $ \gamma$ &$r_m$ &$a_m$&$\sigma_R$&$\sigma_R$/$\sigma_I$ (exp.)\\
\hline
\multirow{4}{*}{$^{31}\rm Mg$} & $1/2_1^{+}$ ($2$p$3$h) & 0.0 & 0.45 & 1 & 3.29 & 0.63 & 1349 & \multirow{4}{*}{1329(11)\tnote{a}} \\ 
& $3/2_1^{-}$ ($3$p$4$h) & 0.66 & 0.50 &12& 3.34 & 0.66 & 1373 & \\
& $3/2_2^{-}$ ($1$p$2$h) & 0.89 & 0.37 &0& 3.21 & 0.59 & 1307 &\\ 
& $3/2_2^{+}$ ($0$p$1$h) & 1.35 & 0.17 &1& 3.17 & 0.53 & 1280 &\\
\hline
\multirow{5}{*}{$^{29}\rm Ne$} & $1/2_1^{+}$ ($2$p$3$h) & 0.0 & 0.45 & 0 & 3.27 & 0.65 & 1325 & \multirow{5}{*}{1344(13)\tnote{b}}\\ 
& $3/2_1^{+}$ ($2$p$3$h) & 0.08 & 0.42 & 0 & 3.27 & 0.65 & 1322 &\\ 
& $3/2_2^{+}$ ($0$p$1$h) & 0.78 & 0.22 & 33 & 3.16 & 0.54 & 1261 & \\
& $3/2_1^{-}$ ($1$p$2$h) & 1.18 & 0.30 & 0 & 3.20 & 0.59 & 1285 &\\
& $3/2_2^{-}$ ($3$p$4$h) & 1.58 & 0.52 & 0 & 3.32 & 0.68 & 1348 &\\
\hline
\multirow{3}{*}{$^{33}\rm Mg$} & $3/2_1^{-}$ ($3$p$2$h) & 0.0 & 0.40 &0&  3.36 & 0.64 & 1393 & \multirow{3}{*}{1399(12)\tnote{a}} \\
& $3/2_1^{+}$ ($4$p$3$h) & 1.15 & 0.52 &14& 3.41 & 0.68 & 1423 & \\
& $3/2_2^{+}$ ($2$p$1$h) & 1.78 & 0.32 &0& 3.31 & 0.60 & 1365 & \\ 
\hline
\multirow{3}{*}{$^{35}\rm Mg$} & $3/2_1^{+}$ ($4$p$1$h) & 0.0 & 0.40 &0& 3.40 & 0.63 & 1427 & \multirow{3}{*}{1443(12)\tnote{a}} \\ 
& $5/2_1^{-}$ ($5$p$2$h) & 0.63 & 0.45 &22& 3.44 & 0.66 & 1444 & \\
& $3/2_1^{-}$ ($3$p$0$h) & 0.79 & 0.32 &0& 3.36 & 0.60 & 1403 & &\\ \hline\hline
\end{tabular}

\begin{tablenotes}
\item[a] $\sigma_R$ from Ref.~\cite{Takechi_Mg_2014}.
\item[b] $\sigma_I$ from Ref.~\cite{takechi_2010}.
\end{tablenotes}
\end{threeparttable}
\caption{\label{table:structure} Properties of $^{29}$Ne and $^{31,33,35}$Mg. The columns represent spin-parities with particle-hole configurations, excitation energies (MeV), deformation parameters $\beta$ and $\gamma$ (degrees), matter radii (fm), diffuseness parameters (fm), and the calculated and experimental total reaction/interaction cross sections (mb) on a carbon target at 240 MeV/nucleon.}
\end{table}

\begin{figure}[htbp]
    \centering
    \includegraphics[width=\linewidth]{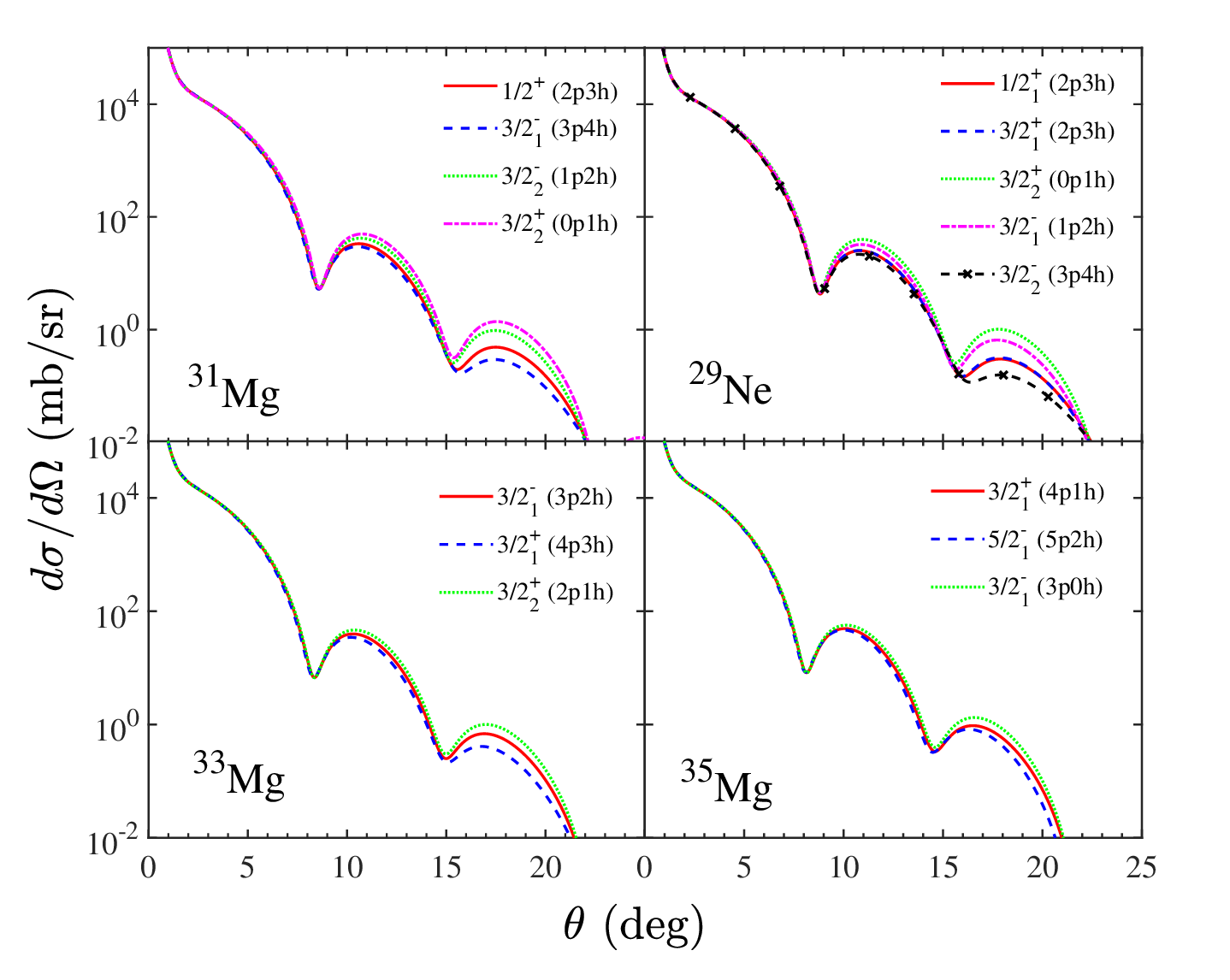}
    \caption{Angular distribution of proton-elastic scattering {cross sections at 800 MeV/nucleon} for the nuclei shown in Table \ref{table:structure}.}
    \label{fig:eldcs}
\end{figure}

Table \ref{table:structure} shows the structure properties of the low-lying states of odd-mass Ne and Mg nuclei in the island of inversion obtained from AMD calculations described in Sec. \ref{subsection:AMD}. {Here, the total reaction cross sections are calculated for carbon-nucleus scattering at 240 MeV/nucleon using the NTG approximation described in Sec. \ref{Glauber.sec}. The NTG approximation incorporates higher-order multiple-scattering effects beyond those included in the OLA, and therefore generally provides a more accurate description of cross sections. The results clearly describe the well-established relationship between deformation, root mean square (rms) matter radii, and the total reaction cross sections $(\sigma_R)$ for each nucleus, i.e., the radii and 
%\RBout{cross sections} 
$\sigma_R$ increase as deformation increases.} Let us now {see how our calculated $\sigma_R$ for each particle-hole configuration compares with the experimental data for these nuclei.}
%\RBout{the total reaction cross sections in detail}. 
We 
%\RBout{start the discussion with}
{first discuss} $^{31}$Mg, which has a well established ground state with  spin-parity, $J^{\pi}=1/2^+$ \cite{Kobayashi16PRC}. The AMD calculations could reasonably describe and reproduce the ground state configuration, and the calculated $\sigma_R$ is also in good agreement with the experimental value. For $^{29}$Ne, as discussed above, the ground state spin-parity is controversial. The total interaction cross section indicates the ground state to be $3/2^-$ (3p4h), which remains consistent even when a different interaction is chosen~\cite{rbarman_2025}. This implies that the $3/2^-$ ground state assignment may be correct for this nucleus.  For $^{33}$Mg, recent experiments show the ground state to be $3/2^-$. However, there have also been debates that it could be $3/2^+$~\cite{nummela_2001}. The AMD calculations agree with the $3/2^-$ assignment of the ground state and predict a 3p2h configuration. Moreover, this choice of the configuration also shows a good agreement with the observed total reaction cross section, further validating this assignment. As for $^{35}$Mg, there is not enough experimental data available for the ground state. One experiment based on observed gamma-ray transitions shows a $5/2^-$ ground state~\cite{gade2011}. The total reaction cross section is consistent with this $5/2^-$ (5p2h) state. However, $3/2_1^+$ (4p1h), which is the ground state in the AMD calculations, also gives a closer value to the observed data. Therefore, this analysis may not be sufficient to conclude the spin-parity of the ground state of $^{35}$Mg. 

Now, let us discuss the angular distribution of proton-elastic scattering cross section. There is a relation between the diffuseness and the cross section at the first diffraction peak. The cross section at the first peak is enhanced if the diffuseness is smaller~\cite{Hatakeyama2018}. This is well reflected in Fig. \ref{fig:eldcs} for the discussed nuclei. Therefore, if the cross section at the first peak is known, one can estimate the diffuseness, and hence, the particle-hole configuration. In this analysis, the cross sections for $^{31}$Mg, $^{29}$Ne, and $^{33}$Mg show significant differences at the first peak. For example, the $3/2_1^-$ (3p4h) state with the largest diffuseness gives a cross section 30\,mb/sr at the first peak, which is $60\%$ smaller than the $3/2_2^+$ (0p1h) state. However, for $^{35}$Mg, the differences in diffuseness are very small, and this nucleus may not be a good candidate for the spin-parity assignment by this method. 

{One may also question the experimental feasibility of determining particle-hole configurations using this method. At present, experimental data for unstable nuclei are not available. However, analyses of existing data for stable nuclei show that the cross sections at the first peak are of the same order of magnitude as those obtained in our calculations, with relatively small uncertainties~\cite{blanpied_PhysRevC.38.2180, Blanpied_PhysRevC.37.1987}. Therefore, if similar measurements can be performed for unstable nuclei, it is expected that different configurations can be distinguished based on the first peak of the proton elastic scattering cross section.}

The recent rapid increase of interaction cross sections in neutron rich calcium isotopes \cite{Tanaka2020} implied a new mechanism for the radius enhancement phenomena which cannot be explained by nuclear deformation. Ref. \cite{Horiuchi2020} showed that this phenomenon observed in calcium isotopes is closely related to the saturation of the central density of nuclei, and it has been shown to be a general phenomenon that can be observed in heavy nuclei such as Lead isotopes \cite{Horiuchi2022d}. This change in internal density is caused by neutron orbitals near the Fermi surface, and it is simultaneously reflected in the density distribution of the nuclear surface \cite{Horiuchi2021b}. By measuring the surface density characteristics by the proton elastic scattering cross section, one can confirm the radius increase phenomenon and deepen our understanding of the saturation of the nuclear density. 

Having analyzed the structural aspects of nuclei in the island of inversion, let us now examine some halo candidates in the region from a reactions perspective.

\subsection{Halos in the island of inversion}
%{In this sub-section, we explore one or two-neutron halo structures that have been observed or predicted in and around the island of inversion, along with their significance in nuclear astrophysics.}

A halo nucleus is one where one or more neutrons are very loosely bound and extend far beyond the rest of the nucleus, forming a kind of “halo” {around the core}. In these nuclei, long low density tails of loosely bound one or two nucleons have been observed. The concept of halo nuclei was first introduced by Hansen and Jonson \cite{HJ87}. This special feature was first observed in the neutron rich nucleus $^{11}$Li \cite{Tanihata1985}, where a large matter radius {was} found through the measurement of interaction cross-section, and later its two-neutron halo structure {was} confirmed. Since then, several one- and two-neutron halo nuclei have been identified in the low mass region. These nuclei generally have small one or two nucleon separation energies (typically, $S_n$, $S_{2n} \leq 1$ MeV). 
%\gsout{The valence nucleon in the case of halo nuclei has low angular momentum, so the hindrance due to the centrifugal barrier to the tunneling of the valence nucleon outside the classically allowed region can be avoided.} 
{However, a small separation energy is not the only prerequisite for halo formation.}
Apart from low separation energies, the valence nucleon(s) in halo nuclei should possess low orbital angular momentum, %\gsout{This is a sort of a requisite as then the hindrance to the tunneling of the valence nucleon(s) outside the classically allowed region due to the centrifugal barrier can be minimized. Indeed, For halo nuclei, the valence nucleon(s) seem(s) to},
typically having $\ell = 0$ or 1 \cite{FJR93, SAGAWA92, RJM92}. {This is because, for a given binding energy, a larger centrifugal (and also Coulomb) barrier results in a more compact wave function with smaller extensions outside the core or the classically allowed region, thus preventing the formation of the halo. This also hints to the rarity of proton halos than neutron halos.}

Most of the two-nucleon halos exhibit a Borromean structure, which refers to a three-body system where none of the two-body subsystems are bound, yet the complete three-body system remains bound~\cite{ZDF93}. In such nuclei, both the root mean square (rms) radius of a single halo nucleon ($r^{\text{rms}}_n$) and the rms separation between the two halo neutrons ($r^{\text{rms}}_{nn}$) are significantly larger than the typical range of nuclear forces. Halo nuclei are characterized by several distinctive experimental signatures, like unusually large reaction and interaction cross sections compared to neighboring isotopes, indicating extended matter distributions. Breakup experiments involving halo projectiles reveal narrow momentum distributions of the outgoing core fragments - a hallmark of weak binding and spatially extended valence nucleons - and enhanced Coulomb breakup cross sections. Increased low-energy dipole strength, known as the soft dipole mode, further supports the existence of diffused halo structures. Some recent studies report enhanced fusion cross-sections for halo nuclei compared to their neighboring isotopes \cite{ARB13, SXL23}. {On the other hand, some previous halo candidates have also been somewhat rejected as two-neutron halos because they were found to have a very substantial $d$-wave component in their ground states. Classic cases include $^{12}$Be, $^{16}$C and $^{20}$C, which were initially considered two-neutron halos, but based on experimental results and recent theoretical studies, were later denied entry into this club \cite{Tanihata2013,Choudhary2023NPA,Jones2023FP}. The pairing between the valence neutrons in these nuclei is strong enough to push them to higher $\ell$ orbitals, thus increasing the centrifugal barrier and reducing or killing off the halo character. Hence, it is all the more crucial that future experiments with RIBs be devised to validate theoretical predictions.} For more details on quantum halos including halo nuclei, one is referred to Refs. \cite{Jensen04RMP,Tanihata2013}. In Ref. \cite{Hammer2023} low-energy properties of halo nuclei are studied using effective field theory.

Over the past decade or {so}, there has been a growing interest in investigating the potential existence of such halo configurations, especially in the region of island of inversion near $N = 20, 28$ \cite{RD09,Chatterjee2013NPA}. Several nuclei have been suggested to have halo structures, however, only a few of these have been established while others await experimental confirmation. In subsections \ref{1n-halo} and \ref{2n-halo}, we focus on examples of one- and two-neutron {Borromean} halo configurations {respectively, that are} found in or near the island of inversion. Subsection \ref{bubble} then shifts attention to another class of exotic nuclear structures, known as {\it bubble nuclei}.

%\subsubsection{Coulomb breakup of 31Ne, 37Mg, 34Na by R. Chatterjee, Gagan, Shubh}\label{1n-halo}
\subsection{One-neutron halos}\label{1n-halo}\label{1n-halo}

%Also astrophysics implications of halo. \\

{Transitioning} from the light to the medium mass region, one-neutron halo structures have been observed in certain isotopes of neon, sodium, and magnesium. In particular, $^{29}$Ne, $^{31}$Ne, $^{34}$Na and $^{37}$Mg, lying at edge of or within the island of inversion, are suggested to exhibit one-neutron halo configurations, as illustrated in Fig. \ref{Fig1.0}. 
%We first concentrate on the Ne isotopes, which have been synthesized up to the drip-line nucleus of $^{34}$Ne \cite{NSA02,Ahn2019,Ahn2022}.
These exotic isotopes offer a valuable platform for theoretical {as well as experimental} investigations, as they help to explore proton-neutron and neutron-neutron interactions, along with many-body correlations near the drip lines. Several experiments \cite{YNS03,DST16,DSA09,Nakamura2009,Nakamura2014,TERRY06,Kobayashi16PRC} have confirmed the emergence of intruder configurations in and around this region. This is due to the inversion {of the} $sd$ and $pf$ shells {of the normal shell model. The $pf$ shell \textit{intrudes} between the $sd$ shell and this intrusion} indicates the deformed structure in the nuclei in this region \cite{Warburton90,ORR1991,DSA09}. The large $B({\rm E}2)$ values and low-lying first excited states also suggest a strong deformation in these nuclei \cite{detraz_1979,Motobayashi1995,Caurier1998, Utsuno1999,YNS03,IWASAKI2001,Gade2007,DSA09,YONEDA2001}. Several studies \cite{Warburton90,CAMPI1975193,Watt_1981,POVES1994} have emphasized that nuclear deformation may also contribute to the enhanced binding energies observed in certain nuclei within this region. A detailed investigation of nuclei close to the neutron drip line {was} conducted in Refs.~\cite{Hamamoto2007,Hamamoto2012}, employing a model that incorporates both spherical and deformed mean-field potentials to describe single-particle motion. These studies suggest that nuclei with neutron numbers in the range $N = 20\text{--}28$ are most likely to exhibit deformation.
Deformation in a nucleus typically results in an increased root mean square (RMS) radius compared to its spherical counterpart. Since the total reaction cross-section ($\sigma_R$) is sensitive to the RMS radii of both the projectile and the target, deformation tends to enhance $\sigma_R$. Experimental measurements revealed large interaction cross-sections---closely approximating $\sigma_R$---for the isotopes $^{29\text{--}32}$Ne~\cite{takechi2010,Takechi2012} and $^{24\text{--}38}$Mg~\cite{Takechi2014c}. These findings indicate a pronounced deformation in $^{29\text{--}32}$Ne~\cite{Minomo2012,Sumi2012,Horiuchi2012}. 

{The emergence of intruder configurations has already been established in neon isotopes~\cite{YNS03,DSA09,Nakamura2014,DST16,TERRY06,Kobayashi16PRC}. Specifically, the isotopes $^{29\text{--}32}$Ne lie at the edge of or well within the island of inversion} making them particularly intriguing candidates for detailed nuclear structure studies. Both experimental and theoretical analyses~\cite{Nakamura2014,Nakamura2009,Horiuchi10,Shubh14NPA,HBK17} have confirmed that $^{31}$Ne exhibits a halo structure accompanied by an intruder neutron configuration. In fact, $^{31}$Ne is the first known halo nucleus in which the ground state has a predominant $p$-wave ($\ell = 1$) neutron plus core configuration, {while also} possessing a supposedly deformed structure~\cite{Nakamura2014,Urata2011,Urata2012, HBK17,Hamamoto2010,Takechi2012}. 
Similar features have been observed in $^{29}$Ne and $^{37}$Mg, which have also been proposed as deformed halo nuclei (with moderate halo characteristics in the case of $^{29}$Ne), where $\ell = 1$ appears as the dominant component in the ground-state configuration~\cite{Kobayashi16PRC,Kobayashi2014,LI2022}. Regarding magnesium isotopes, the analyses in Refs.~\cite{Horiuchi2012,Watanabe2014} suggest that $^{27}$Mg and $^{30}$Mg are predominantly spherical, while $^{25}$Mg, $^{29}$Mg, and $^{33\text{--}38}$Mg are likely to possess deformed structures. These nuclei exhibit large breakup cross sections~\cite{Nakamura2009,Nakamura2014,Kobayashi2014,Kobayashi16PRC} and enhanced interaction/reaction cross sections~\cite{takechi2010,Takechi2012,Takechi2014,Takechi2014c} when compared to their neighboring isotopes.
%\sout{Additionally, small one-neutron separation energies further support the halo character of these nuclei: $0.29 \pm 1.64$~MeV~\cite{JURADO2007} or $0.17 \pm 0.13$~MeV~\cite{Wang2021} or $0.15^{+0.16}_{-0.10}$~MeV~\cite{Nakamura2014} for $^{31}$Ne, $0.96 \pm 0.14$~MeV for $^{29}$Ne~\cite{Wang2021,JURADO2007}, and $0.16 \pm 0.68$ MeV~\cite{Wang_2012} or $0.24\pm 0.11$~MeV~\cite{Wang2021} or $0.22^{+0.12}_{-0.09}$~MeV~\cite{Kobayashi2014} for $^{37}$Mg.} 
%\sout{Furthermore, narrow momentum distributions of the core in Coulomb breakup experiments~\cite{Nakamura2014,Kobayashi2014,Kobayashi16PRC,Shubh15NPA,Mahesh2016} strengthen the evidence for halo structures in these isotopes. Another potential candidate in this region is $^{34}$Na, which has also been suggested to exhibit a one-neutron halo structure~\cite{Gaudefroy12PRL,GSingh16PRC}. It features a small neutron separation energy of $0.17 \pm 0.50$~MeV~\cite{Gaudefroy12PRL} and a ground state with a predominant $\ell = 1$ neutron plus core configuration~\cite{GSingh16PRC}, although experimental confirmation is still pending.}

{Additionally, the small one-neutron separation energies of these nuclei, summarized in Table~\ref{sn_values}, further support their halo character. In particular, nuclei such as $^{31}$Ne, $^{29}$Ne, and $^{37}$Mg exhibit very weak binding, which is a key requisite for halo systems. Furthermore, narrow momentum distributions of the core in Coulomb breakup ~\cite{Nakamura2014,Kobayashi2014,Kobayashi16PRC,Shubh15NPA,Mahesh2016} strengthen the evidence for halo structures in these isotopes. Another potential candidate in this region is $^{34}$Na, which has also been suggested to exhibit a one-neutron halo structure~\cite{Gaudefroy12PRL,GSingh16PRC}. Its small neutron separation energy (see Table~\ref{sn_values}) and a ground state with a predominant $\ell = 1$ neutron plus core configuration~\cite{GSingh16PRC} make it a promising halo candidate, although experimental confirmation is still pending. Experiments conducted very recently at the facility of rare isotope beams (RIB) at Michigan State University for the Coulomb break up of $^{34}$Na and $^{37}$Mg should soon help uncover some of their mysteries \cite{Revel}.}

\begin{table}[ht]
\caption{One-neutron separation energies ($S_n$) of the nuclei included in the present study.}
    \centering
\begin{tabular}{ccc}
    \hline \hline
    Nucleus & $S_n$ (MeV) & References \\
    \hline
    $^{29}$Ne & $0.96 \pm 0.14$ & \cite{Wang2021,JURADO2007}\\
    \hline
    $^{31}$Ne & $0.29 \pm 1.64$ & ~\cite{JURADO2007}  \\
              & $0.17 \pm 0.13$ & ~\cite{Wang2021}\\
              & $0.15^{+0.16}_{-0.10}$ & ~\cite{Nakamura2014} \\
    \hline
    $^{37}$Mg & $0.16 \pm 0.68$ &  \cite{Wang_2012} \\
              & $0.22^{+0.12}_{-0.09}$ & \cite{Kobayashi2014}\\
              & $0.24\pm 0.11$  &   \cite{Wang2021} \\
    \hline
    $^{34}$Na & $0.17 \pm 0.50$ & \cite{Gaudefroy12PRL}\\
    \hline    
    \end{tabular}
    \label{sn_values}
\end{table}

A relative orbital angular momentum of $\ell = 1$ is seen as a characteristic of the neutron halo in all these isotopes. This contrasts with well-established one-neutron halo nuclei such as $^{11}$Be and $^{19}$C, where the ground states predominantly feature an $s$-wave neutron coupled to the core \cite{Nak94,Nak99,Banerjee2000,NUNES1996,Aum2000,Pal03}. Although nuclei near $N = 20\text{--}28$ are expected to be dominated by the $1f_{7/2}$ orbital, its high angular momentum ($\ell = 3$) introduces a substantial centrifugal barrier that inhibits the spatial extension required for halo formation. To facilitate halo structures, significant $s$- or $p$-wave components must mix into the ground state, reducing the barrier \cite{RD09}. This mixing reflects a notable modification of the shell structure, often involving intruder orbitals such as $2s_{1/2}$ or $2p_{3/2}$, and is closely associated with nuclear deformation. 
Thus, halo formation in the medium mass nuclei within the island of inversion is strongly linked to shell evolution which can possibly result due the presence of deformation. Therefore, while studying such nuclei, it becomes essential to consider this {deformation}. Several works \cite{Urata2011,Urata2012,HBK17,Hamamoto2010} have studied these nuclei incorporating deformation, however, here we focus mainly on the implementation of the FRDWBA theory of Coulomb breakup (discussed in Section~\ref{frdwba}), which explicitly accounts for the deformation \cite{Shubh14NPA,Shubh15NPA} and has been successfully applied to all these nuclei \cite{Shubh14NPA,Shubh15NPA,GSingh16PRC,Manju21NPA}.

The Coulomb breakup reaction, wherein a high-energy projectile interacts with the Coulomb field of a heavy target nucleus resulting in the removal of its valence neutron, has emerged as an effective tool for probing the halo structure of neutron-rich nuclei near the drip line (see, for example, Ref.~\cite{BAUR2003}). This reaction mechanism is particularly sensitive to the electric dipole (E1) response of the projectile~\cite{Nak94,Nak99, BAUR2003}, making it a valuable probe of the soft dipole excitations that are characteristic of weakly bound systems.
Among the theoretical frameworks used to describe a breakup process \cite{Hagino22PPNP,Sagawa2025EPJA,Moro12PRC, Capel22FBS, Baye2005PRL,SHYAM1992,Shyam01PRC,Capel03PLB,Pinilla2012PRC,Goldstein2006PRC,Capel2005PRC,Hebborn20PRC,Melezhik1999PRC,Melezhik2001PRC,Rodri08PRC,Deltuva2007PRC,Diego2014PRC,Moro2012PRL,MORO2020,Chen2022JPG,Moro2025EPJA,Yahiro2012PTEP,Esbensen1995,Bertulani1993PR,Bertulani1992b},
%\gs{(We need to mention and give refs. of other works like TDSE, CDCC, DEA, ANC etc.)}, 
the post-form finite-range distorted wave Born approximation (FRDWBA)~\cite{Banerjee2000,Chatterjee00NPA} has carved its own niche. This fully quantum mechanical theory is semi-analytical in nature, incorporating the fragment--target interaction to all orders, while the fragment--fragment interaction is considered perturbatively up to first order. This formalism relies on realistic ground-state wave functions {of the projectile} to model the relative motion {of the core-valence neutron subsystem} %\gsout{between the valence neutron and the core within the projectile} \gs
{comprising it}. One can then gain direct insights into the g.s. structure of the projectile nucleus by comparing theoretically predicted cross-sections with experimental data. Coulomb breakup can also be used to observe another distinct signature of the halo structure, via the calculation of the parallel momentum distributions of the core fragments, which, if narrow, reflect extended spatial distributions of the valence neutrons and are manifested by their large matter radii. Thus, in what follows, we shall start from the wave functions and proceed to discuss breakup cross-sections and parallel momentum distributions, along with other reaction observables such as angular and energy-angular distributions and relative energy spectra, which help in an understanding of the structure of halo candidates, $^{29}$Ne, $^{31}$Ne, $^{34}$Na and $^{37}$Mg.

%\gsout{By comparing theoretical cross-section predictions with experimental data, one can gain direct insight into the ground state structure of the projectile nucleus. A distinct signature of the halo structure is observed in the form of narrow parallel momentum distributions of the core fragments following breakup. Such narrow widths reflect the extended spatial distribution of the valence neutron, which is a hallmark of halo configurations in weakly bound nuclei.}

\begin{figure}[htbp] 
\centering
\includegraphics[trim={0 0 0 0},clip,width=.65\columnwidth]{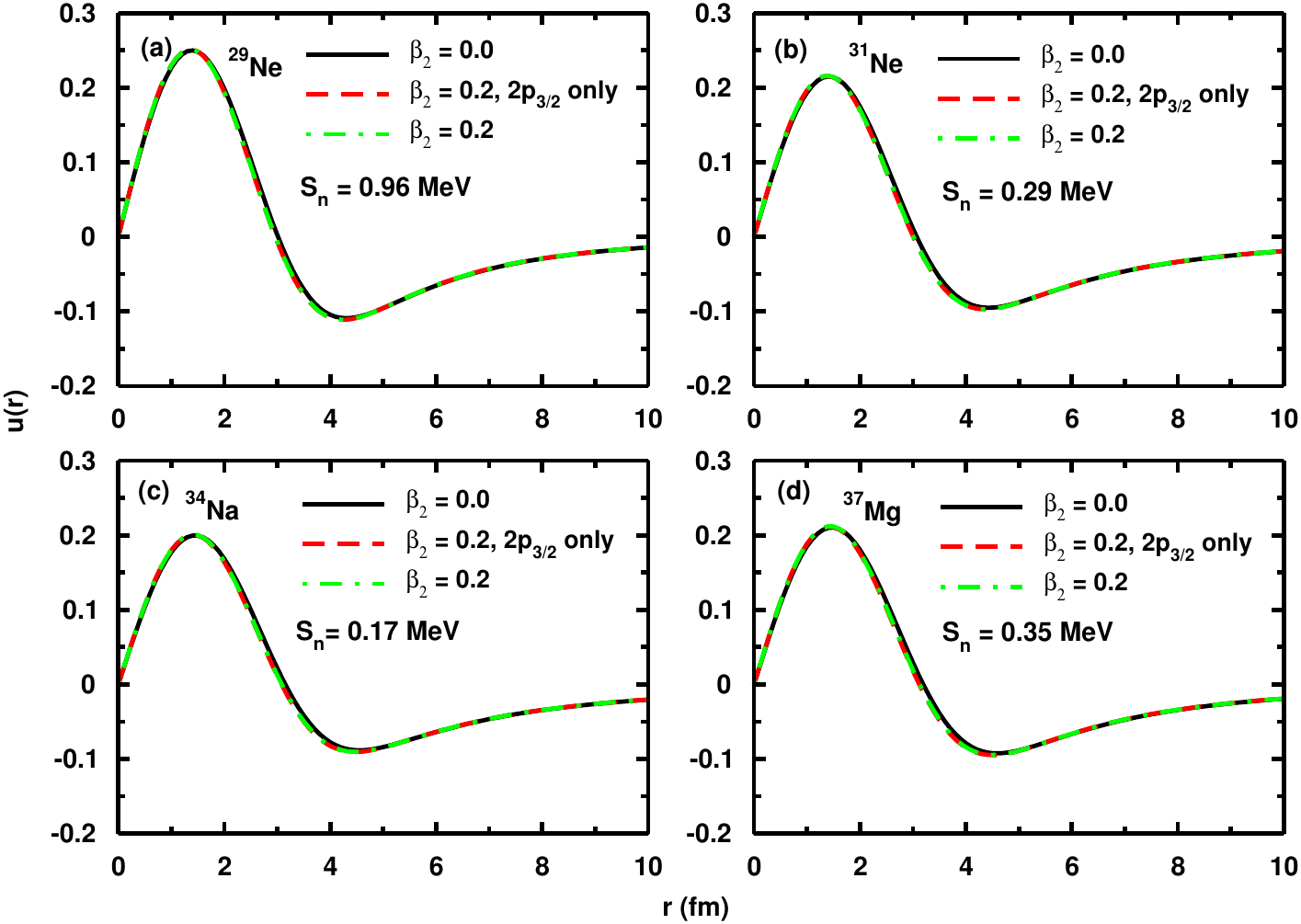}
\caption{\label{fig: def_wfn1} The deformed bound-state wave functions of $^{29}$Ne, $^{31}$Ne, $^{34}$Na, and $^{37}$Mg {calculated by solving Eq.~(\ref{radial})}. The (black) solid curves represent the wave functions at zero deformation. The (green) dot-dashed curves correspond to the wave functions at a deformation of $\beta_2 = 0.2$, including all components associated with $\ell = 1, 3, 5$ and their allowed $j$ values. The (red) dashed lines show the wave functions at $\beta_2 = 0.2$, but restricted to a single component with $\ell = 1$ and $j = 3/2$. For further details, refer to the main text.
%and Ref.\cite{Manju's thesis}.\sh{reference}
}
\end{figure}

%\gsout{In Refs.  the FRDWBA theory of Coulomb breakup reactions has been extended to include the deformation of the projectile by using a deformed Woods–Saxon potential to describe the valence neutron–core relative motion.} 

\subsubsection{The wave functions}\label{wfn}
Recently, the FRDWBA theory of Coulomb breakup reactions \cite{Chatterjee00NPA} was extended to
include the deformation of the projectile by using a deformed Woods–Saxon potential to describe the core-valence neutron relative motion at finite distance \cite{Shubh14NPA,Shubh15NPA}. %\gsout{Hence, this provides} 
This extension provides an elegant theoretical tool to study the Coulomb breakup of neutron-drip line nuclei lying in {the N = 20--28} island of inversion and to investigate the possible correlations between halo formation and the shell evolution and deformation in such weakly bound nuclear systems. Although the wave function obtained using a deformed potential for a given orbital angular momentum $\ell$ contains admixtures of other $\ell$ components with the same parity, it {was} demonstrated in Ref.~\cite{Hama04PRC} (and as has been also discussed in Section~\ref{frdwba}) that for weakly bound valence neutrons, the lowest-$\ell$ component generally dominates the overall composition of the wave function irrespective of the magnitude of the deformation.

\begin{table}[t!]
\caption{Single-particle asymptotic normalization coefficients (ANCs) for the deformed wave functions of $^{29}$Ne, $^{31}$Ne, $^{34}$Na, and $^{37}$Mg at different values of the deformation parameter $\beta_2$. The $S_n$ values in each case are the same as those used in Fig.~\ref{fig: def_wfn1}. Results are shown only for the $\ell = 1$, $j = 3/2$ component at each value of $\beta_2$. } %\gs{Needs revision. Shubh bhaiya please clarify and mention why we have multiple rows for a given $l$ with same beta2 and j. Also say what does the (b) in ANC(b) mean.}}

\centering
\small
\begin{tabular}{@{}lcccccc@{}}
\hline
$\beta_2$ & $\ell$ & $j$ & \multicolumn{4}{c}{ANC ($\text{fm}^{-1/2}$)} \\
\cline{4-7}
 &  &  & $^{29}$Ne & $^{31}$Ne & $^{34}$Na & $^{37}$Mg \\
\hline
0.0 & 1 & 3/2 & 0.787 & 0.328 & 0.234 & 0.386 \\
\hline
0.1 & 1 & 3/2 & 0.783 & 0.326 & 0.233 & 0.384 \\
\hline
%     & 1 & 3/2 &   --   &  --   &  --   &  --   \\
%0.1 & 3 & 5/2 & 0.783 & 0.326 & 0.233 & 0.384 \\
%     & 3 & 7/2 &   --   &  --   &  --   &  --   \\
%\hline
0.2 & 1 & 3/2 & 0.780 & 0.325 & 0.232 & 0.382 \\
\hline
%     & 1 & 3/2 &   --   &  --   &  --   &  --   \\
%0.2 & 3 & 5/2 & 0.780 & 0.325 & 0.232 & 0.382 \\
%     & 3 & 7/2 &   --   &  --   &  --   &  --   \\
%\hline
0.3 & 1 & 3/2 & 0.777 & 0.324 & 0.231 & 0.380 \\
\hline
%     & 1 & 3/2 &   --   &  --   &  --   &  --   \\
%0.3 & 3 & 5/2 & 0.777 & 0.324 & 0.232 & 0.381 \\
%     & 3 & 7/2 &   --   &  --   &  --   &  --   \\
%\hline
0.4 & 1 & 3/2 & 0.775 & 0.323 & 0.231 & 0.379 \\
\hline
%     & 1 & 3/2 &   --   &  --   &  --   &  --   \\
%0.4 & 3 & 5/2 & 0.776 & 0.324 & 0.231 & 0.381 \\
%     & 3 & 7/2 &   --   &  --   &  --   &  --   \\
%\hline
0.5 & 1 & 3/2 & 0.774 & 0.323 & 0.230 & 0.378 \\
\hline
%     & 1 & 3/2 &   --   &  --   &  --   &  --   \\
%0.5 & 3 & 5/2 & 0.775 & 0.324 & 0.231 & 0.382 \\
%     & 3 & 7/2 &   --   &  --   &  --   &  --   \\
%\hline
\end{tabular}
\label{anc}
\end{table}

To examine this point, Fig.~\ref{fig: def_wfn1} shows the ground-state wave functions of the four considered nuclei at a deformation of $\beta_2 = 0.2$, incorporating the mixing of multiple orbital angular momenta ($\ell = 1, 3,$ and $5$) along with their allowed $j$ values (dot-dashed lines), {calculated using the coupled differential equation (\ref{radial}). The $S_n$ values adopted in each case are also specified in the respective panels of the figure, and their choice will be discussed in the next section.} These are compared with the wave functions computed at the same deformation parameter, but restricted to the $p_{3/2}$ component only (dashed lines). For reference, the corresponding spherical wave functions (i.e., at $\beta_2 = 0$) are also shown (solid lines). All wave functions are normalized to unity (see, for example, Ref.~\cite{Shubh17PRC}).
As is evident from the figure, the $p_{3/2}$ component of the wave function clearly dominates, indicating that contributions from other $\ell$ values of the same parity remain small—an observation also reported in Ref.~\cite{Hama04PRC}.

It is also observed that deformation introduces only slight changes in the wave function primarily within the interior and at the surface, whereas in the asymptotic region, variations in $\beta_2$ have a negligible effect.

This observation is further supported by Table~\ref{anc}, which presents the single-particle asymptotic normalization coefficients (ANC), obtained by normalizing the %\shout{ground state wave functions of these nuclei} 
{wave function of the dominant $\ell=1, j=3/2$ component in each case to the corresponding} Whittaker function for various deformation parameters $\beta_2$. %\shout{The ANC values corresponding to the $p_{3/2}$ component at all deformation values are also listed.} 
{The Whittaker function is constructed using the one-neutron separation energy $S_n$ with respect to the ground state of the core, consistent with the single-particle description adopted here.}
%\sout{One can see that the ANC values remain almost unchanged with increasing deformation, indicating that while the wave functions may exhibit slight modifications in the nuclear interior with changing $\beta_2$, they remain essentially identical in the asymptotic region.}
{One can see that the ANC values exhibit only a very weak dependence on deformation, with variations appearing at a negligible level, indicating that while the wave functions may exhibit slight modifications in the nuclear interior with changing $\beta_2$, they remain essentially identical in the asymptotic region. Notably, for weakly bound nuclei, this is the region of interest.}
This justifies the use of a spherical single-particle radial wave function as an approximation to describe the core–valence neutron relative motion in the ground state of the projectile while calculating the Coulomb breakup amplitude via Eq.~(\ref{eq:betafac}). However, it is important to note that deformation effects still enter the FRDWBA formalism through the potential $V_{bc}(\mathbf{r}_1)$, as given in Eq.~(\ref{eq:WSPot}). Within this framework, we now discuss the application of the FRDWBA theory of Coulomb breakup in establishing the halo structure in nuclei near the island of inversion %\gsout{. In particular, we focus on the cases of} 
focusing mainly on $^{29}$Ne, $^{31}$Ne, $^{34}$Na, and $^{37}$Mg~\cite{Manju21NPA,GSingh16PRC,Shubh14NPA,Shubh15NPA}.

\begin{figure}[htbp] 
\centering
\includegraphics[trim={0 0 0 0},clip,width=.75\columnwidth]{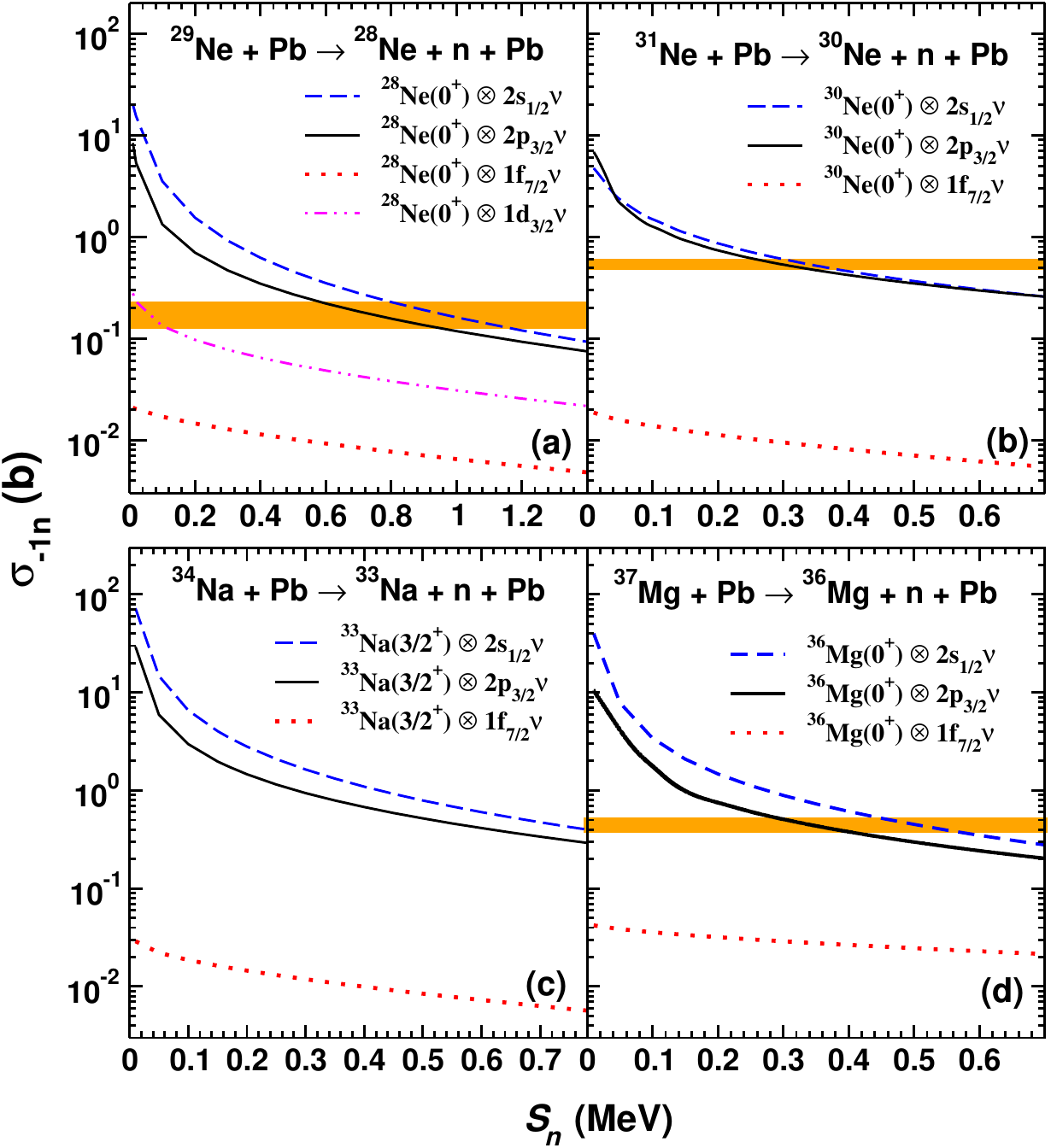}
\caption{\label{css_vs_sn} The total one-neutron removal cross section ($\sigma_{-1n}$) is plotted as a function of neutron separation energy ($S_n$) for $^{29}$Ne, $^{31}$Ne, $^{34}$Na, and $^{37}$Mg undergoing elastic breakup on a Pb target at beam energies of 244, 234, 100, and 244 MeV/nucleon, respectively, obtained using the post-form FRDWBA theory. In all cases, the (black) solid line, (blue) dashed line, and (red) dotted line represent the cross sections for valence neutron configurations in the $p$-, $s$-, and $f$-wave, respectively. The (magenta) dash-double-dotted line shows the result for $^{29}$Ne assuming a $d$-wave ground state for the valence neutron. Experimental data, where available, are indicated by the orange bands. For further details, please see the text.}
\end{figure}

\subsubsection{Neutron removal cross-sections}

Amongst the various reaction observables, we start with the one neutron removal cross-sections. Figure~\ref{css_vs_sn} presents the total one-neutron removal cross-section ($\sigma_{-1n}$) for the Coulomb breakup of $^{29}$Ne, $^{31}$Ne, $^{34}$Na, and $^{37}$Mg on a Pb target at beam energies of 244, 234, 100, and 244~MeV/nucleon, respectively, plotted as a function of the one-neutron separation energy ($S_n$). At such a high beam energy range ($\sim$ a few hundred MeV/nucleon), the final channel fragments emanate with higher velocities and are usually easier to detect. This facilitates measuring Coulomb dissociation observables, such as the angular and energy-angular distributions or the relative energy spectra, which could then be used to put constraints on spectroscopic factors or other structural parameters of a nucleus \cite{GSingh16PRC}. %\gs{(checkout adiabaticity param)}.%\gsout{Coulomb breakups up to this incident energy range can be treated non-relativistically as the differences in the relativistic and non-relativistic momenta of the heavier outgoing fragment (especially for weakly bound light nuclei) are negligible \cite{GSingh13DAE}. However, beyond this range, one must be vigilant.}

{At beam energies of a few hundred MeV/nucleon, relativistic effects may become relevant and, in principle, should be accounted for in a more general framework. However, a fully consistent breakup theory incorporating relativistic dynamics is not yet fully developed. Existing studies, such as those based on modified Klein-Gordon formulations ~\cite{bertulani05}, indicate that in the $100-250$ MeV/nucleon range, relativistic corrections can alter breakup observables by about $10-15$\%. Similar conclusions have been reported in continuum-discretized coupled-channels calculations with eikonal approximation~\cite{ogata_ber09}, where corrections to the nuclear interaction were found to be even smaller.
In view of these results, effects of a similar order may be expected in the present cases, although present experimental uncertainties make their clear identification difficult. Further progress towards a fully relativistic treatment would be valuable, particularly in light of forthcoming high-precision data from next-generation radioactive beam facilities.}

In each case shown in Fig. \ref{css_vs_sn}, calculations were performed for various possible ground state configurations, as indicated in the figure. A comparison with the experimental data — represented by the orange band, where the band width reflects the uncertainty in the measured cross-sections — helps infer the most probable spin-parity ($J^{\pi}$) of the ground states of these nuclei. This analysis is particularly useful in resolving discrepancies in the literature regarding $J^{\pi}$ assignments.
For example, the ground state of $^{31}$Ne has been proposed to be either $3/2^-$ or $1/2^+$ in different studies~\cite{POVES1994,DESCOUVEMONT1999,Minomo2012,Urata2011,Urata2012,Takechi2012,Nakamura2009,takechi2010,Minomo2011,Sumi2012}, {although most recent results seem to rule out the $1/2^+$ \cite{Lu2025PRC}}. Similar ambiguity exists for $^{29}$Ne (with $J^{\pi}$ suggested to be either $1/2^+$~\cite{Hama04PRC,Takechi2012} or $3/2^-$~\cite{Kobayashi16PRC,Utsuno1999}) and for $^{37}$Mg~\cite{Hamamoto2007,Kobayashi2014,Nakada2018,Zhang2023c}, where the valence neutron has been suggested to occupy $2s$ or $2p$ orbitals rather than the $1f$ orbital predicted by the single-particle shell model. 

{We recall that in the FRDWBA architecture}, the structural information of the projectile enters through its bound-state wave function which is generated using a Woods--Saxon potential. The parameters of this potential are adjusted to reproduce the one-neutron separation energy ($S_n$) of the respective nucleus in its assumed ground state. By comparing the calculated breakup cross sections with experimental data {for the given ranges of $S_n$ values} - from Ref.~\cite{Nakamura2009} for $^{31}$Ne, Ref.~\cite{Kobayashi2014} for $^{37}$Mg, and Ref.~\cite{Kobayashi16PRC} for $^{29}$Ne - the possibility of a valence neutron occupying an $f$-wave orbital is clearly ruled out in all three cases. For $^{29}$Ne, the $d$-wave neutron configuration could only reproduce the data at unrealistically small $S_n$ values, significantly lower than the experimentally accepted range. The $d$-wave configuration was also examined (though not shown here) and found to be incompatible with the data for $^{31}$Ne and $^{37}$Mg.

Thus, FRDWBA calculations based on this single reaction observable support a ground state spin-parity of $J^{\pi} = 3/2^-$ for these nuclei, although the possibility of $1/2^+$ cannot be entirely ruled out. For more stringent constraints, additional reaction observables must be considered. Nevertheless, the overlap between the theoretical cross-sections and the experimental bands has provided tighter constraints on $S_n$ values than previously available. For $^{31}$Ne, $S_n$ values of $0.295 \pm 0.055$~MeV and $0.345 \pm 0.055$~MeV were obtained for $p$- and $s$-wave configurations, respectively~\cite{Shubh14NPA}. For $^{37}$Mg, the corresponding values were $0.35 \pm 0.06$~MeV ($p$-wave) and $0.50 \pm 0.07$~MeV ($s$-wave)~\cite{Shubh15NPA}. Similarly, for $^{29}$Ne, the extracted $S_n$ values were $0.80 \pm 0.20$~MeV ($p$-wave) and $0.99 \pm 0.22$~MeV ($s$-wave)~\cite{Manju21NPA}.
Similar conclusions can also be drawn for $^{34}$Na~\cite{GSingh16PRC}, where a dominant $p$-wave configuration was proposed for the ground-state valence neutron, as is evident from the figure. %\shout{In fact, even when equal spectroscopic factors for the three possible orbital angular momentum values of the valence neutron ($\ell = 0, 1,$ and $3$) were considered, the resulting total contribution (represented by the green dot-dashed line) closely matched the individual $\sigma_{-1n}$ calculated with a $p$-wave–dominated configuration.} 
However, in the absence of experimental data, these theoretical predictions have yet to be confirmed. In fact, the ground-state spin-parity of $^{34}$Na has not been experimentally established; shell model calculations reported in Ref.~\cite{Door14PTEP} suggest a $J^{\pi} = 2^-$ assignment, which, unless otherwise stated, is also adopted in the analyses presented in this work.
{Another common trend noteworthy in Fig. \ref{css_vs_sn} is the $1/E$ overall dependence of the one-neutron removal cross-section. This is a direct consequence of the low-lying dipole response, a characteristic feature of halo nuclei which will be discussed in Sec. \ref{sec: relen}.}

\begin{figure}[htbp] 
\centering
\includegraphics[trim={0 0 0 0},clip,width=.50\columnwidth]{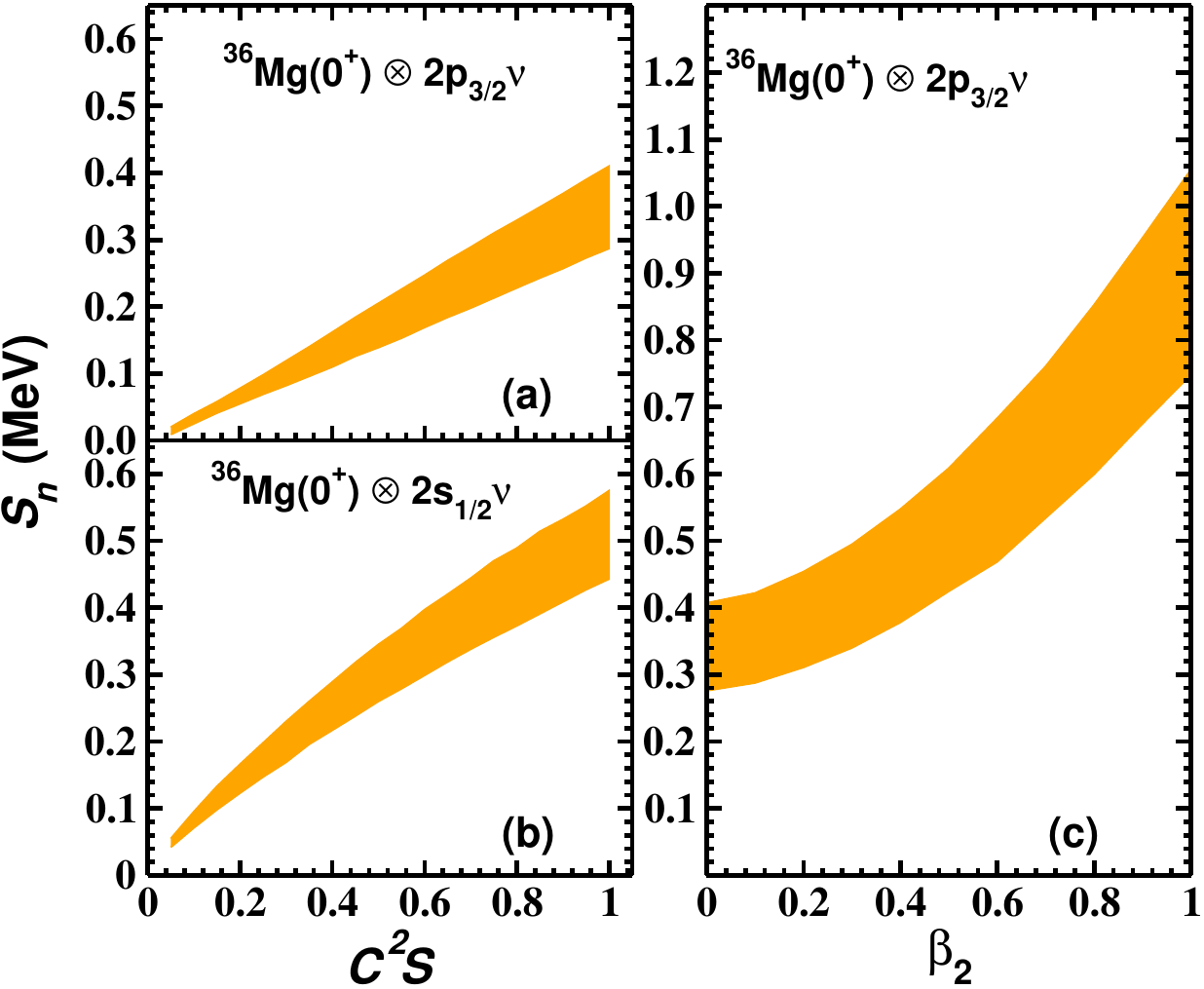}
\caption{\label{sn_vs_beta}
(a--b) One-neutron separation energy ($S_n$) as a function of spectroscopic factor ($C^2S$) for the breakup of $^{37}$Mg on a Pb target, considering two different ground state configurations of $^{37}$Mg.  
(c) $S_n$ deduced from the comparison of theoretical calculations with experimental data as a function of the quadrupole deformation parameter $\beta_2$, corresponding to the $^{36}$Mg($0^+$) $\otimes$ $2p_{3/2} \nu$ configuration of the $^{37}$Mg ground state with $C^2S = 1.0$.}
\end{figure}

In Fig.~\ref{css_vs_sn}, the calculated cross sections are obtained under the assumption that all projectiles are spherical. However, Refs.~\cite{Shubh14NPA,Shubh15NPA,GSingh16PRC,Manju21NPA} have explicitly explored the dependence of the cross sections on the quadrupole deformation parameter ($\beta_2$). It was observed that, in all cases, the cross sections increase with increasing $\beta_2$. Consequently, the limits on the $S_n$ also shift towards higher values. This trend aligns with previous theoretical investigations~\cite{Warburton90,CAMPI1975193,Watt_1981,POVES1994}, which suggest that nuclear deformation can enhance the binding energies of nuclei in the island of inversion due to the mixing of $2\hbar\omega$ $2p$–$2h$ neutron excitations with the $0\hbar\omega$ configurations.

In all the calculations presented above, a uniform spectroscopic factor ($C^2S$) of 1 has been assumed for each configuration. However, significant variations in $C^2S$ values have been reported in the literature. For instance, for the configurations $^{36}$Mg$(0^+)\otimes 2p_{3/2}\nu$ and $^{36}$Mg$(0^+)\otimes 2s_{1/2}\nu$ in $^{37}$Mg, shell model predictions suggest $C^2S$ values of 0.31 and 0.001, respectively~\cite{Utsuno1999}. In contrast, breakup data analyses~\cite{Kobayashi2014} yield $C^2S$ values of $0.42^{+0.14}_{-0.12}$ and $0.40^{+0.16}_{-0.13}$, respectively. These analyses employed the eikonal model of Ref.~\cite{hansen2003} for a Carbon target and the semiclassical Coulomb breakup model of Ref.~\cite{Nakamura2014} for a Lead target. However, the spectroscopic factor for the $^{36}$Mg$(0^+)\otimes 1f_{7/2}\nu$ configuration is not reported in these works. FRDWBA calculations incorporating shell-model spectroscopic factors were presented in Ref.~\cite{shubh2016_c}, which clearly ruled out the possibility of a $1/2^+$ ground state for $^{37}$Mg.

Since the extracted $S_n$ values from Coulomb breakup data are intimately linked to the assumed $C^2S$, {one must} systematically study the sensitivity of $S_n$ to variations in $C^2S$ as well as to changes in the deformation parameter $\beta_2$~\cite{Shubh15NPA}. In Fig.~\ref{sn_vs_beta} (a) and (b), the one-neutron separation energy ($S_n$) is plotted as a function of $C^2S$ for the breakup of $^{37}$Mg on a Pb target [same as in Fig.\ref{css_vs_sn} (d)], considering two different ground-state configurations of $^{37}$Mg. For each chosen value of the spectroscopic factor ($C^2S$), the corresponding one-neutron separation energy ($S_n$) was extracted from the region where the calculated breakup cross section overlaps with the experimental data band, as illustrated in Fig.~\ref{css_vs_sn}. It is observed that $S_n$ increases monotonically with increasing $C^2S$. Additionally, the uncertainty in the extracted $S_n$ grows with $C^2S$, since the flatter portions of the theoretical cross section tend to intersect broader regions of the data band of Fig. \ref{css_vs_sn}, leading to greater ambiguity. 
As an example, for the configuration $^{36}$Mg($0^+$) $\otimes$ $2p_{3/2}\, \nu$ with $C^2S = 0.42$ (used in Ref. \cite{Kobayashi2014}), the extracted $S_n$ is $0.14 \pm 0.03$~MeV, which is notably smaller than the mean value of $0.22$~MeV reported in Ref.~\cite{Kobayashi2014} for the same spectroscopic factor. 

Fig.~\ref{sn_vs_beta}(c) illustrates the variation of the one-neutron separation energy ($S_n$) with the quadrupole deformation parameter $\beta_2$, obtained by comparing theoretical calculations with experimental data. These results are based on the $^{36}$Mg($0^+$) $\otimes$ $2p_{3/2}\, \nu$ configuration for the ground state of $^{37}$Mg, assuming a spectroscopic factor $C^2S = 1.0$. The figure clearly shows a positive correlation between $S_n$ and $\beta_2$, consistent with earlier findings. For example, for $S_n = 0.35$~MeV as considered in Ref.~\cite{Shubh15NPA}, the FRDWBA calculations do not support deformation values $\beta_2 > 0.32$. This result is consistent with the range predicted by the Nilsson model~\cite{Hamamoto2007}, which suggests deformation of $0.30 \leq \beta \leq 0.34$ for a ground state $J^{\pi}$ of $3/2^-$.
Notably, when $\beta_2$ exceeds 0.70, the extracted $S_n$ surpasses the upper bound reported in Ref.~\cite{Wang_2012}. This suggests that, for a ground state dominated by $p$-wave in $^{37}$Mg, the deformation parameter remains reasonable even for the maximum predicted value of the $S_n$.
In contrast, with the s-wave configuration, the $S_n$ was found to remain unchanged with $\beta_2$ ~\cite{Shubh15NPA}. This trend has also been observed for other nuclei considered here when analyzed in a similar framework~\cite{Shubh14NPA,Manju21NPA,GSingh16PRC}.
These comparisons essentially demonstrate the dependence of the one-neutron separation energy ($S_n$) on the projectile's deformation and underscore the importance of having precise knowledge of the spectroscopic factor ($C^2S$) for various configurations in order to reliably determine $S_n$ from Coulomb breakup studies. 

Next, assuming a spectroscopic factor of 1
and using the central value of $S_n$ extracted from the comparison between theoretical calculations and experimental data in Fig.~\ref{css_vs_sn} (for $^{29}$Ne and $^{34}$Na measured values are adopted), we investigate other reaction observables for all the reactions depicted in the figure and analyze the impact of projectile deformation on each of them.
It is important to note that the one-neutron removal cross section data shown in Fig.~\ref{css_vs_sn} do not conclusively rule out the $1/2^+$ spin-parity assignment for the ground state of the nuclei under consideration. This motivates the calculation and measurement of more exclusive observables in the breakup of these nuclei, such as the relative energy spectra of the fragments.

\begin{figure}[htbp] 
\centering
\includegraphics[trim={0 0 0 0},clip,width=.60\columnwidth]{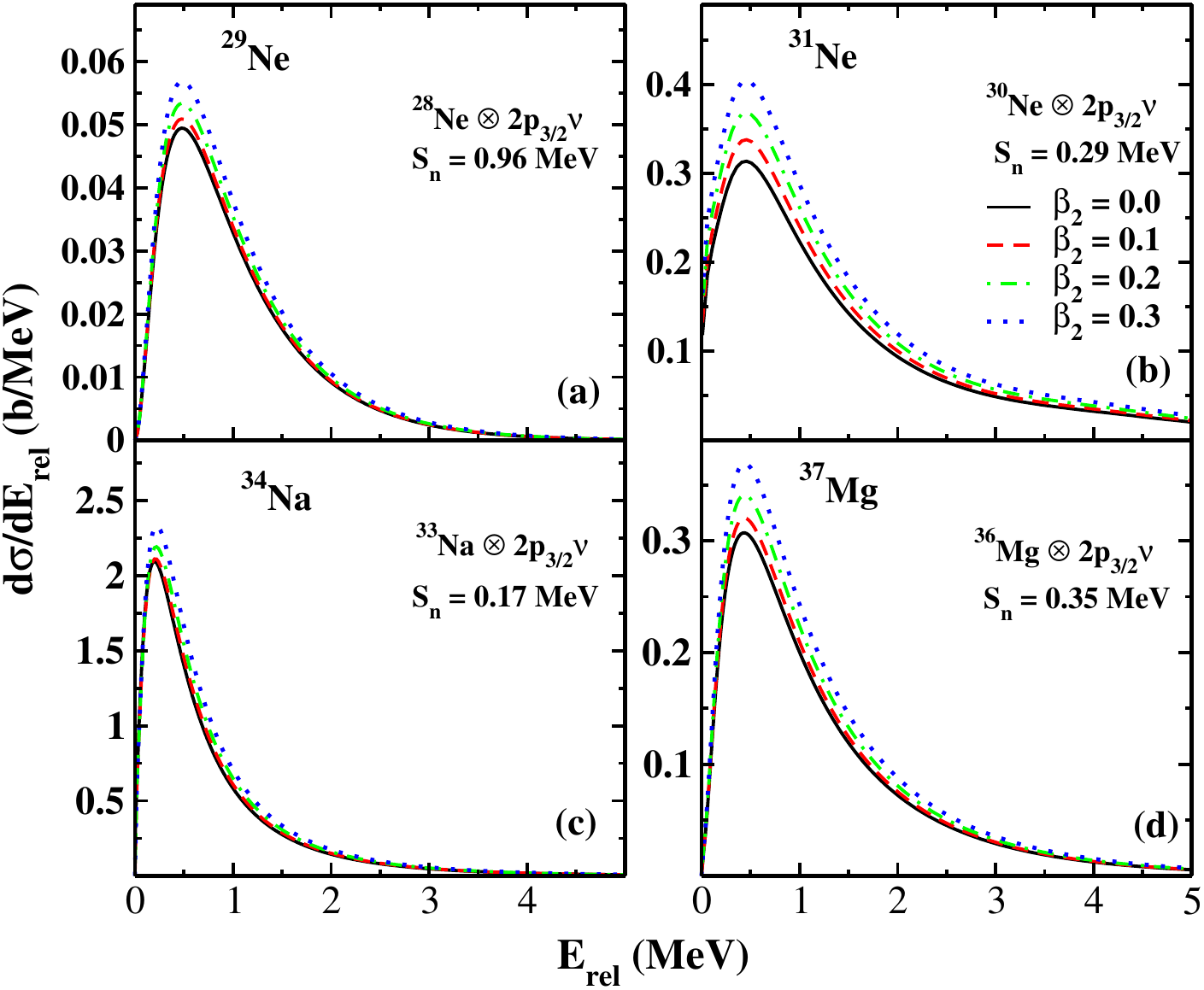}
\caption{\label{fig: relen} Relative energy spectra of the outgoing fragments corresponding to the breakup reactions shown in panels (a--d) of Fig.~\ref{css_vs_sn}, calculated assuming a ground-state configuration in which the valence neutron occupies the $2p_{3/2}$ orbital. The quadrupole deformation parameter $\beta_2$ is varied from 0.0 to 0.3. For further details, refer to the main text.}
\end{figure}

\subsubsection{Relative energy spectra}
\label{sec: relen}

The one-neutron removal cross-sections discussed above are important observables for establishing correlations between experiments and theory, yet their inclusive nature may hide the structural effects that can easily be manifested by measuring other quantities. The relative energy spectra are another observable that are crucial not only for studying the dipole behavior of nuclei, they can also be used to provide approximate rule-of-thumb estimates about the structure of the nucleus by the application of scaling laws \cite{RC08EPJ,Manju19EPJ}.

It is also well established that the peak position of the relative energy spectra ($d\sigma/dE_{rel}$) is sensitive to the configuration of the valence neutron in the projectile~\cite{Banerjee2000,BAUR2003,Nag05,Typ05}. This dependence has been explicitly demonstrated for the Coulomb breakup of $^{31}$Ne and $^{37}$Mg~\cite{Shubh15NPA,shubh_jps}. For identical values of the $S_n$ as well as $C^2S$, the shape and magnitude of the relative energy spectra vary markedly depending on whether the valence neutron occupies a $2s$ or a $2p$ orbital.
Specifically, calculations with an $s$-wave neutron configuration yield cross-sections that are considerably larger near the peak region compared to those obtained using a $p$-wave configuration. The position of the peak also differs between the two: the $p$-wave configuration spectra tend to peak at higher relative energies ($E_{\text{rel}}$) than the $s$-wave case. This clear distinction in spectral behavior further underscores the usefulness of relative energy distributions as a diagnostic tool for probing the ground-state structure of halo nuclei.

For configurations with a non-zero orbital angular momentum component, the height of the peak in a relative energy spectrum is strongly influenced by the deformation of the projectile~\cite{Shubh14NPA,Shubh15NPA,GSingh16PRC,Manju21NPA}. In Fig.~\ref{fig: relen}, we present the relative energy spectra for the reactions analyzed in Fig.~\ref{fig: def_wfn1}, calculated by assuming that the valence neutrons occupy the $2p_{3/2}$ orbital. The quadrupole deformation parameter $\beta_2$ is varied from 0.0 to 0.3. In each case, an increase in $\beta_2$ leads to a noticeable rise in the cross-section, particularly around the peak. 
%\shout{This is perhaps because a larger deformation leads to an increased size of the nucleus}\footnote{This intuitive reasoning depends on the model used and may not always be true.}, \shout{thus manifesting itself also in the relative energy spectra.}
{This behavior may be attributed to the modification of the reaction coupling induced by deformation. In the present calculations, deformation is treated at the linear (first-order) level [Eq. (\ref{eq:WSPot})], which does not alter the effective nuclear size but affects the transition matrix governing the breakup process.}
%\gsout{It is also rather straightforward to observe that nuclei with smaller separation energies have a lower relative energy response.} 
Low-lying relative energy characteristics are a typical feature of halo nuclei \cite{Bertulani1992, Nag05}, being also reflected in the dipole responses, as the former are directly proportional to the latter. In addition, it is worth noting that Coulomb breakup observables are closely related to electromagnetic transition strengths, particularly the dipole (E1) response, which is sensitive to the spatial extension of the wave function through the corresponding sum rules. For loosely bound systems, this extension is primarily governed by the small separation energy, and therefore, exerts significant influence in the observed breakup strength. Furthermore, the peaks lie closer to the origin for a more weakly bound nucleus which is a strong halo candidate, and this is exhibited nicely in Fig. \ref{fig: relen}, especially in the case of $^{34}$Na in panel (c). It is already established that for a given $\ell$, the location of the peak in both the $dB($E1$)/dE_{rel}$ and the relative energy spectrum is influenced by the neutron binding energy of a spherical projectile~\cite{Nag05, RC08EPJ, Typ05, Bertulani1992}. {This further emphasizes that, in loosely bound systems, the asymptotic behavior of the wave function is governed by the separation energy and plays a dominant role.} This proximity of the peaks to the origin can also be used in conjunction with scaling laws to extract relations between the separation energy of a deformed nucleus and its dipole response~\cite{Manju19EPJ}. It is quite likely that even the total dipole response for nuclei in the island of inversion is comparable to each other and is conserved, but needs further investigations.
In light of these possible applications, experimental measurements of the relative energy spectra in the breakup reactions involving these nuclei in the island of inversion would significantly aid in not only squeezing the uncertainties in their ground state configurations and neutron separation energies, a more accurate determination of these properties would, in turn, also improve our understanding of their quadrupole deformation, which plays a crucial role in shaping the peak structures of their relative energy spectra \cite{GSingh16PRC} as well as shell evolution.

\subsubsection{Peak position}

\begin{figure}[htbp] 
\centering
\includegraphics[trim={0 0 0 0},clip,width=.60\columnwidth]{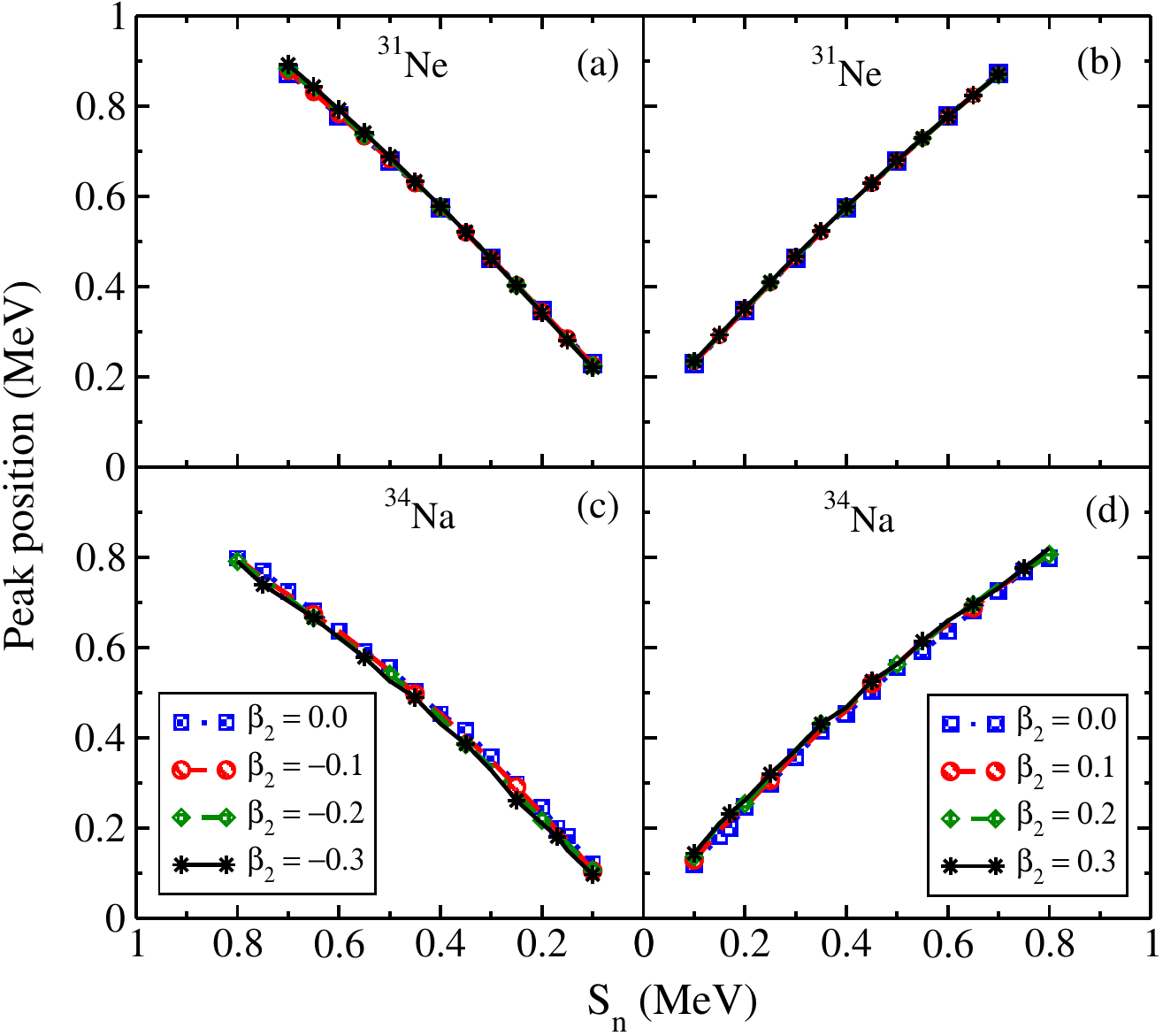}
\caption{\label{fig: peakBeta} Peak positions of relative energy spectra as a function of $S_n$ in the breakup of $^{31}$Ne and $^{34}$Na on $^{208}$Pb at 234\,MeV/nucleon and 100\,MeV/nucleon, respectively. The valence neutrons in the ground states of both $^{31}$Ne and $^{34}$Na are supposed to be filled in the $2p_{3/2}$ orbital. The left panels show the peaks for deformation parameter, -0.3 $\leq \beta_2 \leq$ 0.0, while the right panels display the same for 0.0 $\leq \beta_2 \leq$ 0.3. The average slope of all the curves for $^{31}$Ne and $^{34}$Na is 1.06 and 1.01, respectively. The lines guide the eye to the \textit{almost} straight line trajectory, highlighting the possible applications of scaling laws to deformed, weakly bound nuclear systems near the drip lines.}
\end{figure}

As averred, since scaling laws are known to be valid for weakly bound light mass nuclei, and show good indications of being applicable to medium mass nuclei from the relative energy responses, it would be interesting to see what effects deformation might bring in. For this purpose, the peak positions of the relative energy spectra for the outgoing fragments in the Coulomb breakup of $^{31}$Ne and $^{34}$Na were examined with variation in quadrupole deformation and the results are pictured in Fig. \ref{fig: peakBeta}.

\begin{table}[htbp]
\caption{Slopes of the various curves obtained for different values of quadrupole deformation parameter, $\beta_2$, in panels (c) and (d) of Fig. \ref{fig: peakBeta}. The average slope comes out to be 1.0115.}
\centering
\begin{tabular}{cccccc}
\\[-0.5ex]
\hline\hline\\[-1.5ex]
$\beta_2$ & & & Slope \\  
[0.5ex]\hline\\
-0.3 & & & 1.0054 \\
-0.2 & & & 1.0069 \\
-0.1 & & & 1.0111 \\
0.0 & & & 1.0122 \\
0.1 & & & 1.0186 \\
0.2 & & & 1.0149 \\
0.3 & & & 1.0144 \\
\\[-0.5ex]
%%%%%%%%%%%%%%%%%%%%%%%%%%%%%%%%%%%%%%%
%\\[1ex]
%%%%%%%%%%%%%%%%%%%%%%%%%%%%%%%%%%%
\hline\hline
\end{tabular}
\label{T_Slope}
\end{table}

%\gsout{The minimum values of $S_n$ for $^{31}$Ne and $^{34}$Na were taken to be 0.05\,MeV and 0.01\,MeV, respectively, which explains why the curves for $^{31}$Ne do not start near the origin.} 
The peak positions \textit{nearly} seem to reproduce an identity line, with the average slope for the curves for $^{34}$Na being 1.0115. Even the individual values of all curves obtained for various $\beta_2$ values differ only in the third decimal place (cf. Table \ref{T_Slope}). One might argue that the results for $^{31}$Ne seem better suited for a straight line, but that is perhaps a case of choosing a finer mesh in calculations of the relative energy spectra than for $^{34}$Na. Since it is recognized that there exists a linear relation between the one neutron separation energy and the peak position \cite{RC08EPJ}, this new information can now be explored to obtain a heuristic determination of the $S_n$ value for weakly bound (\textit{deformed}) neutron rich systems \cite{Manju19EPJ}, where the known (or evaluated) separation energies usually carry large uncertainties. Further, the almost identical nature of the slopes for the peaks of relative energy spectra imply that scaling laws should be equally valid for weakly bound systems irrespective of the sign of quadrupole deformation. In other words, the prolate or oblate elongation around the symmetry axis should not affect the deduction of separation energy or even the transition strengths or reduced transition probabilities \cite{Nag05,Typ05,Manju19EPJ}. 

\subsubsection{Parallel momentum distribution of the core fragment}
We now focus on another key observable used to identify the halo character of a nucleus — the parallel momentum distribution (PMD) of the core fragment resulting from the breakup process. {Parallel or longitudinal momentum distributions are especially suitable for studying weakly bound nuclear systems because a small momentum transfer is sufficient for the projectile to dissociate. The relative velocities of the outgoing fragments are then nearly the same which they had in the bound state configuration prior to the breakup. Ever since the early description of widths of parallel momentum distributions by Goldhaber \cite{GOLDHABER1974} they have, in fact, been used quite extensively to also study fragmentation reactions \cite{meierbachtol2012,Reinhold1998PRC}. The perpendicular or the transverse momentum distributions (TMDs), on the other hand, reveal the energy dissipated in the reaction and are quite sensitive to the reaction mechanism. However, a theoretical description of TMDs is comparatively more difficult as they are broadened due
to nuclear and Coulomb diffraction effects \cite{AlKhalili2004}. As the name suggests, a measurement of the TMD needs to be done in a plane perpendicular to the incident beam direction and since it varies also with the target mass and incident beam energy, it is less frequently measured. Nevertheless, within a large angle spectrometer, a precise measurement of the outgoing fragments' positions and angles can make measuring the small transverse component possible \cite{meierbachtol2012}.} 

In contrast to transverse momentum distributions, PMDs exhibit minimal sensitivity to the specifics of the nuclear interaction and to the incident beam energy that initiates the breakup~\cite{greiner1975,orr1995,kelley1995,SHYAM1992,BANERJEE1995}. {This is mainly because PMDs pick up contributions mainly at forward angles, and in the high energy approximation, the target-fragment system $S$-matrices are independent of longitudinal coordinates \cite{Bertulani1992a,Bertulani1993PR}.} It has been reported that the PMD width remains nearly invariant over a broad beam energy range spanning from 50~MeV/nucleon to 2~GeV/nucleon~\cite{kidd1988,mermaz1987}, with significant variations only appearing below 10~MeV/nucleon~\cite{Friedman1983}.
Additionally, empirical models of nuclear fragmentation developed by Goldhaber~\cite{GOLDHABER1974} and Morrissey~\cite{Morrissey1989} indicate that the PMD width is largely independent of the mass of the target nucleus. This observation has been supported by multiple experimental investigations~\cite{sauvan2004,orr1995,orr1992,meierbachtol2012} and reinforced by various theoretical approaches~\cite{Bertulani1992a,BANERJEE1995} focused on fragmentation reactions.

The narrowness of a PMD is directly linked to an extended spatial distribution of the valence neutron, as dictated by Heisenberg’s uncertainty principle. {In a simplified picture, PMDs can be regarded as the square of the Fourier transform of the spatial wave function of the projectile in the ground state \cite{AlKhalili2004,orr1992}.} Empirically, it is well established that the full width at half maximum (FWHM) of the PMD for the charged core fragment in the breakup of typical light halo nuclei such as $^{11}$Be and $^{19}$C is about 44~MeV/$c$ \cite{Chatterjee03PRC,kelley1995}. In contrast, for stable nuclei undergoing similar reactions, the corresponding FWHM is typically around 140~MeV/$c$~\cite{kelley1995,sauvan2004}.

The parallel momentum distribution (PMD) of the core fragment for the reactions considered in this work has been extensively analyzed in Refs.~\cite{Shubh14NPA,Shubh15NPA,GSingh16PRC,Manju21NPA}, including the effects of projectile deformation. The maximum effect of the deformation was found near the peak height, which increases with $\beta_2$, while a slight shift in the peak position was also observed.
A summary of these results is presented in Table~\ref{FWHM}. For the assumed $p$-wave ground state configurations, the calculated PMD widths are found to be significantly narrower (see Table~\ref{FWHM}), and are indicative of a halo structure in these nuclei. It is particularly noticeable that $^{29}$Ne shows a slightly broader width, suggesting a moderate or a relatively weaker halo compared to the others.
While it is anticipated that weakly bound systems will exhibit relatively small FWHM values even in the absence of deformation, a noteworthy observation is the apparent saturation of the FWHM around $\beta_2 \approx 0.4$. Beyond this threshold, further increase in deformation have little to no influence on the PMD width. Most importantly, the FWHM values obtained for $^{31}$Ne, $^{34}$Na, and $^{37}$Mg are comparable to those observed for well known halo nuclei such as $^{11}$Be, reinforcing the interpretation of their halo nature.

%\begin{table}
%\begin{threeparttable}
%\centering
%\vspace{0.2cm}
%\caption{Full width at half maximum of the parallel momentum distribution of core fragment, obtained in Coulomb breakup of $^{29}$Ne, $^{31}$Ne, $^{34}$Na and $^{37}$Mg on a Pb target at the beam energies 244, 234, 100, and 244 MeV/nucleon, respectively. The projectile ground states correspond to the p-state configuration in each case with $C^2S$ taken as 1.0.} 
%\vspace{0.5cm}
%\begin{tabular}{|c|c|c|c|c|}
%\\[-1.0ex]
%\hline\hline
%\\[-1.0ex]
 %        & \multicolumn{4}{c} {FWHM (MeV/c)} \\
%\hline
%$\beta_2$ & $^{29}$Ne \tnote{1}  &  $^{31}$Ne \tnote{2} & $^{34}$Na  & $^{37}$Mg \tnote{3} \\
 %         &($S_n$ = 0.96 MeV) & ($S_n$ = 0.29 MeV) & ($S_n$ = 0.17 MeV) & ($S_n$ = 0.35 MeV)\\
%\hline 
%\\[-1.0ex]
%0.0 & 82 & 51.24 & 36.82 & 54.65\\
%0.1 & 76 & 47.67 & 35.38 & 50.97\\
%0.2 & 71 & 45.09 & 34.76 & 48.03\\
%0.3 & 68 & 43.62 & 33.83 & 45.82\\
%0.4 & 66 & 42.15 & & 44.85 \\
%0.5 & 66 & 42.15 & & 44.61 \\
%\\[-1.0ex]
%\hline
%\hline
%\end{tabular}
%\label{FWHM}
%\begin{tablenotes}
       %\item [1] Measured value is 98(12) MeV/c %\cite{Kobayashi16PRC}
       %\item [2] Measured value is 77(18) MeV/c %\cite{Nakamura2014}
       %\item [3] Measured value is 82(13) MeV/c \cite{Kobayashi2014}
%     \end{tablenotes}
%\end{threeparttable}
%\end{table}

\begin{table}
%\begin{threeparttable}
\centering
%\vspace{0.2cm}
\caption{Full width at half maximum of the parallel momentum distribution of 
core fragment, obtained in Coulomb breakup of $^{29}$Ne, $^{31}$Ne, $^{34}$Na and $^{37}$Mg on a Pb target at the beam 
energies 244, 234, 100, and 244 MeV/nucleon, respectively. The projectile ground states correspond to the p-state 
configuration in each case with $C^2S$ taken as 1.0.} 
\vspace{0.5cm}
\begin{tabular}{|c|c|c|c|c|c|c|c|c|c|}
%\\[-1.0ex]
\hline\hline
%\\[-1.0ex]
         & \multicolumn{5}{c} {FWHM (MeV/c)} & & Measured FWHM \\
\hline
 &  $\beta_2$ = 0.0 & $\beta_2$ = 0.1 & $\beta_2$ = 0.2 & $\beta_2$ = 0.3 & $\beta_2$ = 0.4 & $\beta_2$ = 0.5 & (MeV/c) \\
\hline 
%\\[-1.0ex]
$^{29}$Ne  & 82 & 76 & 71 & 68 & 66 & 66 & 98(12) \cite{Kobayashi16PRC} \\
$^{31}$Ne  & 51.24 & 47.67 & 45.09 & 43.62 & 42.15 & 42.15 & 77(18) \cite{Nakamura2014} \\
$^{34}$Na  & 36.82 & 35.38 & 34.76 & 33.83 & & & \\
$^{37}$Mg  & 54.65 & 50.97 & 48.03 & 45.82 & 44.85 & 44.61 & 82(13) \cite{Kobayashi2014} \\
%\\[-1.0ex]
\hline
\hline
\end{tabular}
\label{FWHM}
%\end{threeparttable}
\end{table}

Experimental FWHM values measured on carbon targets, as reported in Refs.~\cite{Kobayashi16PRC,Nakamura2014,Kobayashi2014}, are included in the footnotes of Table~\ref{FWHM}. In these studies, the PMDs were analyzed using configuration mixing within the $fp$ shell, indicative of a deformed ground state structure, and employed spectroscopic factors from shell-model calculations. These shell-model results suggested a significant $p$-wave contribution in the ground-state configurations of the nuclei examined.
Importantly, wherever FWHM values for specific configurations were available in the literature, they showed reasonable agreement with FRDWBA calculations. For instance, in the Coulomb breakup study of $^{29}$Ne presented in Ref.~\cite{Manju21NPA}, the FWHMs of the PMD for configurations $^{28}$Ne$(0^+)\otimes 2s_{1/2}\nu$, $2p_{3/2}\nu$, $1d_{3/2}\nu$, and $1f_{7/2}\nu$ were calculated (at 
$\beta_2 =0$) as 52, 82, 155, and 207~MeV/$c$, respectively. These results compare well with the corresponding values of 51.5, 84.8, 175.2, and 281.7~MeV/$c$ reported in Ref.~\cite{Kobayashi16PRC}.
Based on these PMD measurements, the authors of Refs.~\cite{Kobayashi16PRC,Nakamura2014,Kobayashi2014} concluded that the ground-state spin-parity of $^{29}$Ne, $^{31}$Ne, and $^{37}$Mg is most likely $3/2^-$, and strongly disfavored a $1/2^+$ assignment. Following on similar lines and consistent with the patterns found for nuclei in the island of inversion, $^{34}$Na, as is shown by the FRDWBA calculations, is a strong halo candidate with most likely a very dominant $p$-wave component in its g.s., and presents a compelling case for experimental verification.

\begin{figure}[htbp] 
\centering
\includegraphics[trim={0 0 0 0},clip,width=.50\columnwidth]{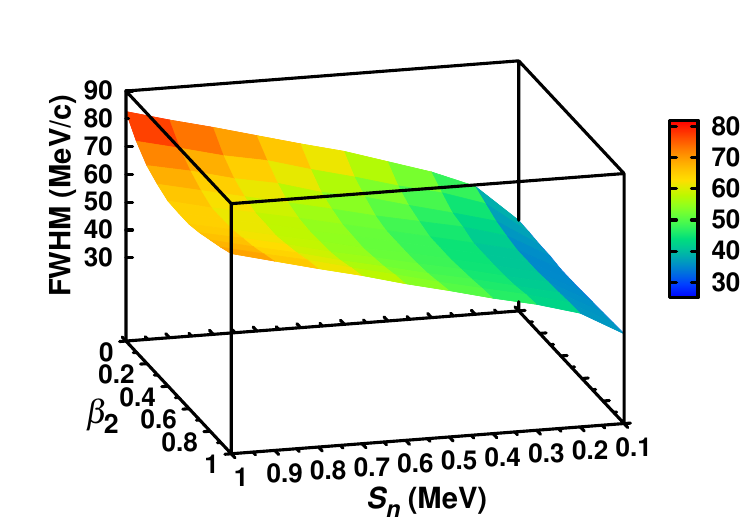}
\caption{\label{3dfwhm} Full width at half maximum (FWHM) of the parallel momentum distribution of $^{36}$Mg, obtained from the Coulomb breakup of $^{37}$Mg on a Pb target at a beam energy of 244~MeV/nucleon, shown as a function of the one-neutron separation energy ($S_n$) and the quadrupole deformation parameter ($\beta_2$). The projectile ground state is assumed to correspond to the configuration $^{36}$Mg($0^+$) $\otimes$ $2p_{3/2}\, \nu$, with a spectroscopic factor $C^2S = 1.0$. Figure adapted from \cite{Shubh15NPA}.}
\end{figure}

In Fig.~\ref{3dfwhm}, we present a detailed analysis of the FWHM of the PMD as a function of the one-neutron separation energy ($S_n$) for various values of the quadrupole deformation parameter ($\beta_2$) in the breakup of $^{37}$Mg on a $^{208}$Pb target at 244\,MeV/nucleon. As expected, the FWHM increases monotonically with increasing $S_n$, irrespective of the $\beta_2$ values. This trend is consistent with the understanding that higher binding energies correspond to more localized neutron wave functions, similar to those in stable nuclei far from the drip line.
Moreover, for most $S_n$ values considered, the variation of FWHM with $\beta_2$ follows a pattern similar to that observed in Table~\ref{FWHM}, where the calculations were performed at a fixed $S_n$. Moving from a spherical to a deformed system, in a phenomenological framework such as this one, tends to increase the spatial extension, manifesting in a lower FWHM. This consistency underscores the robustness of the observed deformation effects on the PMD width.

\subsubsection{Neutron energy-angular distribution}
One of the strengths of the post-form FRDWBA framework lies in its ability to predict multiple reaction observables within a unified theoretical approach. This feature enables a more stringent and consistent investigation of nuclear structure. In this context, we now examine the energy-angular distributions of the neutrons emitted during the breakup process.

\begin{figure}[htbp] 
\centering
\includegraphics[trim={0 0 0 0},clip,width=.60\columnwidth]{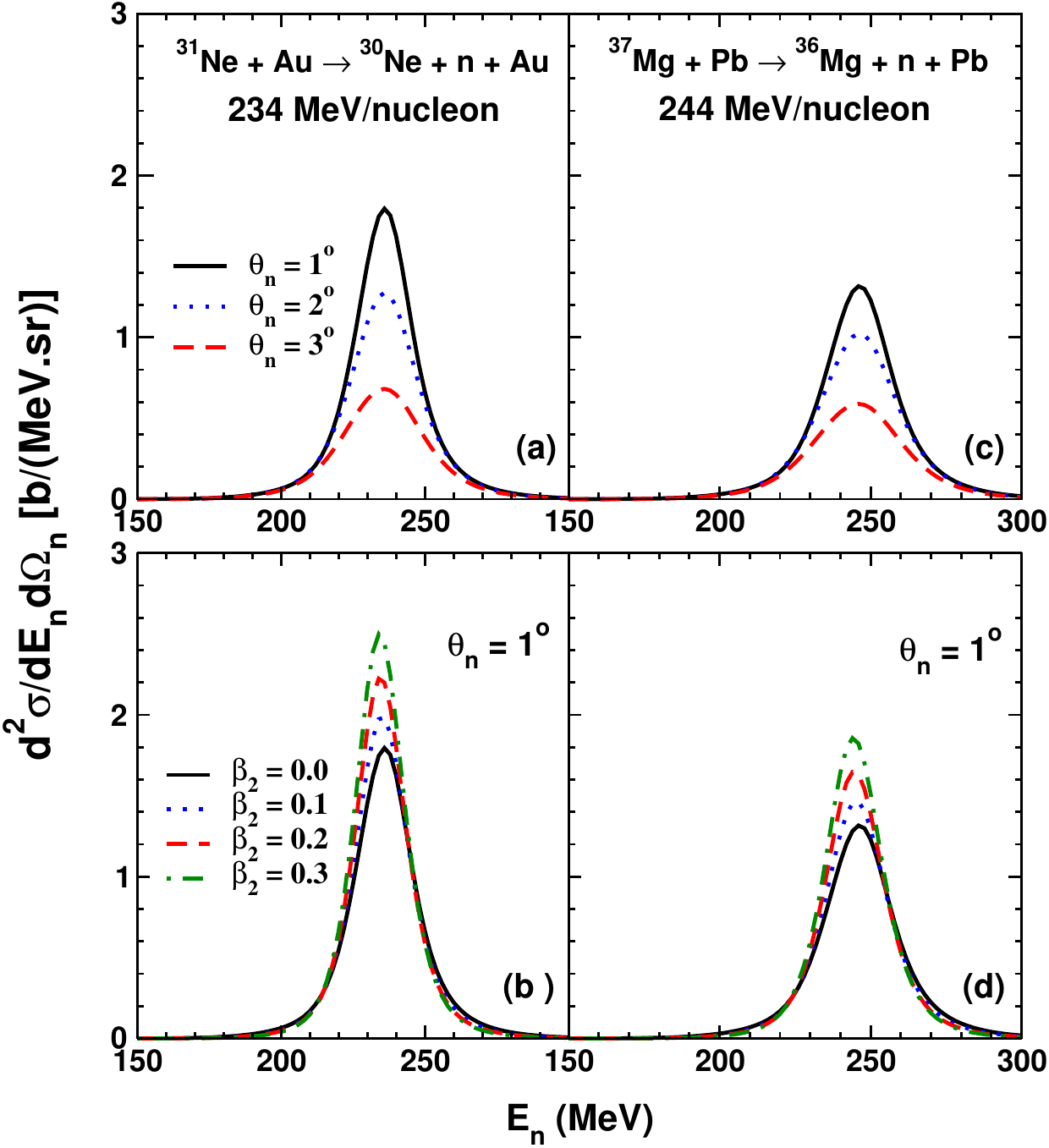}
\caption{\label{eng_ang} Neutron energy-angular distributions for the Coulomb breakup of $^{31}$Ne [panels (a) and (b)] and $^{37}$Mg [panels (c) and (d)] on Au and Pb targets at beam energies of 234 and 244~MeV/nucleon, respectively. Panels (a) and (c) display the calculated distributions at three different neutron emission angles, $\theta_n = 1^\circ$, $2^\circ$, and $3^\circ$ for spherical nuclei, while panels (b) and (d) present the results at a fixed angle $\theta_n = 1^\circ$, but for different values of the quadrupole deformation parameter $\beta_2$.}
\end{figure}

Figure~\ref{eng_ang} shows the double differential breakup cross-sections in terms of the energy-angular distributions, $d^2\sigma/(dE_n\, d\Omega_n)$, plotted as a function of the neutron energy. These results correspond to the breakup of $^{31}$Ne [panels (a) and (b)] and $^{37}$Mg [panels (c) and (d)] on Au and Pb targets at beam energies of 234 and 244~MeV/nucleon, respectively, and are adapted from Refs. \cite{Shubh14NPA,Shubh15NPA}. The nuclear configurations, $C^2S$, and $S_n$ used in these calculations are consistent with those adapted in Fig.~\ref{fig: relen}.

Panels (a) and (c) present the cross-sections computed for three different neutron emission angles, namely $\theta_n = 1^\circ$, $2^\circ$, and $3^\circ$, assuming a deformation parameter $\beta_2 = 0$. It is evident that the peak magnitude of the cross-sections decreases with increasing emission angle. A noteworthy feature of the distributions is that for all the three angles in both nuclei, the peak of the neutron energy spectrum occurs close to the energy corresponding to the beam velocity. This behavior is characteristic of loosely bound systems where, after the breakup, the fragments retain approximately the same velocity as the incoming projectile. %\gs{(SHOULD we say this in terms of velocity or energy itself?)}.
This is in contrast with breakup reactions involving stable nuclei, where post-acceleration effects often shift the neutron peak to energies lower than that of the beam velocity~\cite{Baur72NPA,BAUR1984}. Such a shift is not observed here. The absence of post-acceleration, despite  a full treatment of the Coulomb interaction in both entrance and exit channels, is attributed to the nature of halo nuclei: their low binding energies and the high beam energies involved result in breakup occurring at large distances from the target nucleus. As a result, the interaction between the charged core and the target after the breakup is too weak to induce significant post-acceleration effects~\cite{BANERJEE1993,Banerjee1996,banerjee2002}. This behavior was also observed in the analyses of $^{29}$Ne and $^{34}$Na Coulomb breakup, where post-acceleration was observed to be absent explicitly \cite{GSingh16PRC,Manju21NPA}.

In panels (b) and (d) of Fig.~\ref{eng_ang}, we explore the influence of projectile deformation on the double differential cross-sections for the breakup reactions discussed above, considering a fixed neutron emission angle of $\theta_n = 1^\circ$. It is clearly observed that the magnitude of the cross-section increases with increasing values of the deformation parameter $\beta_2$, with the effect being most pronounced near the peak of the energy distribution. This sensitivity of the cross-section to deformation suggests that measurements of neutron energy-angular distributions can serve as a useful tool to extract information about the ground-state deformation of the projectile nucleus.

\begin{figure}[htbp] 
\centering
\includegraphics[trim={0 0 0 0},clip,width=.60\columnwidth]{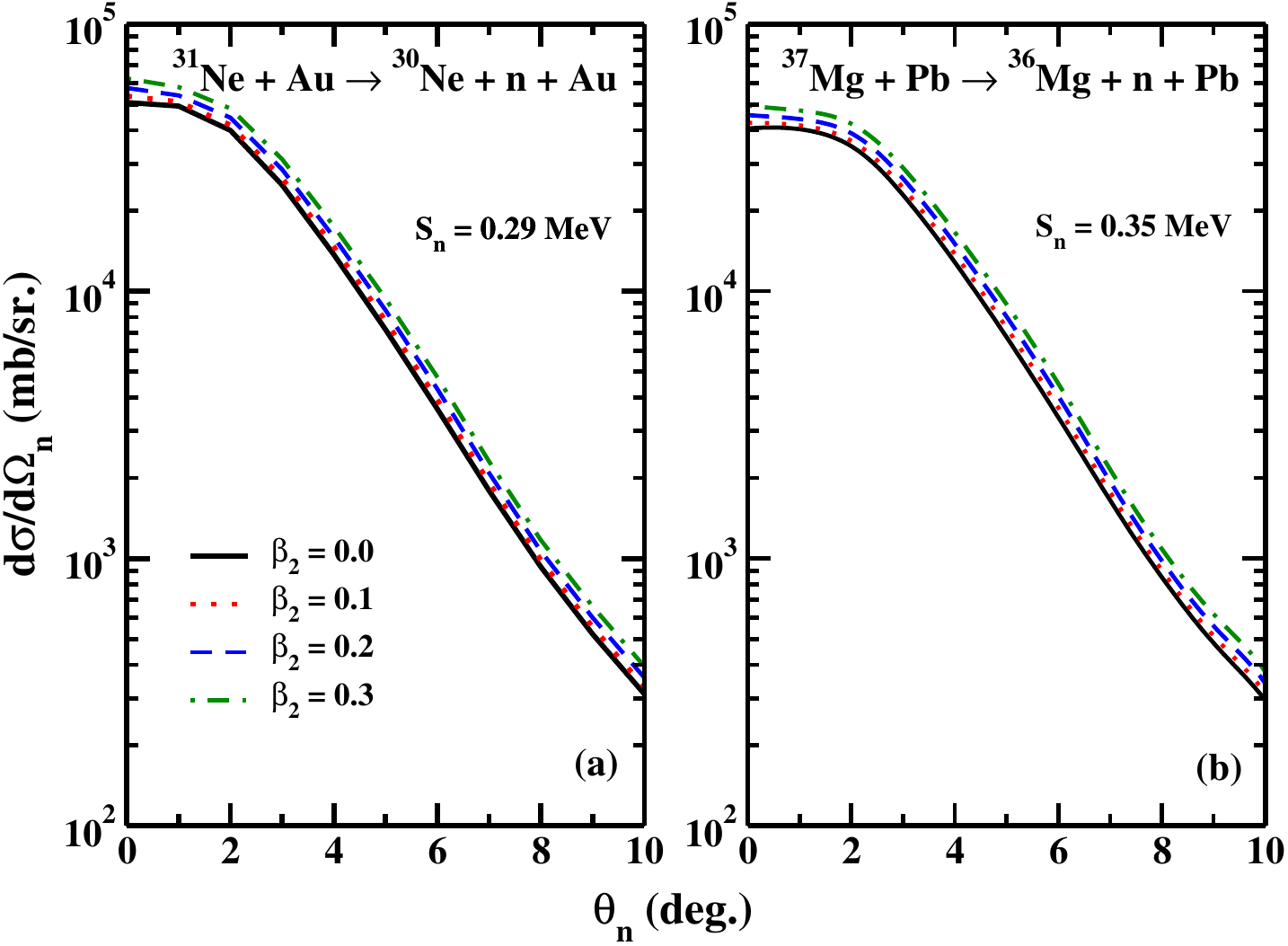}
\caption{\label{ang} Neutron angular distribution for Coulomb breakup of (a) $^{31}$Ne on Au target at 234 MeV/nucleon beam energy (b) $^{37}$Mg on a Pb target at 244 MeV/nucleon beam energy.}
\end{figure}

\subsubsection{Neutron angular distribution }
In Fig.~\ref{ang}, we present the neutron angular distributions corresponding to the breakup processes shown in Fig.~\ref{eng_ang}, using the same values of $C^2S$ and $S_n$ as before. The results are displayed for four different values of the quadrupole deformation parameter $\beta_2$, and the angular range is extended up to $10^\circ$ in both panels. The angular distribution of neutrons emitted in Coulomb breakup reactions provides valuable insight into the momentum distribution of the fragments in the ground state of the projectile~\cite{Esbensen1991}. As such, these distributions can offer further information regarding the halo structure of the neutron in these exotic nuclei.

Additionally, it is important to note that neutron angular distributions, when analyzed in conjunction with relative energy spectra, can help constrain the one-neutron separation energy ($S_n$) of the projectile. Compared to methods relying solely on the position of the peak in the $B(E1)$ strength distribution - which may be affected by higher-order excitations - the angular distribution method is considered to be less model dependent~\cite{Nak99,Nakamura2012}. We emphasize that in the FRDWBA calculations the relative energy spectra account for contributions from the full non-resonant continuum.

From the figure, we observe a sharp decline in the cross-section with increasing neutron angle in the forward direction. The angular distributions remain narrow, particularly for neutron angles below the grazing angle (around $1^\circ$–$2^\circ$ in these reactions), which reflects the narrow parallel momentum distributions and, by Heisenberg's uncertainty principle, indicates a large spatial extension of the valence neutron. One can also observe that the influence of projectile deformation is most prominent at forward angles.

{It is important to mention that, within the present framework, the enhancement of the breakup cross sections with increasing deformation primarily arises from the deformation-dependent dynamical couplings entering through the interaction potential. Since the projectile ground state wave function is approximated by its dominant weakly bound component, static core-deformation effects are not included in a fully self-consistent manner. Therefore, the present treatment differs from particle-rotor or XCDCC-type approaches \cite{Diego2014PRC,Pesudo2017PRL,MORO2020}, where explicit core-excitation channels are consistently incorporated in both the projectile structure and the reaction dynamics.}

Let us now take a break from the one-neutron halos, and delve into the world of Borromean halo systems encountered in the medium mass region.

%\newpage
\subsection{Two neutron Borromean halos}\label{2n-halo}
%\subsubsection{$^{29}$F, $^{31}$F, $^{39}$Na, $^{40}$Mg by Jagjit $^{62}$Ca, $^{72}$Ca by Horiuchi-san}
%\js{In certain neutron-rich nuclei, an exotic structure known as a \textit{Borromean two-neutron halo} can emerge. 
%These are unique, intriguing quantum systems that consist of a bound state formed by a core nucleus and two neutrons, where neither of the two-body subsystems (core-neutron or neutron-neutron) are themselves bound. This counterintuitive configuration leads to an extended matter distribution, dominated by neutrons, and results in an enhanced root-mean-square (rms) radius. This enhanced rms radius is experimentally confirmed by an unusually large reaction cross section ($\sigma_{R}$).}

%\js{In certain neutron-rich nuclei, an exotic structure known as a \textit{Borromean two-neutron halo} can emerge. 
\textit{Borromean two-neutron halos} appear in certain neutron-rich nuclei, where two neutrons are loosely bound to a core, but neither the two-neutron system nor the individual neutron-core systems are bound on their own. This unique arrangement is a sensitive probe of nuclear forces and how the shell structure evolves far from stability, as well as to understand the effects of neutron correlations in weakly bound systems \cite{Hansen1987,JSinghthesis,JCasalthesis, Ogawathesis,Horiuchi2006,Horiuchi_2007,Casal2020,GSingh22PLB,GSingh2024EPJA,Casal19B, Matsumoto2004, Sagawa2015, Hagino2005, Esbensen1997,Pinilla2016,Zhu93,Suz91,Baye2009,Grigorenko2020,MYO2014,Ji2014,Maridi2025}. See also references cited within \cite{Sagawa2025EPJA }. 
%\WH{(and cite earlier studies which can be obtained from these references cited above and below.)} Please add W.Horiuchi, Y. Suzuki, Phys. Rev. C 76, 024311 (2007), 
 
Well known examples of two-neutron {Borromean} halos in the lighter mass region (outside the island of inversion) include $\nucl{6}{He}$ \cite{Aum99, Meister2002,Hagino2005,Horiuchi_2007,Fort2014,Sun21,JSingh2016, JSinghthesis}, %\WH{W.Horiuchi, Y. Suzuki, Phys. Rev. C 76, 024311 (2007) and earlier studies by Russians, Hagino-san, etc.} 
$\nucl{11}{Li}$ \cite{Esbensen1997,Myo2002,Hagino2005,Nakamura2006} %\WH{(Many earlier studies, including Myo-san, Hagino-san, etc.)}, 
$\nucl{14}{Be}$ \cite{Labiche2001}, $\nucl{17}{B}$ \cite{Yamaguchi2004}, $\nucl{19}{B}$ \cite{Coo20, Casal19B,Hiyama2019},%\WH{(Hiyama-san's work)} 
and $\nucl{22}{C}$ \cite{Horiuchi2006, Tanaka2010, Yamashita2011,Ershov2012,Acharya2013,Ogata2013,Kucuk2014,TOGANO2016, JSingh19FBS, Naga2018}. %\WH{(See Naga2018 for relavant references)}. 
Such two-neutron {Borromean} halos are closely tied to changes in nuclear shell structure, often involving the weakening or vanishing of traditional magic numbers due to the intrusion of orbitals from higher shells. A striking case is $\nucl{11}{Li}$, where the classic $N=8$ shell gap collapses and the intruding $2s_{1/2}$ orbital drives the formation of the halo. Recent studies have extended this phenomenon to heavier medium mass systems at the boundary of the island of inversion, such as $\nucl{29}{F}$ \cite{Bagchi2020,JSingh2020,Casal2020,Fortunato2020}, revealing a Borromean halo at $N=20$ and signaling the erosion of this major shell closure via the intrusion of the $2p_{3/2}$ orbital. These discoveries challenge conventional shell-model expectations and provide valuable insights into the evolving structure of nuclei far from stability.

\begin{table}[htbp]
\caption{ Compilation of three-body systems ($^{A}_ZX_N \rightarrow ~^{A}_ZX_{N-2}+n+n$) for $9\leq~Z~\leq~12$. Where $X$, $A$, $Z$, and $N$ are system, mass number, atomic number, and neutron number, respectively. Third column corresponds to neutron core-configurations in naive shell-model picture. $J^\pi$ is the spin-parity of the ground state of $^{A}X^Z_N$ adopted from NNDC. All two-neutron separation energies ($S_{2n}$) for $^{A}_ZX_N$ are compiled from Ref.~\cite{Wang2021}. $nl_j$ are the orbital angular momentum and total angular momentum of the halo orbit for $^{A}_ZX_N$.}
\centering
\begin{tabular}{cccccc}
\\[-0.5ex]
\hline\hline\\[-1.5ex]
Nucleus (Core) & (core-configuration)$_\nu$ &$J^\pi$ & $S_{2n}$& $nl_j$& Ref.\\  
$^{A}_ZX_N$ ($^{A}_ZX_{N-2}$)       &  && (MeV)& & \\ [0.5ex]\hline
\\[-0.5ex]
%%%%%%%%%%%%%%%%%%%%%%%%%%%%%%%%%%%%%%%
$^{29}_{~9}$F$_{20}$ ($^{27}_{~9}$F$_{18}$)   & $(1s_{1/2})^2(1p_{3/2})^4(1p_{1/2})^2$ & $...$ &$1.130\pm0.540$ & $2p_{3/2}$&\cite{Bagchi2020,JSingh2020, Fortunato2020,Casal2020}\\
&$(1d_{5/2})^6(2s_{1/2})^2(1d_{3/2})^2$&&\\[1ex]
%%%%%%%%%%%%%%%%%%%%%%%%%%%%%%%%%%%
$^{31}_{~9}$F$_{22}$ ($^{29}_{~9}$F$_{20}$)   & $(1s_{1/2})^2(1p_{3/2})^4(1p_{1/2})^2$ & $...$ &$-0.550\#\pm0.100\#$ & $2p_{3/2}$&\cite{Wang2021,GSingh22PRC}\\
&$(1d_{5/2})^6(2s_{1/2})^2(1d_{3/2})^4$&&bound\cite{Ahn2019}\\[1ex]
%%%%%%%%%%%%%%%%%%%%%%%%%%%%%%%%%%%
$^{39}_{11}$Na$_{28}$ ($^{37}_{11}$Na$_{26}$)   & $(1s_{1/2})^2(1p_{3/2})^4(1p_{1/2})^2$ & ... &$-0.700\#\pm0.280\#$ \cite{Wang2021} & $2p_{3/2}$&\cite{Ahn2022,Zhang2023, JSingh2024, JSingh2025arxiV}\\
&$(1d_{5/2})^6(2s_{1/2})^2(1d_{3/2})^4$&&bound\cite{Ahn2022}\\
&$(1f_{7/2})^6$&&\\[1ex]
%%%%%%%%%%%%%%%%%%%%%%%%%%%%%%%%%%%
$^{40}_{12}$Mg$_{28}$ ($^{38}_{12}$Mg$_{26}$)   & $(1s_{1/2})^2(1p_{3/2})^4(1p_{1/2})^2$ & $0^+$ &$0.670\pm0.710$ & $2p_{3/2}$&\cite{Baumann2007,Macchiavelli2022, Yuan2024,JSingh2024,JSingh2026ACPB}\\
&$(1d_{5/2})^6(2s_{1/2})^2(1d_{3/2})^4$&&\\
&$(1f_{7/2})^6$&&\\[1ex]
%%%%%%%%%%%%%%%%%%%%%%%%%%%%%%%%%%%
$^{62}_{20}$Ca$_{42}$ ($^{60}_{20}$Ca$_{40}$)   & $(1s_{1/2})^2(1p_{3/2})^4(1p_{1/2})^2$ & $...$ &... & $3s_{1/2}$&\cite{Horiuchi2022a, Hove2018, Hagen2013}\\
&$(1d_{5/2})^6(2s_{1/2})^2(1d_{3/2})^4$&&\\
&$(1f_{7/2})^8(2p_{3/2})^4(2p_{1/2})^2$&&\\
&$(1f_{5/2})^6$&&\\[1ex]
%%%%%%%%%%%%%%%%%%%%%%%%%%%%%%%%%%%
$^{72}_{20}$Ca$_{52}$ ($^{70}_{20}$Ca$_{50}$)   & $(1s_{1/2})^2(1p_{3/2})^4(1p_{1/2})^2$ & $...$ &... & $3s_{1/2}$&\cite{Horiuchi2022a,Hove2018,Hagen2013}\\
&$(1d_{5/2})^6(2s_{1/2})^2(1d_{3/2})^4$&&\\
&$(1f_{7/2})^8(2p_{3/2})^4(2p_{1/2})^2$&&\\
&$(1f_{5/2})^6,(1g_{9/2})^{10}$&&\\[1ex]
%%%%%%%%%%%%%%%%%%%%%%%%%%%%%%%%%%%
\hline\hline
\end{tabular}
\label{Tborro}
\end{table}

%\js{Recent studies have shown that this phenomenon is not limited to light nuclei. In $\nucl{29}{F}$, the traditional magic number $N=20$ weakens due to the intrusion of the $2p_{3/2}$ orbital, facilitating halo formation: a finding that parallels the shell-gap erosion seen in $\nucl{11}{Li}$
The nucleus $\nucl{29}{F}$ has been identified as the heaviest known two-neutron Borromean halo till date \cite{Bagchi2020,JSingh2020,Casal2020,Fortunato2020,Luo2021}. This observation stems from observations of the disappearance of the $N=20$ shell gap on the low-$Z$ side of the island of inversion (as highlighted in the red circle on the green block in Fig.~\ref{Fig1.0}). The observation of a Borromean halo in $\nucl{29}{F}$ has sparked considerable theoretical interest in exploring whether similar structures exist beyond this point. In particular, interest has recently focused on the low-$Z$ side of the $N=28$ shell closure, where certain sodium (Na) and magnesium (Mg) isotopes \cite{JSingh2024} (highlighted by red circles on the orange block in Fig.~\ref{Fig1.0}) as well as the nucleus $\nucl{31}{F}$ \cite{Masui2020,GSingh22PRC,Michel2020}, have emerged as strong candidates for exhibiting two-neutron {Borromean} halo structures. This exploration extends even further to more massive systems, including isotopes of Ca \cite{Hagen2013,Hove2018,Horiuchi2022a}, as well as predicted halo structures in isotopes of Al, Si, P, and S \cite{Li2024}. A summary of the systems considered in this review is provided in Table~\ref{Tborro}. These developments highlight a broader trend: the emergence of halo phenomena in heavier and more complex nuclear systems, driven by the evolving shell structure far from stability. They also underscore the importance of Borromean systems as sensitive probes of nuclear forces and shell evolution in extreme neutron-rich environments.

\begin{figure}[h!]
\centering
\includegraphics[width=0.9\linewidth]{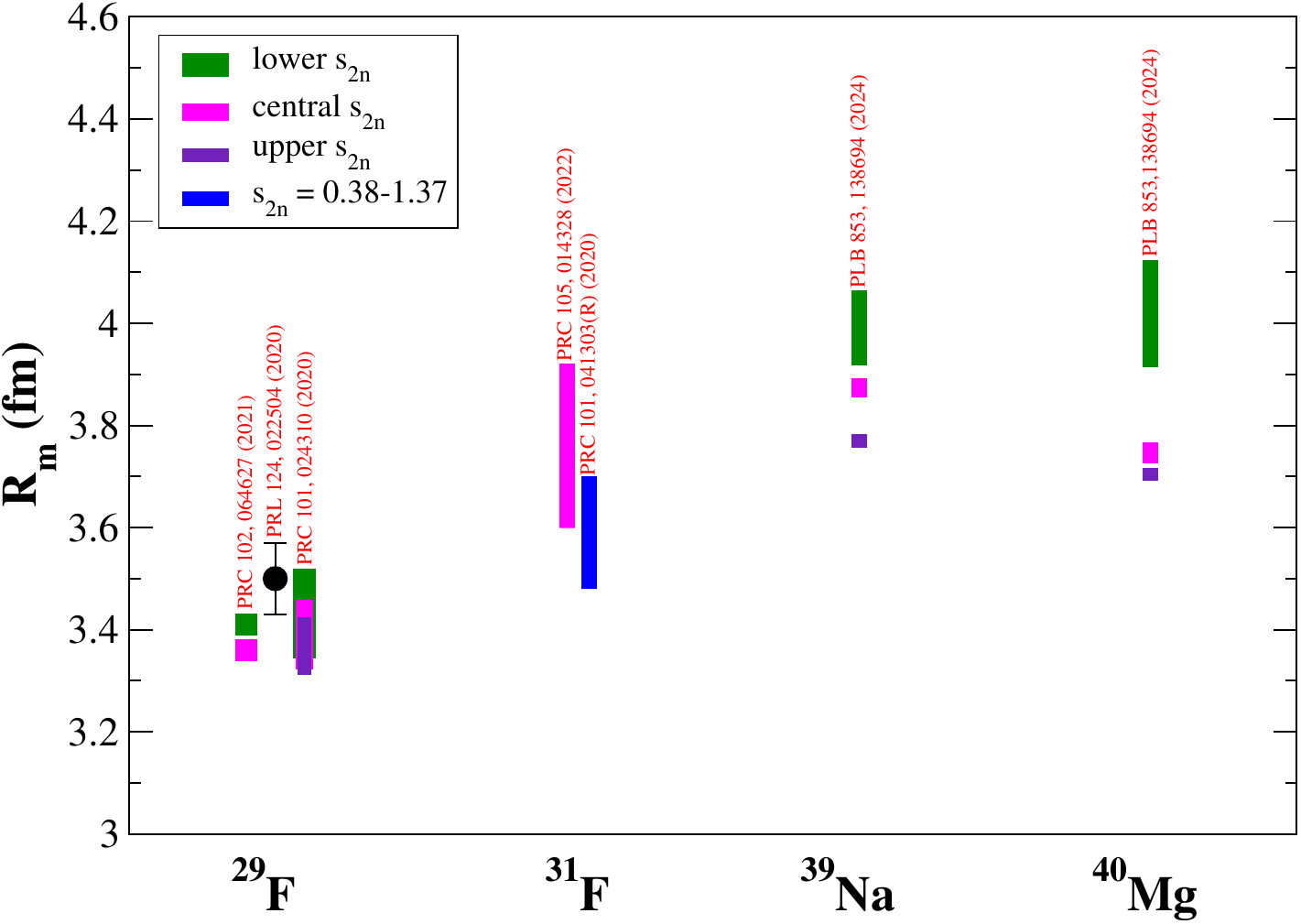}
\caption{The matter radii ($R_m$) for various isotopes of F ($^{29}$F and $^{31}$F), Na ($^{39}$Na), and Mg ($^{40}$Mg). The experimental value shown in the black circle is taken from Ref.~\cite{Bagchi2020}.} 
\label{Fig1.14}
\end{figure}

The spectral properties of core$+$neutron (core$+n$) subsystems play a crucial role in determining the structure of Borromean three-body nuclei. This is because the core$+n$ interaction enters explicitly into the three-body Hamiltonian. Therefore, it is essential to construct a core$+n$ potential that describes the low-lying continuum spectrum of systems such as $\nucl{28}{F}$, $\nucl{30}{F}$,  $\nucl{38}{Na}$, $\nucl{39}{Mg}$, $\nucl{61}{Ca}$, and $\nucl{71}{Ca}$. In all of these cases, the separation between the core and the valence neutron is not well defined. For simplicity, an inert-core approximation is commonly adopted. In this approach, any effects due to internal core rearrangements or core-valence exchange are effectively absorbed into the parameters of the angular momentum dependent potential. This method has also been applied in previous three-body studies, such as the modeling of the $\nucl{14}{Be}+n$ subsystem in the description of $\nucl{16}{Be}$ \cite{Lovell2017}. Additionally, the spin of the unpaired proton in odd-$Z$ isotopes is typically neglected, as the focus is solely on neutron degrees of freedom. This simplifies the construction of the three-body ground-state wave function, which is treated as a $0^+$ state. In cases where experimental data is scarce or unavailable, the core$+n$ interaction is modeled using a Woods-Saxon potential with both central and spin-orbit components. For specific details on the mathematical form of the Woods-Saxon geometry and the associated parameters used in various systems, the reader is referred to the original sources. For the $\nucl{28}{F}$ case, see Table~1 of Ref.\cite{JSingh2020} and the discussions in Refs.\cite{Fortunato2020, Casal2020}. The parameters for $\nucl{30}{F}$ are given in Table~1 of Refs.\cite{GSingh22PRC,Masui2020}. For $\nucl{38}{Na}$ and $\nucl{39}{Mg}$, relevant data can be found in Table~1 of Ref.\cite{JSingh2024}, while for $\nucl{61}{Ca}$, and $\nucl{71}{Ca}$, the core$+n$ potential details are discussed in Sec.~III-A of Ref.\cite{Horiuchi2022a}.

The second binary interaction in the three-body Hamiltonian, in addition to the core$+n$ potential, is the neutron-neutron (nn) interaction, $V_{nn}$. For systems such as $\nucl{29}{F}$, $\nucl{31}{F}$,  $\nucl{39}{Na}$, $\nucl{40}{Mg}$,  the Gogny-Pires-Tourreil (GPT) potential \cite{Gogny1970}, which includes central, spin-orbit, and tensor components, is employed. This choice is consistent with earlier three-body studies \cite{Casal2013, Lovell2017, FACE2004} and widely used in the literature. In contrast, for $\nucl{31}{F}$ in the work by Masui et.al, \cite{Masui2020}, and for $\nucl{62}{Ca}$ and $\nucl{72}{Ca}$ \cite{Horiuchi2022a}, the Minnesota interaction \cite{Thompson1977} is used. {Although the GPT interaction formally incorporates a richer operator structure and partial-wave dependence than the purely central Minnesota force, the limited number of partial waves included in the present three-body basis means that only part of this angular-momentum dependence is effectively explored, reducing the practical difference between the two interactions in the calculations.}

In addition to binary core$+n$ and $V_{nn}$ potentials, it is customary to introduce a ternary potential ($V_{3b}$) to account for possible effects that
are not explicitly included in the above mentioned two-body interactions. To reproduce the experimental or target ground state energy, a phenomenological three-body force is introduced. This is typically modeled as a simple Gaussian potential, with its strength adjusted freely to match the desired two-neutron separation energy $s_{2n}$. It is important to note that the specific choice of the $V_{nn}$ interaction has a relatively minor impact on the ground-state properties of the three-body systems, as long as the interaction reproduces low-energy $nn$ scattering observables reasonably well \cite{Zhu93}. This has been confirmed through comparisons using alternative $nn$ potentials, such as those in Refs. \cite{Garrido1997,Garrido2004}. Although these alternatives tend to produce slightly less binding, requiring correspondingly stronger three-body forces to recover the same energy, the resulting wave functions are nearly identical to those obtained with the GPT interaction in terms of radii and partial-wave structure.

%\js{Considering the limited information available, we model the core+$n$ interaction as a Woods-Saxon potential including only central and spin-orbit terms,
%\begin{equation}
%    V_{{\rm core}+n}(r) = \left(-V^{(l)}_0+V_{ls}\lambda_\pi^2 \vec{l}\cdot\vec{s}\frac{1}{r}\frac{d}{dr}\right)\frac{1}{1+\exp[(r-R_c)/a]}\,,\label{eq:WS}
%\end{equation} where $V^{l}_0$ is in general $l$-dependent, and $R_c=r_0A_c^{1/3}$  with $A_c$  the mass number of the core. The spin-orbit interaction is written in terms of the Compton wavelength $\lambda_\pi=1.414$~fm. Following Ref.~\cite{Horiuchi10}, the spin-orbit strength is taken to follow the systematic trend \cite{BOHR} and has the value $V_{ls}=$ ..... MeV for ..... The value $r_0=1.25$~fm is as originally suggested for $\nucl{31}{Ne}$ ~\cite{Horiuchi10,JSingh2020}, and we examine different scenarios, ..... In Set ..., $V_0$ is chosen to be $l$-independent. In set .. it is adjusted to fix the ... ground-state resonance of .. at $0.129$\,MeV, corresponding to the prediction of Ref.~\cite{Fossez2016}. This set gives a \textit{``normal"} shell-model scenario, with an additional $p_{3/2}$ resonance appearing at higher energy, $...$\,MeV.}  \gs{One can see that Eq. \ref{eq:WS} closely resembles Eq. \ref{eq:WSPot}.}
%\subsubsection{$^{29}$F}
%$^{29}$F~\cite{Bagchi2020, JSingh2020, Fortunato2020, Casal2020}
%\subsubsection{$^{39}$Na}
%$^{39}$Na~\cite{Zhang2023,Zhang2023b, JSingh2024}
%\subsubsection{$^{40}$Mg}
%$^{40}$Mg~\cite{JSingh2024}

With all these ingredients in place the core$+n$ potential, the neutron–neutron interaction, and the phenomenological three-body force we compute the ground state matter radii for $\nucl{29}{F}$, $\nucl{31}{F}$,  $\nucl{39}{Na}$, $\nucl{40}{Mg}$ within the hyperspherical formalism by using an analytical transformed harmonic oscillator basis \cite{JCasalthesis}. For $\nucl{31}{F}$, results within the cluster orbital shell model framework \cite{Masui2020} are also shown. The results for different choices of the two-neutron separation energy ($s_{2n}$) are shown in Fig.~\ref{Fig1.14}. It is important to note that setting the three-body potential strength $V_{3b}=0$ typically leads to an overbinding of the system. Therefore, a repulsive or attractive three-body term ($V_{3b}>0$ or $V_{3b}<0$) is required to reproduce $s_{2n}$ values consistent with available evaluations. For instance, in the case of 
$\nucl{29}{F}$, the target $s_{2n}$ values range from 0.40 to 2.09 MeV (with central value 1.44 MeV); for $\nucl{31}{F}$, values span from 0.38 to 1.37 MeV (with central value 0.15 MeV).  Similarly, for  $\nucl{39}{Na}$, the range is from 0.01 to 0.82 MeV, and for $\nucl{40}{Mg}$, the adopted values are 0.01, 0.67, and 
1.38 MeV. These values reflect significant uncertainties in the evaluated separation energies, as reported in Ref.~\cite{Wang2021}.  Given this, we systematically study the sensitivity of the matter radius and reaction cross sections to changes in $s_{2n}$, by varying the strength of the three-body interaction accordingly.

As shown in Fig.~\ref{Fig1.14}, the calculated matter radii agree well with available experimental data, especially for $\nucl{29}{F}$. This agreement lends credibility to the three-body models and supports the predictive power of the frameworks. Consequently, our predictions for $\nucl{31}{F}$,  $\nucl{39}{Na}$, $\nucl{40}{Mg}$ provide valuable guidance for future experimental campaigns targeting these exotic systems.
%\js{With all these ingredients, the three-body results for the ground state matter radii of $\nucl{29}{F}$, $\nucl{31}{F}$, $\nucl{39}{Na}$, $\nucl{40}{Mg}$, $\nucl{62}{Ca}$, and $\nucl{72}{Ca}$  using the scenarios for the core$+n$ potentials are shown in Fig~\ref{Fig1.14}. As already discussed, there are large uncertainties in the evaluated $\stn$ value \cite{Wang2021} of ...... Given these uncertainties, we discuss the sensitivity of the matter radius and reaction cross sections of the ground state with $\stn$, using different values of the three-body potential strength. For $V_{3b}=0$ leads to an overbinding of the system, so we need to choose $V_{3b}>0$ to get $\stn$ corresponding to the lower, central and upper limits of the evaluation. Note that the lower limit would result in an unbound state. However, the experimental findings support a bound ground state, so we have considered a barely bound case as well. For $\nucl{39}{Na}$, due to the lack of experimental information and systematic information on $\stn$, it seems reasonable to keep the range of $V_{3b}$ values determined for $\nucl{40}{Mg}$ and make predictions. With this prescription, the shallowest case becomes unbound for all sets. To restrict the situation to bound states, we modify $V_{3b}$ in that case so that $\stn$ is $0.010$MeV, since we are interested in giving predictions for a bound halo nucleus, which could be tested experimentally. We predict $\stn$ for $\nucl{39}{Na}$ between $0.010$-$0.824$ ($1.828$)\,MeV without (with) $V_{3b}=0$ cases.

\begin{figure}[ht]
\centering
\includegraphics[width=0.9\linewidth]{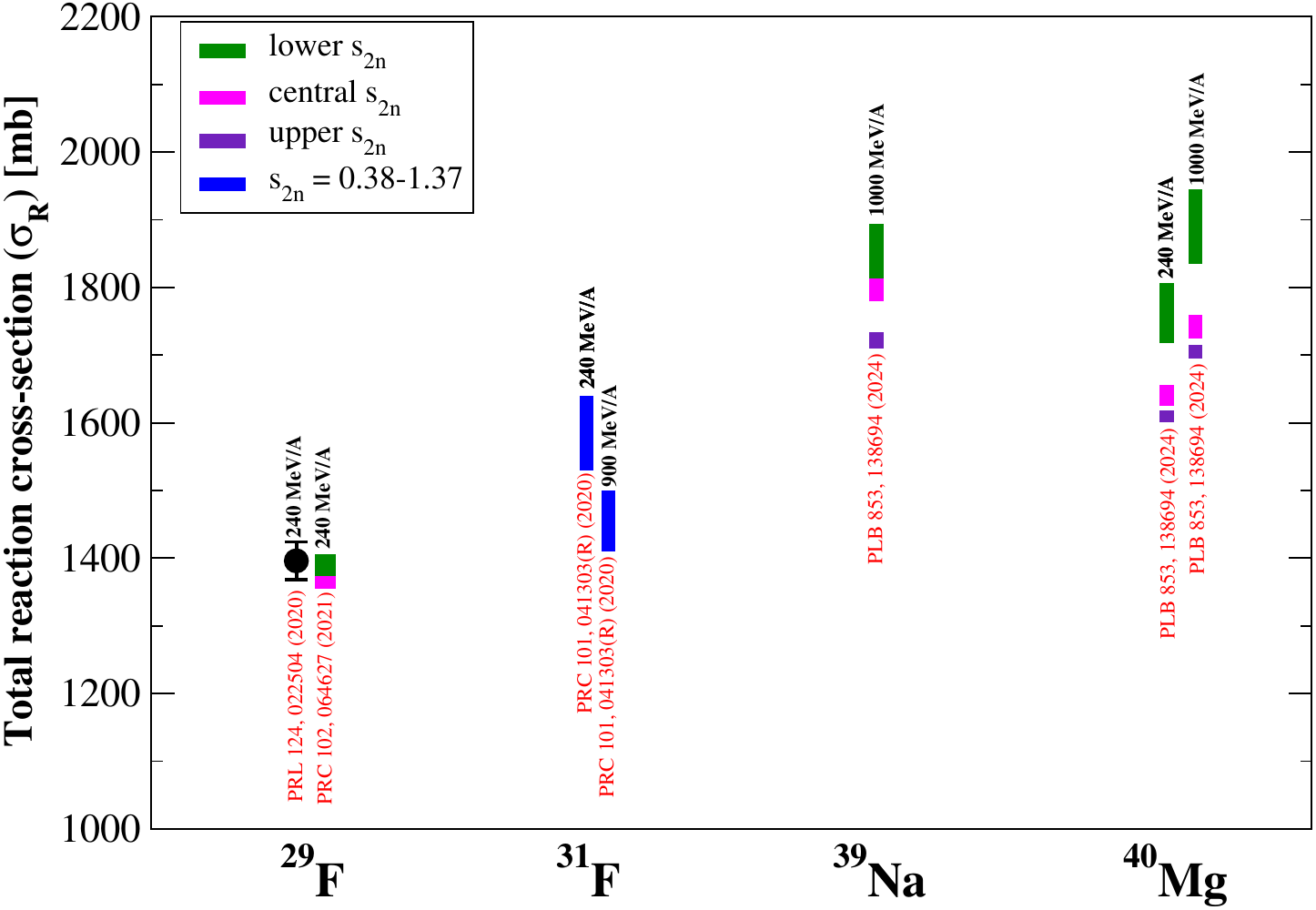}
\caption{The total reaction cross section ($\sigma_R$) for various isotopes of F ($^{29}$F and $^{31}$F), Na ($^{39}$Na), and Mg ($^{40}$Mg) at different incident energies. The experimental value shown in the black circle is taken from Ref.~\cite{Bagchi2020}.} 
\label{Fig1.14b}
\end{figure}

As also shown in Fig.~\ref{Fig1.14}, the expected trend is clearly observed, ground states with lower $s_{2n}$ (i.e., more weakly bound states) exhibit larger matter radii, while more deeply bound configurations correspond to smaller radii. This behavior holds consistently across all the nuclei studied. More notably, the sensitivity of the matter radius to $s_{2n}$ becomes particularly pronounced in cases where shell inversion occurs. Nevertheless, due to the relatively heavy mass of the core in these systems, the overall matter radii remain quite similar across different structural scenarios. Even when a significant neutron halo is present, the variation in total matter radius for a fixed $s_{2n}$ remains modest. This limited variation suggests that distinguishing between different structural configurations based solely on measured radii may be experimentally challenging. Therefore, complementary observables such as those extracted from nucleon knockout or transfer reactions are highly valuable. These reactions can provide insight into the partial-wave composition of the ground state, offering a way to constrain theoretical models and discriminate between competing wave function scenarios discussed in this work.

Experimentally, a very obvious way to determine whether a nucleus is a halo nucleus is to look for an enhanced reaction cross section. Thus, we examine the total reaction cross section ($\sigma_R$) by employing the conventional Glauber theory.
The nucleon-target formalism is utilised, and the nucleon-nucleon profile function is applied in all cases. The additional theoretical inputs required for this reaction model include the density distributions of both the projectile and target nuclei. For this, we first generate a harmonic-oscillator (HO) type density distribution for core nuclei (such as $\nucl{27}{F}$, $\nucl{29}{F}$, $\nucl{37}{Na}$, and $\nucl{38}{Mg}$) that reproduces the measured total reaction cross section where available by assuming simple shell model configurations.% that reproduces the measured total reaction cross section of $\nucl{38}{Mg}$$+$$\nucl{12}{C}$ by Takechi \textit{et al.,}  at $240$MeV/nucleon ($1535\pm 21$ mb) by assuming a simple shell model state $(1f_{7/2})_\nu^8\otimes(1d_{5/2})_\pi^4$. }

The density for the three-body system was constructed by simply adding the $2n$ densities calculated within a three-body model to the core density. No center-of-mass correction was applied in these calculations, but we believe this is a safe strategy for a system with such a heavy core. Using this prescription, we predict the $\sigma_R$ for different nuclei at different incident energies, shown in Fig.~\ref{Fig1.14b}. As can be seen from Fig.~\ref{Fig1.14b}, the predicted values of $\sigma_R$ for all systems show significant enhancement for shallower cases for different choices of incident energy. The calculated $\sigma_R$ agree well with available experimental data for $\nucl{29}{F}$. This shows that we can use these reaction cross sections as an indication of the melting of the $N=20$ and $28$ shell closures.

{Furthermore, two-neutron Borromean halo nuclei can be more comprehensively characterised through fully exclusive measurements that extend beyond total reaction cross sections and directly probe neutron-neutron ($n$-$n$) and core-neutron correlations within appropriate reaction frameworks. In particular, coincidence experiments in nucleon-knockout reactions, such as ($p,pn$), and in Coulomb dissociation measurements, can provide event-by-event reconstruction of the three-body final-state kinematics, enabling observables sensitive to the relative motion of the valence neutrons with respect to the core. Within established reaction models and the spectator approximation, momentum-space correlations between the removed neutron and the core can be related to the underlying spatial opening-angle distribution, offering a direct probe of di-neutron versus cigar-like configurations. Complementary sensitivity is provided by dipole strength distributions extracted from Coulomb breakup measurements, which constrain the average $n$-$n$ opening angle and hence the degree of spatial correlation. Such exclusive approaches, combined with improved reaction theory descriptions, are expected to clarify the spatial localisation of pairing correlations, in particular their enhancement in the low-density surface region of neutron-rich Borromean systems. For an overview of these recent developments, the reader is referred to Section-9 of Ref.\cite{Moro2025EPJA} and the references therein.}

\subsection{Bubble nuclei} \label{bubble}

We now turn our attention to another exotic feature in medium-mass nuclei, that of `bubble nuclei'. They are characterized by a significant reduction in the density of their centers, which challenges the traditional quantum liquid model of atomic nuclei. The concept of a bubble was first introduced by Wilson in late 1940's, where he theorized that nuclei be collected in a hollow shell rather than a sphere, much like a soap `bubble' \cite{Wilson1946PR,Wilson1950PR}. Of course, it was later found that for bubble nuclei, the shell is not hollow and that the density is slightly reduced at the center. Within the liquid drop model, Fullerene like structures made of $\alpha$ particles were proposed by Walter Greiner for $Z$ = 120 \cite{Greiner2001}, and for a range of parameters depicting a soliton-like structure, it is speculated that such an anatomy for bubble nuclei could also house a core-like structure where all the near magic number of protons and neutrons are held within a cage, surrounded by a neutron gas \cite{Misicu2018Book}. Toroidal shapes have also been proposed as possible manifestations of bubble nuclei \cite{Misicu2024Part,Wong1985PRL}. The reality of bubble nuclei was later investigated in several studies \cite{Anon1974Nat,Shukla2014PRC,Grasso2009PRC,Decharge1999PLB,Saxena2018PLB,Choudhary2024EPJ,Ren2024PLB,Choi2022PRC}. Eventually, experimental verification of the presence of a bubble in $^{34}$Si \cite{Mutschler2017NatPhys} established the concept on a firmer footing, which was also corroborated by theoretical calculations \cite{Duguet17PRC}. This unusual structure arises because nucleons occupy higher angular momentum orbits, which peak at the nuclear surface, rather than low angular momentum orbits that concentrate in the interior. As a result, these nuclei exhibit sharper surfaces with reduced diffuseness. There is a clear relationship between the occupation of orbital angular momentum states and the nuclear surface diffuseness \cite{Choudhary2020}.
 
Using the close relationship between the internal density and the surface density, we can obtain information about the bubble structure of the nucleus where the center density is remarkably low \cite{Choudhary2020}. The size of the bubble is usually measured by a depletion fraction (DF), which relates the decrease in the density at the center of the nucleus to the maximum density (near the surface). It is defined as \cite{Saxena2018PLB,Shukla2014PRC}:

\begin{align}
    DF = \frac{\rho_{max} - \rho_c}{\rho_{max}},
\end{align} 
where $\rho_{max}$ and $\rho_c$ represent the maximum and central densities, respectively. This DF is sensitive to quantal effects, because the density fluctuation is related to filling up of the single particle energy levels near the Fermi surface. It is well known that only the $s$-orbitals ($\ell = 0$) have a non-zero radial wave function peaking at the origin, which should result in an increased density at the center. Thus, the bubble structure of light nuclei is caused by the deficiency of the number of $s$-orbitals occupying near the center of the nucleus. The lack of occupation of $s$-orbitals or low angular momentum orbitals pushes them up in energy as they are not favored by a self-consistent mean field potential. This shifting of the $s$-orbitals to higher energy levels reduces the interior density and pushes the matter outward slightly increasing the radius \cite{Shukla2014PRC}. In addition to this condition, the $s$-orbitals near the Fermi surface must be enclosed by high $\ell$ orbitals (the higher, the better) that are well separated in energy from nearby single particle states to avoid strong dynamical correlations \cite{Saxena2018PLB}. In other words, the maximum of the occupied levels must occur at large distances from the center. Therefore, the surface distribution of the bubble nucleus is expected to have a more distinct nuclear surface than the surrounding nuclei. For example, to establish the bubble structure in $^{22}$O, it is essential to analyze the systematic proton elastic scattering cross-section for oxygen isotopes. We remark that a global analysis of the nuclear bubble structure is given in Ref. \cite{Ebata2025}.
It is necessary to investigate the occupancy of orbitals near the Fermi surface for obtaining information on such nuclear structure. This is reflected in the density distribution near the nuclear surface.

{Ref. \cite{Choudhary2020} uses $^{28}$Si as a test case to establish how bubble structure manifests in reaction observables. Two HO configurations are mixed — one with an empty $1s$ orbit (maximum bubble) and one with a filled $1s$ orbit (non-bubble) — with a mixing parameter $\alpha$ controlling the degree of depletion. A generalized bubble parameter G, which can take both positive and negative values, is introduced to quantify this. When proton-nucleus reaction probabilities are examined, the internal density depression is found to be essentially invisible at small impact parameters due to strong absorption, consistent with black-disc behaviour. However, the bubble density has a characteristically sharper nuclear surface, and this surface difference imprints itself on the elastic scattering differential cross section: the more bubble-like the nucleus, the larger the cross section at the first diffraction peak. This establishes a practical hadronic probe for the bubble structure.}

Unlike halos, which are yet to be found in heavy nuclei, bubble structures have been predicted and are prevalent throughout the nuclear chart in low, medium, heavy, superheavy and hyperheavy regions \cite{Saxena2018PLB,Shukla2014PRC,Misicu2024Part,Choudhary2024EPJ}. However, like halos, bubbles can be deformed. In general, pairing correlations and deformations are supposed to weaken the bubble structure, as then the $s$-orbitals get mixed with higher angular momentum orbits, making it difficult for them to be empty. Nevertheless, it is possible that the $s$-wave mixing due to the deformation is quite weak and a bubble character might still manifest. As an example, for a spherical nucleus with $N$ or $Z$ = 14, the 1d$_{5/2}$ orbital is fully filled while the 2s$_{1/2}$ is fully empty, promoting a bubble. Then for introductions of small to moderate deformations, the mixing of these $s$- and $d$-orbitals is rather small and the bubble structure might still remain. $^{24}$Ne, $^{32}$Si and $^{32}$Ar is strong candidates for deformed bubbles in the light mass region \cite{Shukla2014PRC}.

Bubble structure is also known to affect the spin-orbit potential. The strength of the potential is proportional to the gradient of the central nuclear field and thus, a depletion in density at the center pushes the spin-orbit strength to go opposite in sign compared to its surface contribution \cite{Shukla2014PRC}. Further, bubble formation, typically, lowers the Coulomb energy of the nucleus \cite{Perera2022PRC}. For astrophysics point of view, bubble candidates like $^{20}$N are known to influence neutron capture reaction rates due to their vividly different diffuseness \cite{Choudhary2024EPJ}.

To detect possible bubble nuclei electron scattering on an unstable nucleus is a powerful tool. However, direct measurement of the nuclear charge distribution using electron scattering is somewhat limited but total reaction cross-sections and proton elastic scattering cross-sections are advantageous in that they can be applied to a wide range of mass regions if the beam intensity is sufficient \cite{Choudhary2020}. Systematic cross-section measurement can be used as a method of nuclear spectroscopy, and the information on nuclear size properties can be used as a new measure of nuclear structure. {Charge exchange reactions, such as ($p,n$), ($n,p$), or even ($^3$He,$t$) and ($t$,$^3$He), could also be a viable option for studying bubble nuclei because they probe the isovector (neutron-proton asymmetric) components of the nuclear density and response \cite{Liu2022PRC}.}

\section{Astrophysical relevance of reaction theories} \label{sec: astro}

We now progress to analyzing the astrophysical significance of medium mass nuclei, seen from a perspective of applications of nuclear reactions. It is known that neutron rich isotopes in the low and medium mass region serve as stepping stones for the formation of heavy elements in the various astrophysical scenarios \cite{Terasawa_2001,Sasaqui2005AJ,ARCONES2017PPNP}. Along a given reaction network, these exotic nuclear structures play a critical role in determining the abundance of various isotopes, particularly for the neutron capture processes of nucleosynthesis, viz., the $s$- and the $r$-process. Between them, the $r$-process primarily takes place in extreme astrophysical environments, such as core-collapse supernovae and neutron star mergers \cite{Abbott2017PRL,Cowan21RMP}. It is responsible for most of the nuclei heavier than iron, away from the valley of stability. Since it is significantly dependent on the neutron capture rates, the efficiency of the $r$-process is heavily influenced by the structure of participating exotic nuclear systems. Shell structure modifications, $\gamma$-decay strengths or properties like halo characters, deformations, central density depletion (bubble), etc., alter reaction rates and thereby affect elemental abundances \cite{Chatterjee2020AA,rbarman_2023,Barman_2024}. Nevertheless, our information repositories lack sufficient details about such properties and traits for each and every nucleus in question.

The lacunae in our knowledge database for nuclear attributes require that exotic nuclei be studied systematically so that their properties, as inputs to reaction networks and abundance calculations, be known precisely. However, considering neutron capture reactions specifically, they usually occur in the $T_9$ temperature range\footnote{In astrophysics, temperatures are measured and denoted by $T_3$, $T_6$, $T_9$, $T_{10}$ and so on, where $T_9$ = $10^9$\,K, $T_{10}$ = $10^{10}$\,K, etc.}, which roughly equates to 0.1\,MeV in the energy regime. As such, it is extremely difficult to conduct direct capture experiments at this low energy range, prompting the search for alternative approaches. This is where reaction theories come in handy. Various indirect methods have been developed to handle the problem \cite{Ber16JPc,Casal2013,Bert86NPA,Trib2014RPP}. An elegant modus operandi among these is the Coulomb dissociation technique \cite{Bert86NPA,Shyam1999} that has been successful in studying various $(n,\gamma)$ reactions under the umbrella of the FRDWBA theory \cite{Dan24EPJ,GSingh17PRC,Shubh24FBS,Dan19PRC,Shyam1999,Ban08PRC,Ban98PRC,Neelam15PRC,Chatterjee2020AA}. Therefore, we now briefly describe the process of extracting the $(n,\gamma)$ radiative capture reaction rates using FRDWBA. This approach will then be used to analyze the sensitivity of nuclear physics inputs to nucleosynthesis calculations.

%\textbf{ Introduction to Exotic Nuclear Structures and Their Role in Nucleosynthesis}
%Exotic nuclear structures play a critical role in determining the abundance of elements in astrophysical scenarios, particularly in the $r$-process \cite{Cowan21RMP}. The formation of heavy elements relies on neutron capture reactions, where neutron-rich isotopes serve as stepping stones along the reaction network. This compilation of research investigates various neutron-rich nuclei, their structural properties, and their implications for nucleosynthesis.
%The r-process primarily takes place in extreme astrophysical environments, such as core-collapse supernovae and neutron star mergers. It involves successive neutron captures followed by beta decays, moving isotopes towards heavier elements. The efficiency of the r-process is significantly influenced by the nuclear structure of isotopes near the neutron drip line. Properties such as halo structures, deformation, and central density depletion (bubble nuclei) alter reaction rates and thereby affect elemental abundances.

Let us revisit the triple differential cross-section for the \textit{a} + \textit{t} $\longrightarrow$ \textit{b} + \textit{c} + \textit{t} reaction, which is given by Eq. \ref{tdcrs} above and is written as 
\begin{equation}
\frac{d^3\sigma}{dE_{b}d\Omega_{b}d\Omega_{c}} = \frac{2\pi}{\hbar v_{at}}\rho{(E_{b},\Omega_{b},\Omega_{c})}\sum_{\ell,m}|\beta_{\ell m}|^{2},
\label{eq:tdcrs1}
\end{equation}

where $v_{at}$ is the relative velocity between the projectile and the target in the initial or entrance channel, while $\rho{(E_{b},\Omega_{b},\Omega_{c})}$ is the three-body final state phase space factor \cite{Fuchs82NIM}. As averred, $\beta_{\ell m}$ being the reduced transition amplitude from the initial to the final state and \textit{$\ell$} is the angular momentum of the wave function that has a projection \textit{m}. In the bra-ket notation, $\beta_{\ell m}$ is written as \cite{Chatterjee2018},

\begin{equation}
\beta_{lm}(\textbf{q}_{b},\textbf{q}_{c};\textbf{q}_{a}) = \langle \chi_{b}^{(-)}(\textbf{q}_{b},\textbf{r})
\chi_{c}^{(-)}(\textbf{q}_{c},\textbf{r}_{c})\vert
 V_{bc}(\textbf{r}_{1})\vert
\phi_{a}^{lm}(\textbf{r}_{1})\chi_{a}^{(+)}(\textbf{q}_{a},\textbf{r}_{i})\rangle.
\end{equation}

%Here, $\textbf{q}_{i}$'s are the wave vectors in Jacobi coordinate system and $\chi_{i}$'s are the pure Coulomb distorted wave of the ith particle (\textit{i=a,b,c}). $V_{bc}$ is the Woods-Saxon potential without deformation and $\phi_{a}$ is the ground state wave function of the projectile(\textit{a}). %Above reduced transition amplitude can be written in integral form also. 
%From the above equation we can easily find the relative energy spectra by integrating over solid angle of \textit{b} and \textit{c} and which is written as
Using Eq. \ref{tdcrs}, one can obtain the relative energy spectrum by multiplying with the appropriate Jacobian \cite{Ban08PRC}.
\begin{equation}
\frac{d\sigma}{dE_{rel}}\ = 
\frac{2\pi}{{\hbar}\textit{v}_at}
%\int\dfrac{\rho_{(E_{b},\Omega_{b},\Omega_{c})}}{2l+1}%
\int\sum_{l,m}|\beta_{lm}|^{2}\frac{1}{2l+1}{d\Omega_{bc}d\Omega_{at}}\frac{\mu_{bc}\mu_{at} p_{bc}p_{at}}{\hbar^6} \label{eq:rel}
\end{equation}
Taking the above reaction as a pure Coulomb breakup, the relative energy spectrum $({d\sigma}/{dE_{rel}})$ can then be related to the photodisintegration\footnote{Or photo-dissociation and which, in terms of nuclear physics, can also be called photo-nuclear. For the emission of a neutron, it can also be interchangeably called photo-neutron.} cross-section as \cite{Bert86NPA},

\begin{equation}
\frac{d\sigma}{dE_{rel}} = \frac{1}{E_{\gamma}}\sum_{\pi,\lambda}{\sigma^{\pi\lambda}_{(\gamma,n)}}{n_{\pi\lambda}},\label{eq:rel_cap}
\end{equation}

where $n_{\pi\lambda}$ is the virtual photon number with $\pi$ = \textbf{E} or \textbf{M} with $\lambda$ being the order of polarity. The $\gamma$ - energy is
$E_{\gamma}= E_{rel} + S_n$, where $(S_n)$ is the one neutron separation energy and $E_{rel}$ is the relative energy between fragments \textit{b}-\textit{c} in the final channel.
%Sum of one neutron separation energy $(S_n)$ with relative energy between fragments is written as $E_{\gamma} (= E_{rel} + S_n)$.
For a single multipole and one type of transition, only one term will remain in the above equation. 
Knowing the photodisintegration cross-section, the time reversed radiative capture cross-section can be computed using the principle of detailed balance \cite{Bert86NPA} as,
%Now the time reversal process of photo-disintegration is known as radiative capture, which can be studied using the principle of detailed balance knowing the photodisintegration cross-section, the radiative capture cross-section is given by

\begin{equation}
\sigma_{(n,\gamma)} = \frac{2(2\textit{j}_a+1)}{(2\textit{j}_b+1)(2\textit{j}_c+1)}\frac{\textit{k}_\gamma^2}{\textit{k}^2}{\sigma_{(\gamma,n)}}, \label{eq:cap}
\end{equation}
where $\textit{j}_{i}$'s are total spins $\textit{(i=a,b,c)}$, $\textit{k}_\gamma$ is wave number of photon while ${\hbar^2}k = {2\mu_{bc} E_{rel}}$ with $\mu_{bc}$ being the reduced mass of the \textit{b}-\textit{c} system.

It is pertinent to note that the condition of dominance of a single multipole for the application of CD as an indirect method may not be too restrictive. If there are contributions from additional multiples, one could always measure or calculate the relevant Coulomb breakup cross section on a different (heavy) target and extract the photodisintegration cross sections corresponding to the other multipoles. After all, the virtual photon number would depend only on the projectile-target combination and, of course, the beam energy of the projectile.

Finally, the reaction rate for the radiative capture reaction   $(\textit{b} + \textit{c}  \longrightarrow  \textit{a} + \gamma)$ is,

\begin{equation}
R=N_A \langle\sigma_{(n,\gamma)}\textit{v}\rangle, 
%(1+\delta_b_c)^-^1
\end{equation}
where $N_A$ is the Avogadro number. $\langle\sigma_{(n,\gamma)}\textit{v}\rangle$ is known as the reaction rate per particle pair and is averaged over the Maxwell-Boltzmann velocity distribution \cite{RolfsBook}. We can write it in unit of $cm^3$ $s^{-1}$ as

\begin{equation}
\langle\sigma_{(n,\gamma)}\textit{v}\rangle=\sqrt{8/\pi\mu(k_BT)^3}\int_{0}^{\infty}\sigma_{(n,\gamma)}(E_{rel})E_{rel}exp(-E_{rel}/k_BT)dE_{rel},
\end{equation}
where, $k_B$ is the Boltzmann constant and $T$ is the temperature. The possibility of computing radiative capture involving excited states of a nucleus from Coulomb breakup calculations has been studied in Ref. \cite{Neelam15PRC}. For more details, one is referred to \cite{Ban98PRC,GSingh17PRC,Dan19PRC,Dan24EPJ,Shubh24FBS}. 

%\textbf{Sensitivity of Nuclear Physics Inputs to Nucleosynthesis Calculations}

%The review \gs{RC and MD} on the effect of exotic nuclear structure in determining element abundances underscores the need for improved reaction models incorporating deformation, halo effects, and detailed nuclear configurations. A study on the A $\approx$ 30 mass region, including Na, Mg, and Al isotopes, demonstrates that neutron capture reactions dominate at lower temperatures ($T_9 = 0.62$), pushing isotopes toward the neutron drip line. These findings emphasize the importance of structure-dependent reaction rates in determining final abundance patterns.

%\gs{We should merge the following with what is written above.}\\

\subsection{Sensitivity of reaction rates to structure inputs}
With the theoretical framework of calculating reaction rates via the lens of the FRDWBA, let us try and analyze the sensitivity of nucleosynthesis projections to nuclear structure inputs. As already stated, the importance of considering precise nuclear physics inputs like neutron capture rates, beta-decay rates, mass models, including structure information that goes into these models such as level density, exact separation energies, spin-parities, $\gamma$-strength functions, for accurate nucleosynthesis studies is clearly understood by the community \cite{Pogliano23PRC,GORIELY97AA,Larsen19PPNP,arcones_2011,Mumpower2016PPNP,Sasaqui2005AJ,Terasawa_2001,ARCONES2017PPNP}. In fact, the Photo-Absorption of Nuclei and Decay Observation for Reactions in Astrophysics (PANDORA) project \cite{Tamii2023EPJA} has been initiated to study the photo-nuclear reactions of light and medium mass nuclei ($A <$ 60) to limit the discrepancies between different experimental and theoretical models for observables like photoabsorption
cross-sections and decay branching ratios. Such studies highlight that uncertainties and small changes in the structural parameters can significantly result in vastly distinct rates and impact abundance calculations. It is interesting to see, from a nuclear reaction point of view, the origin of some of these differences to understand the sensitivity of nuclear astrophysics outcomes to some fundamental structural parameters. To that effect, we try and see the impact of these different structural parameters on the same observable (i.e., the capture cross-section) and explore the transmittance of these behaviors to the final reaction rates. This is illustrated beautifully in Fig. \ref{fig: cap_var}. The variations in the three parameters chosen for the study in this figure were chosen to represent the possible uncertainties in the g.s. structure of the weakly bound exotic nucleus $^{34}$Na that lies near the neutron dripline in the island of inversion and is, therefore, an ideal candidate for the study.

Hence, in panels a1), b1) and c1) of Fig. \ref{fig: cap_var}, we show the radiative capture cross-section, obtained from Eq. (\ref{eq:cap}), for the $^{33}$Na$(n,\gamma)$$^{34}$Na radiative capture reaction using the elastic Coulomb dissociation of $^{34}$Na on $^{208}$Pb at 100\,MeV/nucleon beam energy. In the right column, panels a2), b2) and c2) then show the corresponding reaction rates deciphered from these capture cross-sections. The mentioned reaction is analyzed for capture cross-section and reaction rate behavior with variations in the one neutron separation energy, quadrupole deformation and the ground state spin-parity of $^{34}$Na.

It is evident from the left panels of Fig. \ref{fig: cap_var} that even slight variations in these structural parameters can cause significant changes in the obtained capture cross-sections. Of utmost interest is panel a1) of this figure where the cross-sections seem to flip the pattern. For a majority of the spectrum of the relative or the center of mass energy, the higher the $S_n$ value, the higher is the cross-section. Conventionally, it would have made sense to assume that a more tightly bound nucleus should have a higher probability of formation and thus, have a higher capture cross-section. However, this higher relative energy region is not the portion that contributes the most to the subsequent reaction rates: it is the low relative energy region below and around 1\,MeV that is significant when computing the reaction rates for weakly bound nuclei \cite{GSingh17PRC,Shubh17PRC,Dan19PRC,Neelam15PRC}. The behaviour of the capture cross-section is completely opposite in this low energy region, with higher $S_n$ values resulting in lower cross-sections. {This is because the cross-section is related inversely to the $S_n$ via Eq. \ref{eq:rel_cap} (recall, that this is valid for a single multipole dominated reaction; see also the pioneering work by Bertulani \textit{et al.} in Ref. \cite{Bert86NPA}). Qualitatively, when a nucleus is more weakly bound (i.e., has a lower separation energy), it has a higher tendency to dissociate easily, and thus, at low relative energies of the constituting fragments, should also have a higher capture probability that is manifested by a higher capture cross-section. The analytic cause of this flip is explained in Ref. \cite{GSingh17PRC}, where it is shown that all components of Eq. \ref{eq:rel_cap} play a role in determining the nature and position of the flip. The third panel in Fig. \ref{fig: cap_var}, panel b1) shows the variation with $\beta_2$ while the fifth displays the effect of different{\footnote{By `different', we mean a simple coupling of the $^{33}{\rm Na}(3/2^+)$ core-spin with a $2p_{3/2}\nu (l=1, j=3/2)$ neutron leading to \textit{possible} $J^\pi = 0^-, 1^-, 2^-, 3^-$ for $^{34}{\rm Na}$.}} $J^{\pi}$} on the cross-sections. The changes are easily deciphered by the naked eye, with higher $\beta_2$ and $J^{\pi}$ values bearing higher cross-sections. Larger deformations are known to impact interaction cross-sections and thus, it is suitable that they also result in larger capture cross-sections. On the other hand, the comparatively larger differences in the curves in panel c1) are a direct consequence of the total spin factors in Eq. (\ref{eq:cap}). {Of course, one must also keep in mind that the information about the spin of the projectile also enters via the radial part of the wave function (cf. Eq. \ref{radial}).

\begin{figure}[htbp]
\centering
\includegraphics[width=0.81\linewidth]{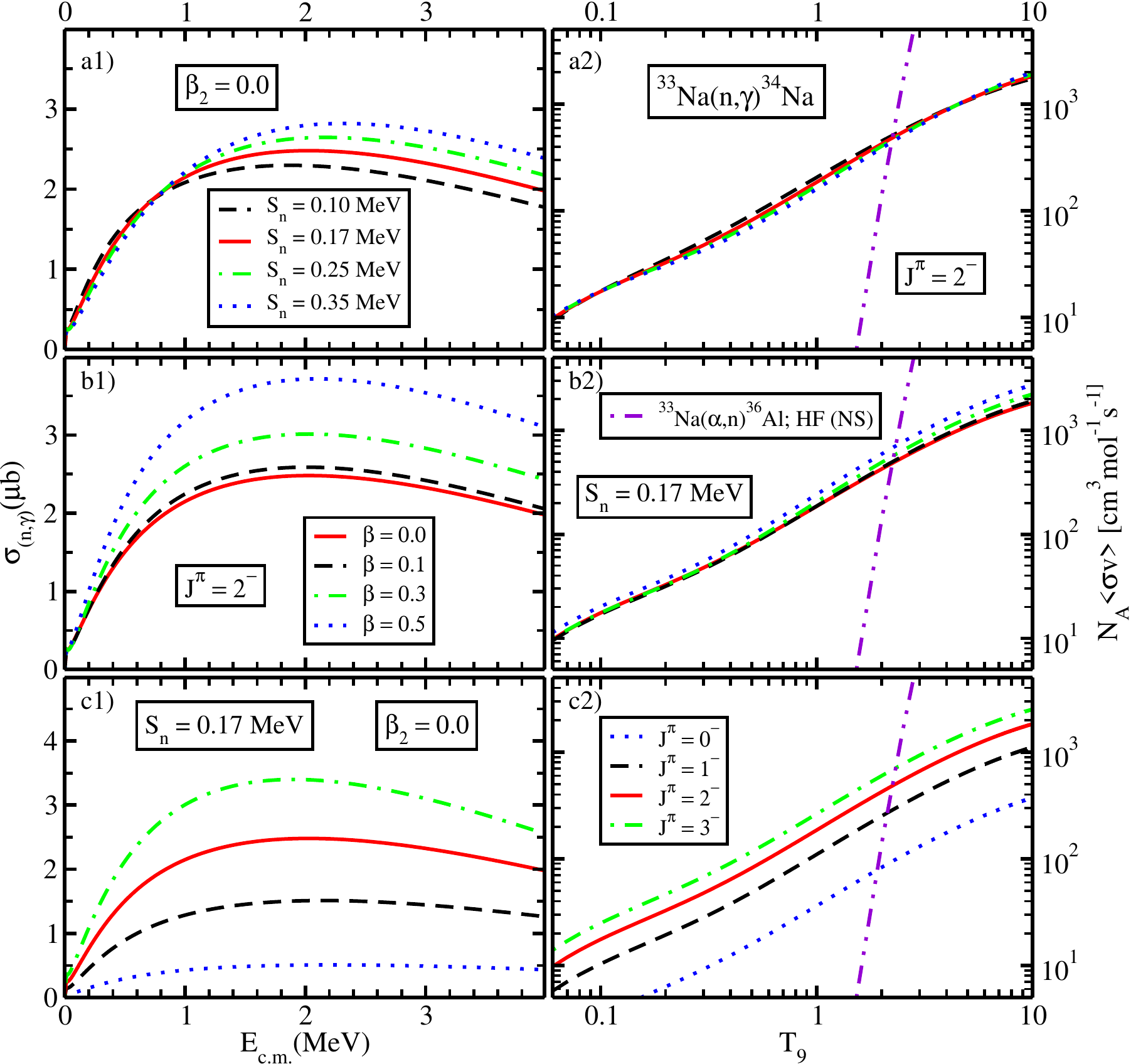}
\caption{\label{fig: cap_var} The variation of the capture cross-section and reaction rate due to differences in the structural parameters of a weakly bound nucleus. The study is done for the $^{33}$Na$(n,\gamma)$$^{34}$Na radiative capture reaction using the elastic Coulomb dissociation of $^{34}$Na on $^{208}$Pb at 100\,MeV/nucleon beam energy. Panels a1), b1) and c1) show the capture cross-sections while the parallel panels a2), b2) and c2) exhibit the corresponding reaction rates. The parameters kept constant for a particular case are mentioned in each parallel panel set a), b) and c). Clearly, capture cross-sections are highly sensitive to the structure information that goes in, and this sensitivity propagates to the rate calculations showing it is imperative that precise nuclear structure inputs be used for reaction rate calculations. To study the nucleosynthesis direction, the reaction rate for the $\alpha$-capture reaction $^{33}$Na$(\alpha,n)$$^{36}$Al obtained by Hauser-Feshbach method is also shown by the double-dot-dashed line in the right column. The low rate of the latter at the equilibrium temperature indicates that nucleosynthesis for the Sodium isotopic chain goes towards the drip line.}
\end{figure}

One may argue that the g.s. spin-parity and one neutron separation energy of $^{34}$Na are relatively known compared to other exotic nuclei lying near the drip line ($J^{\pi}$ being supposedly 2$^-$ and $S_n$ being 0.17(35)\,MeV \cite{Door14PTEP,Gaudefroy12PRL}), however the unique value proposition of Fig. \ref{fig: cap_var} is to show that variations in these structural parameters affect capture cross-sections quite appreciably. These significant changes would then invariably propagate when calculating the reaction rates from these cross-sections \cite{GSingh17PRC}. This is seen aptly in right panels a2), b2) and c2) of Fig. \ref{fig: cap_var}. The behavior of the reaction rates (expectedly) follow exactly the same pattern as the capture cross-sections, albeit from the low relative energy region. That is why in panel a2) we have a more tightly bound nucleus having a lower reaction rate, in panel b2) a nucleus with more deformation has a higher rate corresponding to a higher cross-section. Finally, panel c2) implies that the higher the spin, the higher the reaction rate.

{Thus, capture cross-sections are highly sensitive to the structure information that goes in, and this sensitivity propagates to the rate calculations showing it is imperative that precise nuclear structure inputs be used for reaction rate calculations. It is clearly seen} that even though the changes in the capture cross-sections due to variations in the structural parameters are minor, the corresponding changes that manifest in the reaction rates are appreciably large, even bearing an order of magnitude difference in some cases like the $J^{\pi}$. If one was then to compute the abundances with these rates, evidently the differences would be further amplified. Hence, properly constrained structural inputs are a key to nuclear reactions and the ensuing reaction rates, and subsequently, to nucleosynthesis studies. Specifically, limiting the uncertainties in reaction rates is of paramount importance for $r$-process nucleosynthesis \cite{Larsen19PPNP}, the impact of these uncertainties being greater for weakly bound systems where even the basic information like g.s. spin-parities or separation energies are very uncertain. This also calls for studying each individual nucleus separately, from the structure, reaction as well as an astrophysics point of view since each of them could be a catalyst for any number of nucleosynthesis events. 

This also prompts us to compare various capture reaction rates by the same isotope to determine the nucleosynthesis path flow. Hence, in Fig. \ref{fig: cap_var}, we also compare the radiative neutron capture rate with $\alpha$-capture reaction rate by $^{33}$Na. This comparison of the $^{33}$Na$(n,\gamma)$$^{34}$Na and $^{33}$Na$(\alpha,n)$$^{36}$Al rates enables a comprehension of the course that the reaction network might take, providing a way either for dripline Sodium isotopes like $^{34}$Na and $^{35}$Na to be possibly more abundant than its isotopes lying closer to the valley of stability, or to push the formation of heavier elements with higher charge disrupting the neutron capture flow. From the figure, it is amply clear that at an equilibrium temperature of 0.62T$_9$ and mass density $\rho$ = 5.4 $\times 10^2$ g/cm$^3$ (under a neutrino-driven wind model where the main path of the $r$-process reaction network goes through extremely neutron rich nuclei \cite{Terasawa_2001}), the $^{33}$Na(n,$\gamma$) reaction rate (computed through FRDWBA) highly dominates over the $^{33}$Na$(\alpha,n)$ reaction rate (obtained from Hauser-Feshbach estimates \cite{Rauscher2001}). This should push the isotopic abundance of Sodium towards the neutron dripline, perfectly in line with the predictions of the neutrino-driven wind model \cite{Terasawa_2001}. Similar calculations were also done for the $^{18}$C$(n,\gamma)^{19}$C and $^{36}$Mg$(n,\gamma)^{37}$Mg reactions using the Coulomb breakup of $^{19}$C and $^{37}$Mg, respectively \cite{Dan19PRC, Shubh17PRC}. However, it was found that neutron capture rates from $^{18}$C and $^{36}$Mg were also much higher than their respective $\alpha$-capture counterparts, accumulating Carbon and Magnesium towards the neutron dripline, as against previous predictions of an $\alpha$-capture induced divergence towards higher charge nuclei \cite{Terasawa_2001}. 

A noteworthy aspect of the analyses in Refs. \cite{Dan19PRC,Shubh17PRC} was the comparisons of the $(n,\gamma)$ rates from the Hauser-Feshbach model with those obtained via FRDWBA. The difference in the reaction rates was stark, more than an order of magnitude at the equilibrium temperature, suggesting the requirement of not only using specific nuclear structure inputs for precise rate calculations, but also employing a microscopic (or semi-microscopic) theory to overcome the uncertainties associated with a statistical method like the Hauser-Feshbach. Having said that, one must be aware that while the Hauser-Feshbach cross-sections have contributions only from the compound nuclear formations and decay mechanisms, the Coulomb dissociation approach using the FRDWBA reproduces only the direct capture non-resonant component. However, in principle, both components can easily coexist and should be considered simultaneously if and when they are of the same order of magnitude. 

FRDWBA has also been applied successfully in the lower mass neutron rich region to study the $^{11}$B$(n,\gamma)^{12}$B and $^{10}$Be$(n,\gamma)^{11}$Be radiative capture reactions \cite{Shubh24FBS,Dan24EPJ}. The former is an important reaction in proton-boron fusion reactors \cite{McKenzie23JFE} while the latter involves $^{11}$Be, the archetypical halo \cite{Moro06NPA,Moro12PRC,MORO2020,Capel03PLB,Capel22PLB}. On the other hand, investigations with $^{19}$C highlight the impact of the halo structure of the nucleus for the $^{18}$C$(n,\gamma)^{19}$C reaction rate, whereas analysis with $^{20}$N examines its bubble structure and its effect on the $^{19}$N$(n,\gamma)^{20}$N reaction \cite{Choudhary2024EPJ}. The $^{19}$N$(n,\gamma)^{20}$N reaction is critical to the Fluorine abundances in the $r$-process \cite{Roder2016PRC}. To study the bubble characteristic and consequently, its effect on this reaction, the FRDWBA theory was coupled with the AMD method, which provided the g.s. wave function of $^{20}$N taking into account its deformation and reduced density profile at the center. Comparison with (n,$\gamma$) reaction rates obtained from a phenomenological Woods-Saxon modeled $^{20}$N, at different temperatures, helped in highlighting the effect of bubble structure on the overall rates. Thus, FRDWBA has emerged as a reliable tool to study the (n,$\gamma$) of weakly bound neutron rich nuclei. This is primarily because of its simple requirement of just the g.s. wave function of the projectile nucleus that can be obtained from a number of methods, both microscopic or phenomenological.

\subsection{Abundance calculations}

A significant challenge in modeling the $r$-process involves the reaction rates of exotic, neutron-rich nuclei. Most astrophysical network calculations use rates derived from the statistical Hauser-Feshbach model. This approach creates large discrepancies for exotic nuclei because it assumes a high density of nuclear states, whereas in reality, these weakly-bound systems have low level densities where direct capture processes likely dominate~\cite{Xu_2014}. Since the final $r$-process abundances are sensitive to the reaction rates of these light, neutron-rich species~\cite{Terasawa_2001}, incorporating a proper nuclear structure description is critical for improving the accuracy of network calculations. 

To investigate the effect of these rates, we perform a network calculation for a localized network consisting of Na, Mg, and Al isotopes (see Fig. \ref{fig:net_Na_Mg_Al}) in the region of the island of inversion. We consider the conditions of a neutrino driven wind (NDW) model in a core-collapse supernova (CCSN), following Ref~\cite{Terasawa_2001}. The NDW emerging from the core-collapse supernovae (CCSNe) is believed to be a promising site for the $r$-process nucleosynthesis~\cite{Terasawa_2001,arcones_2011, Wanajo_2012}. The NDW is generated during the initial phase of the expansion when charge particle reactions are the primary contributors to the nucleosynthesis. For a short expansion timescale, the charge particle reactions cannot produce enough seed nuclei, and the neutron-to-seed ratio is enhanced, resulting in a synthesis of heavier $r$-process elements. However, modern simulations of the NDW under realistic conditions suggest that CCSNe can only support a weak $r$-process~\cite{Roberts_2010, Arcones_2013, Shibagaki_2016}.

The abundance evolution {for the reaction network described in Fig. \ref{fig:net_Na_Mg_Al}} is calculated under thermodynamic conditions corresponding to a CCSN~\cite{Terasawa_2001}, specifically at a temperature near the $r$-process freeze-out, with the results shown in Fig. \ref{fig:abundance_Na_Mg_Al}.
\begin{figure}[htbp]
    \centering
    \includegraphics[width=0.6\textwidth]{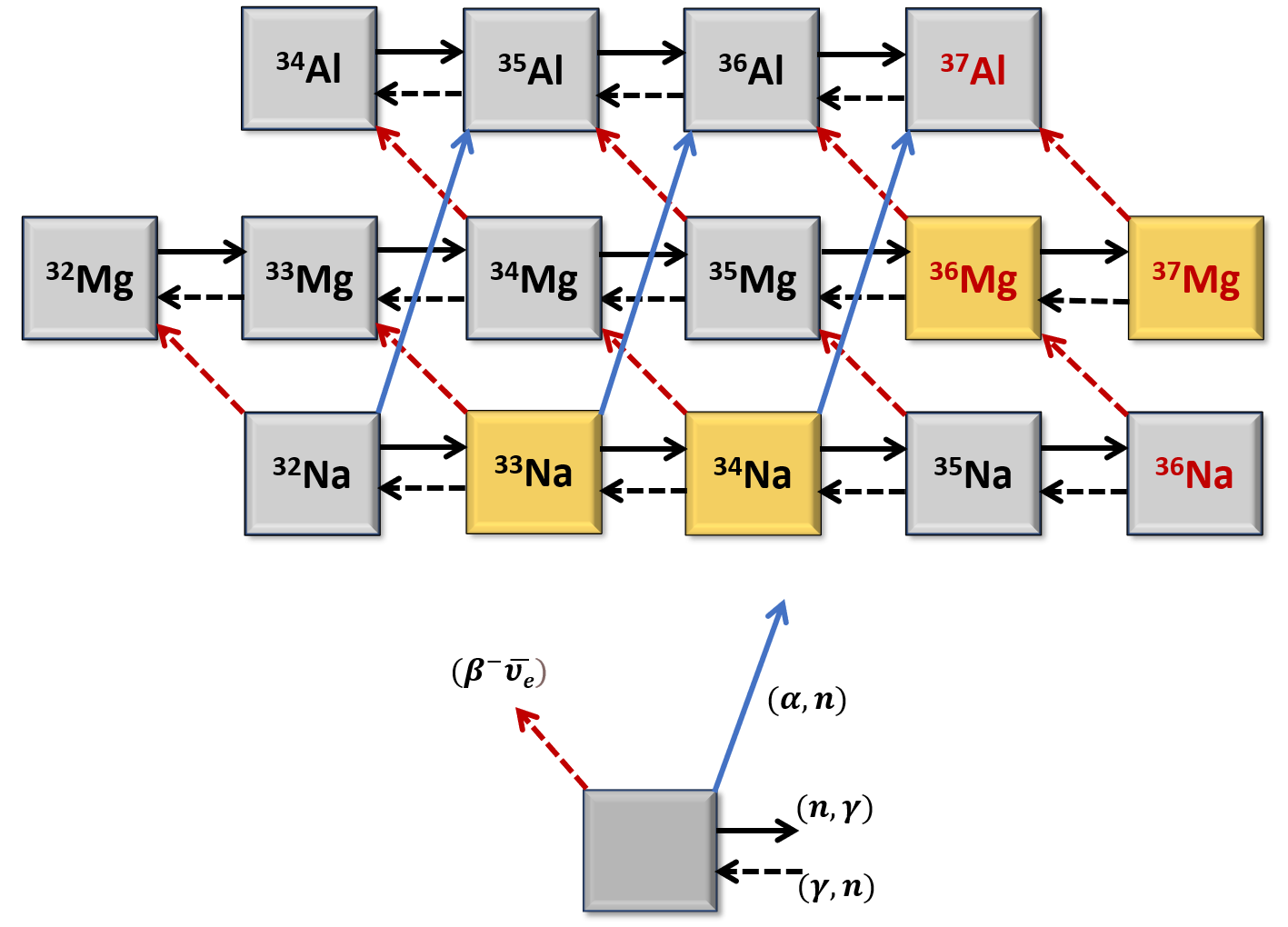}
    \caption{Limited network consisting of Na, Mg and Al isotopes.}
    \label{fig:net_Na_Mg_Al}
\end{figure}
\begin{figure}[htbp]
    \centering
    \includegraphics[width=\textwidth]{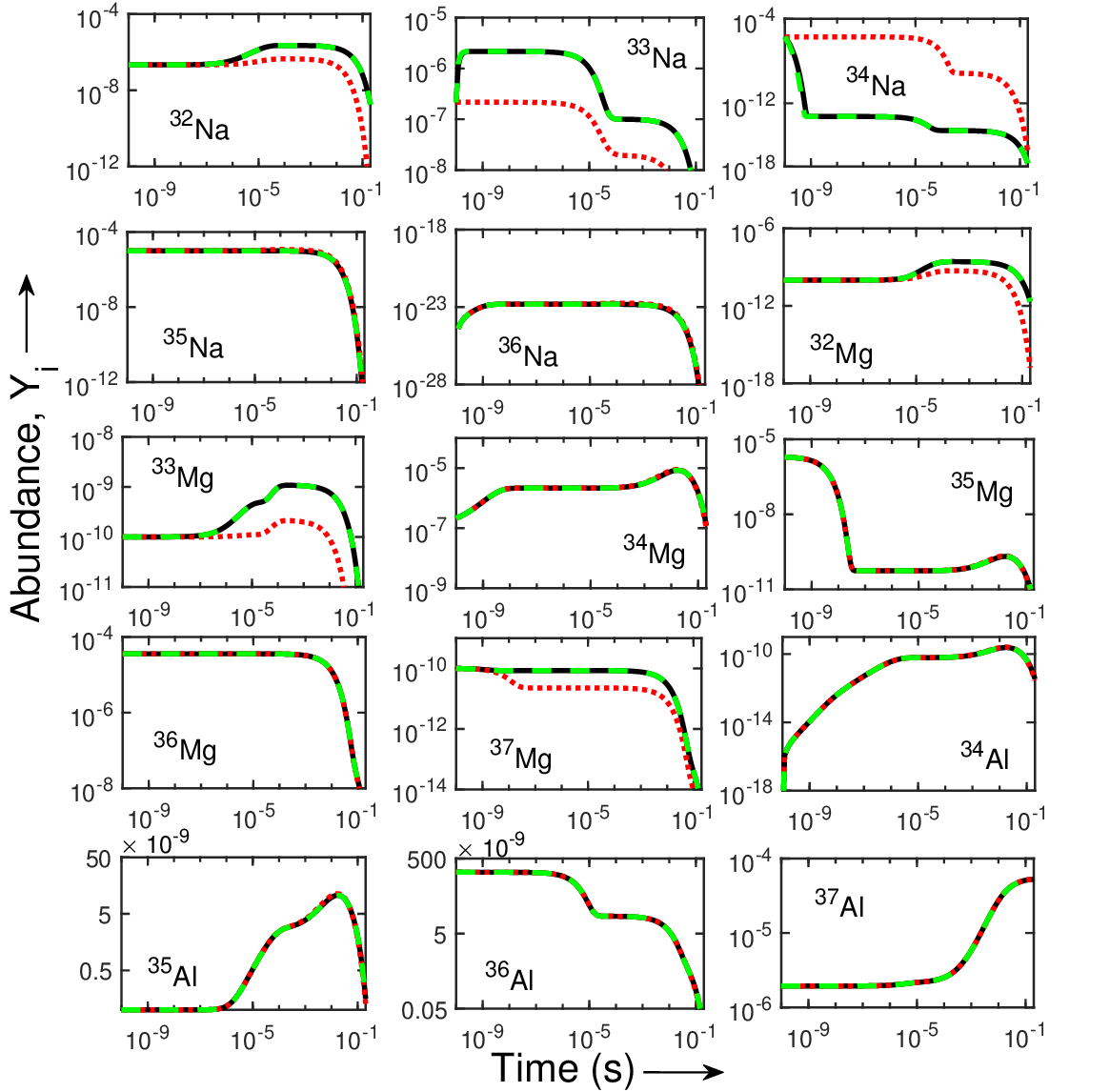}
    \caption{Abundance evolution of network shown in Fig. \ref{fig:net_Na_Mg_Al} at $T_9=0.62$. The solid black lines and the dashed green lines represent the abundances calculated using the FRDWBA rates {with deformation parameters $\beta_2 = 0$, and $\beta_2=0.5$, respectively, together with the JINA-REACLIB rates.} The dotted red lines represent the abundances calculated using only the JINA-REACLIB rates. More details are given in the text. }
    \label{fig:abundance_Na_Mg_Al}
\end{figure}
At first, the network is evolved using $(n,\gamma)$, $(\gamma,n)$, and $(\alpha,n)$ reaction rates from the JINA-REACLIB database~\cite{Cyburt_2010}, indicated by the dotted red lines. {Then for two reaction channels, i.e., for the $^{33}$Na$(n,\gamma)^{34}$Na and $^{36}$Mg$(n,\gamma)^{37}$Mg reactions, rates (including also the inverse photodisintegration rates) were computed using the FRDWBA theory. Subsequently, these FRDWBA rates for these channels are used to replace the corresponding JINA-REACLIB estimates within the given network.} 
%\gsout{two reaction channels, $^{33}$Na$(n,\gamma)^{34}$Na and $^{36}$Mg$(n,\gamma)^{37}$Mg, along with their inverse photodisintegration rates, are replaced with values obtained from the FRDWBA calculations. }
Both spherical ({$\beta_2=0$}, solid black lines) and deformed ({$\beta_2=0.5$}, dashed green lines) configurations are considered. The $\beta$-decay rates are taken from experimental data, wherever available; otherwise, they are obtained from shell model calculations~\cite{Suhonen:2007vjh1, rbarman_2023}.

{As seen in Fig.~\ref{fig:abundance_Na_Mg_Al}, the FRDWBA rates with and without deformation do not produce significant differences in the abundances, indicating that the abundances are not very sensitive to deformation at this temperature.} {However, a comparison with the results obtained using only JINA-REACLIB data shows that} the inclusion of FRDWBA-based reaction rates leads to noticeable deviations in the predicted abundances, particularly in nuclei directly influenced by the modified rates. This further highlights the sensitivity of nucleosynthesis predictions to underlying nuclear structure, especially in regions near the neutron drip line where traditional statistical models may be less reliable. A noticeable change is in the abundance patterns of isotopes in the vicinity of $^{34}\rm Na$, due to a huge difference in REACLIB and FRDWBA rates for the  $^{34}$Na$(\gamma, n)^{33}$Na reaction~\cite{rbarman_2023}. As shown in Table II of Ref.~\cite{rbarman_2023}, the FRDWBA decay constant for $^{34}$Na$(\gamma, n)^{33}$Na is higher by almost 8 orders of magnitude, and thus, $^{34}$Na is used up immediately as compared to the REACLIB results, where the $^{34}$Na abundance remains almost constant up to $10^{-5}$ s after the evolution starts. This results in higher $^{33}$Na production, as shown by solid black lines in Fig. \ref{fig:abundance_Na_Mg_Al}. This effect propagates to $^{33}$Mg, $^{32}$Na, and $^{32}$Mg via $\beta$-decay, and $(\gamma, n)$ reactions, affecting their respective abundances. Thus, incorporating these exotic structures can change the abundances and potentially alter the $r$-process pathways. {A similar study, along with a sensitivity test, was performed for a reaction network consisting of exotic neutron-rich C-N-O isotopes~\cite{Barman_JPG_2024}, where FRDWBA rates for the $^{18}$C$(n,\gamma)^{19}$C and $^{19}$N$(n,\gamma)^{20}$N reactions, along with their corresponding photodisintegration channels, were incorporated into the network to investigate the impact of the halo structure of $^{19}$C and the bubble structure of $^{20}$N on the $r$-process abundance evolution. These results emphasize the need for realistic structure inputs in $r$-process nucleosynthesis calculations.}
%In a full-scale calculation, with more than 7800 nuclei, the impact of these rates, however, seems to diminish~\cite{}.

We now discuss the $r$-process abundances calculated incorporating five $(n,\gamma)$ rates involving the exotic structures of $^{20}$N, $^{29}$Ne, $^{31}$Ne, $^{34}$Na, and $^{37}$Mg,
and their reverse reactions, calculated using FRDWBA theory, as inputs to the reaction network code SkyNet~\cite{Lippuner_2017}. Figure \ref{fig:final_YvsA} shows the final mass-integrated abundances {obtained after an evolution till $10^9$s. The abundance calculations are performed using two different sets of reaction rate inputs: one employing all reaction rates from the JINA-REACLIB database, and the other obtained by replacing the rates of the selected nuclei with the corresponding values calculated within the FRDWBA framework. The results are} compared with the solar system $r$-process data.
This comparison makes sense because the Sun (an intermediate-aged population I star, roughly 4.6 billion years old) is supposed to be made up of matter that is nearly 9-10 billion years old, given that the age of the observable universe is approximately 13.8 billion years. The elemental abundance of the solar system, particularly that of the Sun{\footnote{The Sun holds more than 99 percent of the matter in the solar system, and due to its proximity, comprehensive knowledge about its chemical composition exists.}}, thus provides a fundamental benchmark for cosmic elemental abundances \cite{Lodd2009,Lodd2020}.

\begin{figure}[htbp]
    \centering
    \includegraphics[width=\textwidth]{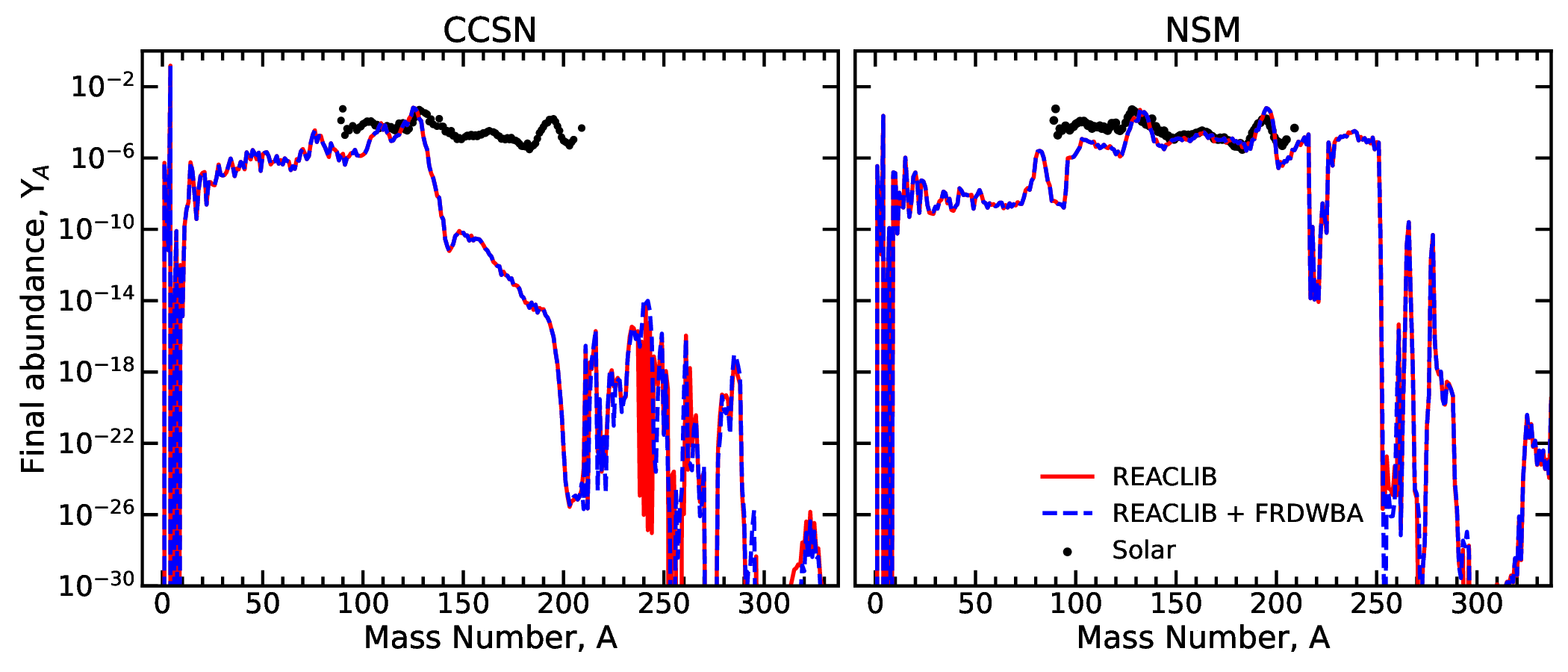}
    \caption{{Comparison of the final mass-integrated abundances obtained using only the JINA-REACLIB reaction rates (solid red line) and those obtained by incorporating the FRDWBA rates together with the JINA-REACLIB rates (dashed blue line), shown for two astrophysical scenarios: CCSN and NSM. The black points indicate the solar abundances.}}
    \label{fig:final_YvsA}
\end{figure}
\begin{figure}
    \centering
    \includegraphics[width=\linewidth]{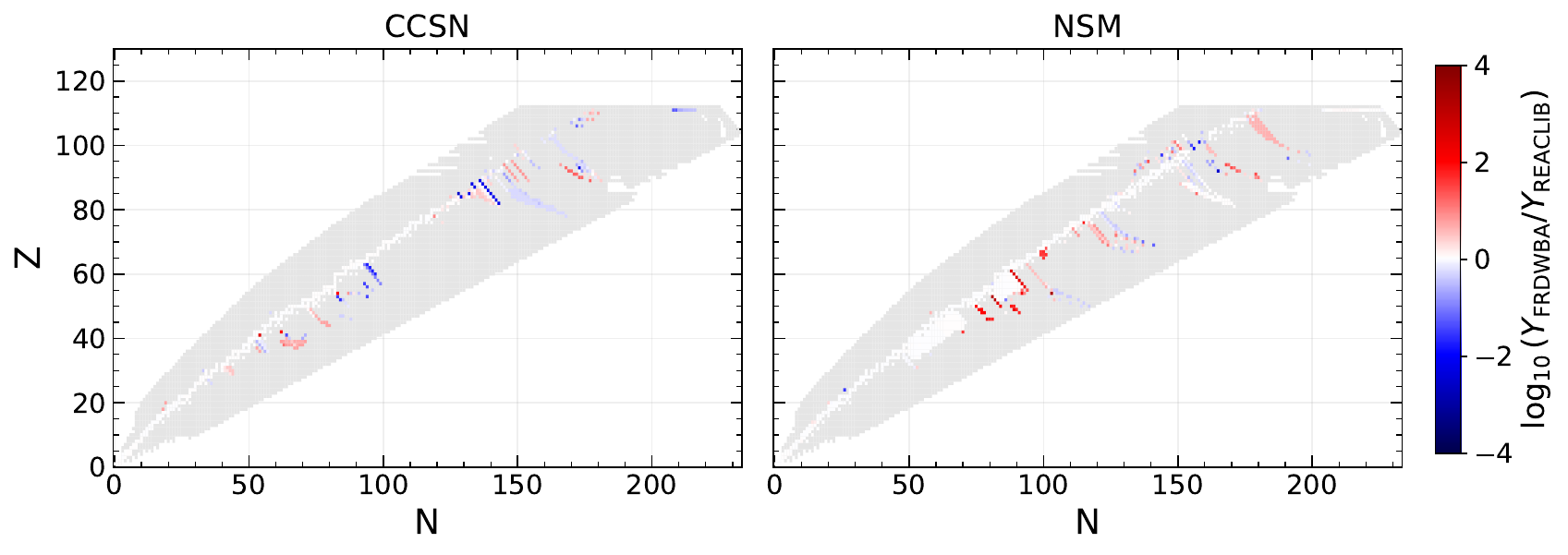}
    \caption{The logarithmic abundance ratios shown for the CCSN and NSM cases. Each square represents a nucleus, and the color scale indicates the logarithmic ratio between the abundances obtained by incorporating the FRDWBA reaction rates for the selected nuclei while retaining all other rates from the JINA-REACLIB database $(Y_{\mathrm{FRDWBA}})$, and those obtained using only JINA-REACLIB rates $(Y_{\mathrm{REACLIB}})$. The ratios are shown only for nuclei with abundances greater than $10^{-30}$.}
    \label{fig:abun_diff}
\end{figure}
We consider two astrophysical environments for the $r$-process: {core-collapse supernova (CCSN)} and {neutron star merger (NSM)}. The $r$-process nucleosynthesis in these sites are primarily characterized by three parameters, the electron fraction, $Y_e$, the entropy, $s$, and the dynamical expansion timescale, $\tau$. Among these, the electron fraction $Y_e$ is arguably the most critical parameter. It determines the neutron-richness of the medium. A lower $Y_e$ value means the material is highly neutron-rich, and a hallmark condition for successful r-process nucleosynthesis. The entropy is another essential quantity that dictates the compositions in nuclear statistical equilibrium (NSE) and controls the neutron-to-seed ratio. A high entropy condition is important for a successful $r$-process, which is a characteristic feature of nucleosynthesis in CCSN~\cite{Arcones_2013}. The expansion timescale dictates how long the neutrons are exposed in the material to produce the $r$-process elements. Thus, an optimal choice of these three parameters is vital to successfully describe nucleosynthesis due to the $r$-process.

In CCSN, $Y_e$ tends to be relatively high (typically close to 0.5) due to the influence of charged-current weak interactions. For a $Y_e=0.4$, winds with entropies 
$150-300\,\rm k_B \hspace{0.5mm} baryon^{-1}$, and expansion timescales as small as 30\,ms are expected to produce a full r-process~\cite{Otsuki_2000, Arcones_2013}. However, these conditions are difficult to observe in realistic nucleosynthesis scenarios. For an NSM, $Y_e$ of dynamical ejecta can range between $0.05-0.5$, depending on if the ejected matter is undergoing weak interactions~\cite{Shibata2019}. In a tidally driven ejecta outflow expanding in a very short dynamical timescale of a few milliseconds, the composition can be very neutron rich with $Y_e\sim0.1$ and $s\sim10$\,$\rm k_B \hspace{0.5mm}baryon^{-1}$ and is capable of producing a full $r$-process nucleosynthesis~\cite{Radice_2018}.

For the CCSN calculation, a high entropy NDW condition with $Y_e=0.42$, $s=142$\,$k_B \rm baryon^{-1}$, $\tau$=5.1\,ms, at a temperature of 9\,GK has been considered following Ref.~\cite{Terasawa_2001}, while for the NSM case, a tidally driven ejecta outflow following Ref.~\cite{Radice_2018} has been adapted. While the {CCSN} condition yields a complete $r$-process pattern in Ref.~\cite{Terasawa_2001}, the SkyNet results, which is based on a larger reaction network with self-consistent evolution, indicate that this condition is insufficient to reproduce the full $r$-process.

{The comparison between the abundances obtained using the modified rates and those calculated with the original JINA-REACLIB rates indicates that the overall final abundance pattern remains largely unchanged. Nevertheless, small deviations appear in the heavier mass region. This occurs mainly because the nuclei for which the reaction rates were modified are relatively less abundant compared to the stable nuclei that dominate the final abundance distribution. However, even these few modified rates introduce noticeable differences across the nuclear chart. To highlight this effect, Fig.~\ref{fig:abun_diff} presents the logarithmic ratio of abundances obtained using FRDWBA rates (for the selected exotic nuclei) combined with JINA-REACLIB rates ($Y_{\mathrm{FRDWBA}}$) relative to those obtained using only JINA-REACLIB rates ($Y_{\mathrm{REACLIB}}$), illustrating the extent to which the FRDWBA rates can alter the resulting abundance distribution. We observe differences in both astrophysical scenarios of up to a few orders of magnitude, particularly in the heavier mass region. Similar effects were also observed in an earlier study with different astrophysical conditions, where the inclusion of FRDWBA rates for five exotic nuclei led to noticeable modifications in the abundance flow toward heavier mass nuclei~\cite{Barman_2024}. This underscores the importance of proper nuclear structure and reaction inputs for estimating reaction rates in network calculations, especially in regions where statistical model estimates are unreliable.} 

%It is also interesting to note that the changes in abundances emerging due to the modified rates significantly affect the $\beta$-decay pathways, as evident from the trend seen in Fig.~\ref{fig:abun_diff}, indicating that such structural inclusion in the network calculation also induce changes in $\beta$-sensitive regions.} 
%%%%%%%%%%%% end of the body
\section{Conclusions and perspectives}
Unlocking the mysteries of atomic nuclei at the drip-lines is a major focus in nuclear physics. These
exotic nuclei are pushing the boundaries of our understanding and revealing new features, especially the phenomenon of nuclear halos, bubbles, and multi-neutron emissions. This review has attempted to highlight how the interplay of nucleon-nucleon interactions and weak binding erodes traditional magic numbers, leading to deformed ground states, extended matter radii, and unconventional density distributions. Through the antisymmetrized molecular dynamics method, we have captured the microscopic underpinnings of these features without any \textit{a priori} assumptions, allowing seamless descriptions of cluster correlations, shell-like configurations, and triaxial deformations in exotic nuclei. The integration with reaction theories—the Glauber model for interaction cross sections at high energies and the FRDWBA for Coulomb dissociation—has proven invaluable in translating these structures into experimentally and theoretically accessible observables.

Applications of these state of the art procedures not only validated the erosion of $N$=20 and $N$=28 shell closures but also illuminated the B-IoI as a crucible for shell evolution, where $pf$-intruding orbitals drive collectivity and halo formation. The astrophysical implications were seen to be equally compelling: sensitivity analyses revealed that extended density tails amplified neutron-capture cross sections in $r$-process paths, directly influencing the production of heavy elements in neutron star mergers. It is, therefore, a reasonable belief that abundance calculations incorporating proper exotic structures can potentially resolve longstanding discrepancies in simulated yields for A$\sim$ 80–120 isotopes, bridging gaps between observations from gravitational wave events like GW170817 \cite{Abbott2017PRL} and the available theoretical models.

To conclude, the neutron rich medium mass $N$ = 20 - 28 island of inversion has been a major player in breaking and expanding the barriers of our understanding of exotic nuclei. Exotic phenomenon like halos, deformation and density distributions have been observed and studied for several decades now \cite{Tanihata1985, Shyam85AP, BOHR, Bert86NPA}, but more questions remain unanswered then the problems solved. For example, a large area of the nuclear chart remains as yet unexplored. Halos have been confirmed from $^6$He (lightest) to $^{37}$Mg (heaviest yet), but many more have been predicted. Could halos actually exist beyond the $pf$-shells and how will the shell structure evolve in such a scenario? How will this halo formation vary for heavier nuclei, especially $A>80$? Speculations are already afoot about the existence of \textit{4n-halos} in neutron rich systems (where $^8$He is a case in point for light mass systems). How would the resulting neutron skin-thickness impact their density profiles and thus, structure? These excess neutrons that cause the neutron `skin' are known to be crucial for understanding the nuclear force in extreme conditions like the neutron stars \cite{Chiru2016}. Now that it is conjectured that even a doubly magic, historically spherical $^{208}$Pb nucleus is deformed \cite{Henderson2025PRL}, how would deformation evolve when traversing the nuclear chart? Could bubble nuclei be more prevalent in the medium and heavy mass regions? What would be the astrophysical implications of such exotic nuclei? Studying medium and heavy exotic nuclear systems would not be easy either due to their weakly bound nature and ephemeral lifetimes. Hence, what are the experimental techniques that could be used or developed to analyze them?

To put things into perspective, the field stands at the threshold of unprecedented advances, where future efforts should be focused on advancing few-body structure and reaction models by explicitly incorporating core-excitation and deformation effects for projectiles exhibiting three-body and more complex (four-body) structures. A step in this direction could be the development of universal few-body models that not only include these intricate structural effects but also bridge the gap between few-body and many-body approaches. Such advancements will strengthen the predictive power of few-body theories in exploring rare isotopes and expand their applicability to interdisciplinary domains, including Efimov physics, ultracold atoms, few-body resonances, and clustering phenomena. One could start by prioritizing extensions of AMD to higher mass nuclei, with $A>40$, coupled with eikonal models incorporating medium effects for more accurate few-body dynamics in exotic nuclei. Such extensions can be used as microscopic inputs to the FRDWBA reaction theory that, in turn, can be used to calculate various inclusive and exclusive reaction observables \cite{Chatterjee2013NPA, Dan2021, Choudhary2024EPJ}. 
{Future studies of deformation effects and intruder configurations in exotic nuclei would also benefit from complementary probes beyond Coulomb breakup reactions on strongly absorptive targets. In particular, gamma-coincident measurements, transfer reactions, knockout reactions, quasi-free scattering, and proton-induced reactions at modern radioactive ion beam facilities can provide additional constraints on single-particle structure, core excitations, and configuration mixing near the drip lines. The combined use of these probes is expected to further improve our understanding of deformation-driven shell evolution and its astrophysical implications.}

Experimental synergies with next-generation facilities— such as HIE-ISOLDE-CERN (Switzerland), SPIRAL2-GANIL (France), RIBF-RIKEN (Japan), FRIB-MSU (USA), FAIR-GSI (Germany), {TRIUMF (Canada),} and SPES-LNL (Italy), among others— should enable precision measurements of longitudinal momentum widths and gamma-correlated breakup, targeting elusive dripline candidates like $^{28}$O, $^{31}$F and $^{40}$Mg, thus expanding our information database in the medium mass region. As a systematic investigation of both total reaction and elastic scattering cross sections across nuclei in various isotopic chains is often desired, the TRIP-S3CAN and MESA projects, which are initiatives at RIBF, RIKEN, aim to perform the measurement of interaction cross sections and elastic scattering. This can unveil the nuclear size, their shell evolution and possible new exotic features across the unexplored regions of deformation. In fact, the analyses presented in this review using the AMD and Glauber model are directly relevant to this experimental goal. Modern developments in theoretical descriptions also call for the need to quantify the uncertainties in the various reaction approaches. This can be achieved by machine learning models or neural networks, but is a non-trivial task requiring significant computational costs. Machine learning to nuclear structure have revealed promising results \cite{Acharya2025PRL,Hu2022NP,Jiang2024PRC,Jiang2025PRC,Acharya2023FP,Muli2025PRL}, but implementations to nuclear reactions are rarer and more difficult to execute \cite{Maldonado2025PRC,Odell2024PRC,Hagino2025PRC}. {Uncertainty quantification via Bayesian inferences require very high-dimensional parameter spaces (typically $\sim 10^4 - 10^6$). Emulator techniques have offered a way forward, providing a model construction that is orders of magnitude computationally cheaper and faster. For example, the eigenvector continuation method works well for uncertainty quantification for the bound states \cite{Sarkar2021PRL,Konig2020PLB}. However, developing reduced basis emulators for reaction approaches is tricky and open questions remain \cite{Lei2026PRC,Cata2026PRC,Furnstahl2020PLB}.} FRDWBA shows a lot of promise {in that regard as it involves minimal adjustable parameters, besides being an elegant} tool to calculate (n,$\gamma$) reaction rates for light and medium mass weakly bound neutron rich nuclei near the driplines, and as such, merging it with a robust neural network {that has constrained uncertainties} should also provide excellent theory support to initiatives like the PANDORA project \cite{Tamii2023EPJA}. Hybrid approaches merging machine learning techniques with variational methods could further expedite generator coordinate mixing, facilitating full-chart extrapolations. The initiation of the deblurring approach using the Richardson-Lucy algorithm \cite{Nzab2023PRC,Tam2025PRC,Nzab2023PLB} is a welcome step that not only is useful theoretically, but also promises better implications for the interpretation of experimental results. Ultimately, these developments promise to map the neutron dripline comprehensively, unravel the limits of nuclear binding, and refine nucleosynthesis simulations, deepening our grasp of matter's genesis from Big Bang nucleosynthesis to cosmic cataclysms.
%%%%%%%%%%%%%%%%%%%%%%%%%%%%%%%%%%%%%%%%%%%%%%%%
\section{Acknowledgments}
This work was supported by the Indo-Japan Cooperative Science Programme (IJCSP)-2023 under the DST-JSPS Bilateral Programs No. JPJSBP120247715 and No. DST/INT/JSPS/P-393/2024(G) [RC, MK, WH]. The work was in part supported by JSPS KAKENHI Grants No. 23K22485, No. 25K07285, No. 25K01005 [WH, MK], and by the U.K. Science and Technology Funding Council (Grant No. ST/V001116/1 and No. ST/Y000323/1) [JS]. [S] acknowledges SERB, DST, India, for a Ramanujan Fellowship (RJF/2021/000176) and {a seed grant from Dr. B. R. Ambedkar NIT Jalandhar.} [RB] acknowledges RIKEN for support via the International Program Associate (IPA) fellowship.

We would like to express our sincere gratitude to our collaborators: R. Shyam, P. Banerjee, Manju, Munna Dan, V. Choudhary, J. Casal, L. Fortunato, N. R. Walet, W. Satula, Nicholas Keeley, Antonio M. Moro and the late Andrea Vitturi, with whom we have had the pleasure of working on these topics.
\bibliography{ref}
\end{document}